\documentclass[11pt]{article}
\usepackage{eurosym}
\usepackage{amsfonts}
\usepackage{amssymb}
\usepackage{graphicx}
\usepackage{amsmath}
\usepackage{makeidx}
\usepackage{indentfirst}
\usepackage[T1]{fontenc}
\usepackage[utf8]{inputenc}

\newcounter{resultnum}[section]
\newcounter{conclusionnum}[section]
\newcounter{conditionnum}[section]
\newcounter{conjecturenum}[section]
\newcounter{examplenum}[section]
\newcounter{exercisenum}[section]
\newcounter{lemmanum}[section]
\newcounter{notationnum}[section]
\newcounter{theoremnum}[section]
\newcounter{definitionnum}[section]
\newcounter{corollarynum}[section]
\newcounter{remarknum}[section]
\newcounter{propositionnum}[section]
\newcounter{acknowledgementnum}[section]
\newcounter{algorithmnum}[section]
\newcounter{axiomnum}[section]
\newcounter{casenum}[section]
\newcounter{claimnum}[section]
\newcounter{summarynum}[section]
\newcounter{problemnum}[section]
\begin{document}

\title{Batalin-Fradkin-Vilkovisky quantization of Einstein gravity with
off-diagonal solutions encoding Ho\v{r}ava type generating functions}
\date{December 27, 2025}
\author{ {\textbf{El\c{s}en Veli Veliev }}\thanks{%
email: elsen@kocaeli.edu.tr and elsenveli@hotmail.com} \\
{\small \textit{\ Department of Physics,\ Kocaeli University, Kocaeli,
41001, T\"{u}rkiye }} \and 
\textbf{Sergiu I. Vacaru} \thanks{%
emails: sergiu.vacaru@fulbrightmail.org ; sergiu.vacaru@gmail.com } \vspace{%
.1 in} \\
{\small \textit{Astronomical Observatory, Taras Shevchenko National
University of Kyiv, Kyiv 01601, Ukraine}} \\
{\small \textit{Department of Physics, California State University at
Fresno, Fresno, CA 93740, USA}} \\
{\small \textit{\ Department of Physics, Kocaeli University, Kocaeli, 41001,
T\"{u}rkiye }}
 \vspace{.1 in} }
\maketitle

\begin{abstract}
We develop and apply the Batalin–Fradkin–Vilkovisky (BFV) formalism for the covariant quantization of generic off-diagonal solutions of the Einstein equations in general relativity (GR). In the classical regime, such nonholonomic configurations are formulated entirely within GR and are characterized by nonlinear symmetries of generating functions, running cosmological constants, integration functions, and effective matter sources. These constructions are further extended to quantum gravity (QG) models involving effective local Lorentz symmetry violations and anisotropic scaling, as realized in Ho\v{r}va–Lifshitz (HL)–type theories. The classical geometric framework is formulated on Lorentz manifolds endowed with nonholonomic 2+2 and  3+1 splitting structures and subsequently generalized to quantum configurations determined by HL-type generating functions. The 2+2 dyadic splitting, incorporating connection distortions, provides a systematic method for constructing exact and parametric classical and quantum solutions described by generating functions and effective sources depending on all spacetime coordinates, physical constants, and anisotropic scaling or deformation parameters. The complementary 3+1 splitting allows for a consistent implementation of the BFV quantization procedure. We demonstrate the renormalizability of off-diagonal quantum HL-type deformations of GR. The resulting classical and quantum nonholonomic BFV models represent viable candidates for asymptotically free theories of gravity and may provide a mechanism for resolving unitarity issues in QG. In appropriate classical limits, the framework reproduces physically relevant off-diagonal GR solutions with or without locally anisotropic scaling, offering potential applications to nonlinear classical and quantum phenomena in accelerating cosmology and dark energy and dark matter physics. 

\end{abstract}

\tableofcontents



\section{Introduction, motivations and objectives}

\label{sec1}

The Batalin-Fradkin-Vilkovisky (BFV) scheme of quantization \cite%
{fv75,bv77,ff78} was developed as a systematic approach to the quantization
of constrained relativistic systems, including gravitational and matter
field theories. The BFV formalism has become one of the fundamental tools
for constructing consistent models of quantum gravity (QG) and has
stimulated significant progress in modern mathematical and theoretical
physics. Comprehensive reviews of geometric and quantum aspects relevant to
the present work are provided in \cite{fr12,barv23,bbd24,vacaru25,vapny24}.%
\footnote{%
The abbreviation "BV"\ is also used in the literature for closely related or
extended quantization methods. A complete chronological account of the
extensive literature on BFV and its applications is beyond the scope of this
paper.} 

\vskip4pt In this work, we apply the BFV quantization methods and develop
the corresponding formalism within the framework of string theory and
general relativity (GR); see the standard monographs \cite%
{blum12,misner73,kramer03} and references therein. Among the various
directions in modern gravity related to modified gravity theories (MGTs) and
quantum gravity (QG), we focus on two main aspects:

\begin{enumerate}
\item The first class of nonassociative MGTs is defined by star-product
R-flux deformations originating from string/M-theory \cite{blum16,asch17}
and formulated in a nonholonomic form \cite{partner06,partner02}, which
allows the construction of general off-diagonal exact and parametric (on the
string and Planck constants) solutions. These models exhibit asymptotic
safety properties \cite{vapny24,weinberg79,niedermaier06} and can be
quantized using the BFV approach within algebraic quantum field theory \cite%
{fr12,vacaru25}. It is worth noting that the parametric nonholonomic
geometric constructions developed in \cite%
{partner02,partner06,vapny24,vacaru25} include GR as a particular
associative and commutative limit, where various classes of physically
relevant solutions can be quantized through suitable modifications of the
BFV formalism.


\item The second class of MGTs involves Ho\v{r}ava-Lifshitz (HL) gravity
models \cite{horava1,horava2,lifschitz} and their recent BFV quantizations 
\cite{fr12,barv23,bbd24}. These constructions assume that at high energies,
QG exhibits anisotropic scaling of time and space, violating Lorentz
invariance and introducing higher-order derivatives only in spatial
directions. Originally motivated by condensed matter and phase-transition
physics \cite{lifschitz}, this approach yields local, unitary, and
potentially renormalizable models consistent in the ultraviolet (UV) regime. 
\end{enumerate}

Paper \cite{vacaru10} analyzed modified dispersion relations in Hořava–Lifshitz gravity and string/brane models, demonstrating that locally anisotropic effects are naturally encoded in Finsler brane constructions. This approach extends to nonassociative classical and quantum gravity \cite{blum16,asch17,partner02,partner06,vapny24,vacaru25}, formulated on (co)tangent Lorentz bundles endowed with R-flux–induced twisted star products, which effectively define noncommutative eight-dimensional phase-space geometries \cite{vacaru18,partner06}. By imposing suitable nonholonomic constraints, associative sectors can be consistently reduced to four-dimensional Lorentz manifolds, with Einstein gravity recovered as a special case. Using the anholonomic frame and connection deformation method (AFCDM)\footnote{A general technique for decoupling and integrating off-diagonal nonlinear partial differential equations.}, we constructed exact off-diagonal solutions of the (modified) Einstein equations in a broad class of modified gravity theories and in general relativity \cite{vv25,vacaru26,vacaru18,partner02}.


There are three important motivations (M) to elaborate on QG models by
considering general classes of solutions of (modified) Einstein equations:

\begin{enumerate}
\item[M1: ] 
BFV quantization can be implemented explicitly for certain classes of gauge and BRST symmetries, including consistent regularization and renormalization schemes involving physical coupling constants. Physically relevant solutions in GR and MGTs are characterized by specific nonlinear symmetries, sets of interaction and integration constants, and corresponding boundary, asymptotic, or Cauchy conditions. For linearized classical models, quantization can be consistently formulated using the associated Lagrangian or Hamiltonian densities for gravitational and matter field interactions. These geometric and quantum methods are particularly effective for exact or parametric solutions with diagonal metrics and high degrees of symmetry, such as spherical, cylindrical, toroidal, or certain Killing-type configurations. In contrast, for generic nonlinear interactions and more general off-diagonal solutions, quantization approaches based on abstract Lagrangian and Hamiltonian formulations remain largely undeveloped, inefficient, and affected by significant conceptual and technical challenges.


\item[M2: ]
The AFDM and related techniques enable the construction of exact and parametric off-diagonal solutions in general relativity and modified gravity theories, including models endowed with nonlinear connection structures. Such solutions are physically relevant for locally anisotropic cosmologies, quasi-stationary black holes and wormholes, as well as dark matter and dark energy scenarios formulated within GR. Off-diagonal configurations are determined by generating functions, effective sources, and nonholonomic constraints, and they naturally encode dynamically induced anisotropic spacetime scaling. Standard perturbative and canonical quantization methods are inadequate for these systems, necessitating a synthesis of advanced geometric methods with appropriately adapted quantization schemes. Quantum properties -- such as renormalizability, asymptotic safety, and nonlocal effects -- can be consistently defined, but they crucially depend on the underlying geometric data and physical input. At the quantum level, generating functions may acquire Ho\v{r}ava--Lifshitz - type behavior, facilitating controlled quantum deformations of GR. Although effective Lagrangian and Hamiltonian formulations can be constructed, quantization must preserve the intrinsic nonlinear symmetries and generating structures. This requires a systematic extension of the BFV formalism adapted to nonholonomic off-diagonal configurations in GR.
 

\item[M3: ]
 Renormalizable Ho\v{r}ava–Lifshitz (HL) quantum gravity models \cite{barv23,bbd24} are usually analyzed using the background field method on flat backgrounds. Extending these techniques -- together with the BFV formalism -- to curved backgrounds requires a substantial reformulation due to the increased geometric and gauge-theoretic complexity. In \cite{vapny24,vacaru25}, we demonstrated that the BV approach \cite{fr12} admits a nonholonomic generalization suitable for the quantization of off-diagonal solutions in nonassociative and noncommutative deformations of GR and other MGTs. This framework enables a synthesis of the AFCDM  \cite{vv25,vacaru26,vacaru18,partner02,partner06} with modern renormalized BFV and HL techniques \cite{svw11,br11,brs12,cgs14,bbd22,bbd23,bbd24,barv23,barv17}. Within this nonholonomic geometric and quantum formalism, quasi-stationary and cosmological GR solutions can be off-diagonally deformed, quantized via a generalized BFV procedure, and rendered renormalizable. Conversely, renormalized off-diagonal quantum configurations can be constrained to recover physically relevant GR solutions in the classical limit. In quasi-classical and locally anisotropic cosmological regimes, such configurations naturally encode quantum corrections and HL-type structures. 
\end{enumerate}


\vskip4pt 
The main goal of this work is to develop a unified synthesis of the AFCDM for GR and MGTs with a generalized BFV formalism encoding HL - type configurations. This framework enables us to establish the renormalizability of a broad class of physically relevant off-diagonal solutions in GR, defined by generating functions and effective sources that encode locally anisotropic (space–time) scaling behavior.

\vskip4pt 
At the classical level, we adhere to the orthodox paradigm of GR, performing all constructions on Lorentz manifolds. Our \textbf{central hypothesis} is that  \emph{a self-consistent approach to quantum gravity must incorporate generic nonlinear off-diagonal interactions of gravitational and matter fields. At high energies, such interactions manifest as effective locally anisotropic, phase - transition - like regimes, analogous to those in HL gravity, which admit BFV quantization. In the quasi-classical limit, these regimes dynamically generate the Einstein or other effective nonholonomic theories modeled on Lorentz manifolds.}

\vskip4pt 
For suitable classes of generating functions supplemented by terms describing local violations of Lorentz invariance, these phases exhibit anisotropic space–time scaling characteristic of HL gravity \cite{horava1,horava2,lifschitz}. In this work, HL deformations are introduced at the quantum level and employed as a quantization mechanism, while the corresponding classical limits are described by nonholonomic configurations within GR. More generally, both classical and quantum off-diagonal deformations encode a rich spectrum of nonlinear structures, including pattern-forming, quasi-periodic, and solitonic configurations \cite{partner02,partner06,vapny24,vacaru25,vacaru26}. Such structures may be nonassociative and/or noncommutative, yet at appropriate energy scales HL-type configurations can be consistently embedded within the GR framework.\footnote{These solutions cannot be diagonalized by coordinate transformations on finite spacetime regions; in 4-dimensional Einstein gravity, the corresponding metrics involve six independent degrees of freedom with coefficients generally depending on all spacetime coordinates. Explicit constructions using only the Levi-Civita (LC) connection are technically impossible; however, applying the AFCDM with suitable auxiliary connections derived from the same metric structure allows one to obtain explicit solutions, from which LC configurations can be extracted via appropriate nonholonomic constraints.}.


\vskip4pt 
Following the above hypothesis and employing a geometric formalism with double nonholonomic fibrations on Lorentz manifolds,\footnote{This framework involves nonholonomic (equivalently, anholonomic, i.e., non-integrable) dyadic 
 2+2 variables, connection distortions, and an Arnowitt–Deser–Misner (ADM) 3+1 splitting \cite{misner73}.} we construct effective projectable HL - type quantum gravity (QG) theories determined by specific classes of solutions in GR. Within this approach, the classical paradigm of GR remains intact at the quasi-classical level. The corresponding quantum constructions and projections involve nonholonomic “hatted” variables that have not been considered in previous treatments of BFV and HL theories. We argue that locally anisotropic configurations, characterized by different scalings of temporal and spatial coordinates, are dynamically generated by off-diagonal nonlinear quantum interactions and nonholonomic geometric flow evolution. Such effective quantum Ho\v{r}ava gravity models can be shown to be asymptotically free in the ultraviolet (UV) limit (involving effective 2+1 configurations), admit fixed points of the renormalization-group flow, and provide a natural coordinate framework for the analysis of asymptotic freedom \cite{vapny24,weinberg79,niedermaier06,fr12,barv23,bbd24,vacaru25}. The resulting quantum HL configurations can always be nonholonomically constrained to reproduce physically important classes of solutions in GR \cite{vv25,vacaru26}.


\vskip4pt 
From a skeptical and pragmatic perspective, a fully self-consistent quantum formulation of GR cannot be achieved using standard Lagrangian or Hamiltonian approaches alone. Such frameworks are intrinsically tailored to perturbatively renormalizable quantum theories and are most effective for classical systems with (at most) quadratic variables, where perturbative quantization procedures of quantum mechanics and quantum field theory apply. Within this setting, physically relevant nonlinear effects can be addressed only as small quantum deformations, leading to perturbative polarizations and renormalizations of physical constants. These methods become inefficient, or even inapplicable, when confronted with genuinely nonlinear theories, including solitonic hierarchies, nonperturbative regimes of QG, or nonlinear phenomena arising in modern cosmology, dark energy (DE), and dark matter (DM) models.

\vskip4pt 
At the classical level, GR admits a rigorous axiomatic and mathematical formulation on Lorentz manifolds, based on the Levi-Civita (LC) connection or certain canonical alternative variables. However, in the absence of direct observational or experimental input, there is currently no canonical prescription for extending GR or MGTs to Planck-scale quantum regimes. Moreover, a mathematically rigorous quantization framework -- supported by an appropriate nonlinear functional analysis -- capable of treating generic nonlinear theories such as GR is still lacking.

\vskip4pt 
In this work, we argue and demonstrate that a nonholonomic synthesis of the AFCDM, together with extensions of quasi-stationary and cosmological GR configurations to quantum HL-type settings, enables the application of the BFV formalism. This approach allows for the consistent quantization of generic off-diagonal solutions in GR. Importantly, at the classical level, our construction does not modify GR into a classical HL theory. Rather, it shows that certain nonlinear gravitational configurations can be quantized in such a way that they exhibit a well-defined quantum behavior of HL type.
\vskip4pt 
The (modified) Einstein equations constitute highly nonlinear systems of partial differential equations (PDEs), making the construction of generic off-diagonal solutions technically demanding. Such solutions depend not only on physical and integration constants but also on generating functions, effective sources, and matter fields. As a result, they naturally describe quasi-stationary or cosmological configurations exhibiting branching phenomena, including phase-space transitions, solitonic hierarchies, and filamentary structures. Many of these configurations admit effective Lagrange -- Hamilton or almost symplectic descriptions, which in turn allow for alternative geometric interpretations of black holes, wormholes, and locally anisotropic cosmologies.

\vskip4pt 
Rather than emphasizing formal Lagrangian or Hamiltonian formulations at an abstract level, a more constructive strategy is to investigate the quantization of specific off-diagonal solutions that encode physically relevant and, in principle, observable gravitational, dark energy, and dark matter effects. The classical properties of such solutions—determined by their generating data, nonlinear symmetries, and anisotropic scaling behavior—govern their compatibility with quantum methods. When effective classical and quantum geometric structures support a self-consistent formulation with potential physical applications, one can go beyond formal path-integral expressions for quantum gravity and develop a diagrammatic formalism. In this approach, nonholonomic canonical “hatted” variables incorporate off-diagonal classical backgrounds and enable controlled nonlinear superpositions and solution-generating symmetries for broad classes of physically relevant configurations.

\vskip4pt 
In this work, we develop a synthesis of the AFCDM with nonholonomic extensions of the BFV quantization approach for HL–type theories. In \cite{vapny24,vacaru25,partner06}, a nonassociative geometric and quantum flow formalism was constructed for asymptotically free MGTs using BV quantization in canonical variables, allowing generic off-diagonal solutions to be encoded as quasi-classical limits of the corresponding quantum models. Here, we concentrate on establishing the renormalizability of broad classes of off-diagonal solutions in general relativity, for which locally anisotropic HL-type configurations are determined by generating functions and effective sources. We further outline how locally anisotropic cosmological solutions, encoding effective (2+1)-dimensional HL dynamics, can be quantized within a nonholonomic BFV framework.

\vskip4pt The main objectives of this paper are organized as follows. In
Section \ref{sec02}, we introduce the geometry of Lorentz manifolds equipped
with nonholonomic 2+2 and 3+1 fibrations, including corresponding
distortions of the LC connection (Obj1). We present a canonical off-diagonal
metric ansatz with 2+2 decomposition, which allows decoupling and
integration of the nonholonomically distorted Einstein equations. We also
demonstrate how HL configurations can be effectively modeled within GR using
appropriate classes of generating functions and sources. 

In Section \ref{sec03}, we perform the BFV quantization of off-diagonal
solutions in GR with nonholonomic HL structures. Starting from classical
nonholonomic ADM configurations, we construct effective HL models and then
quantize them using canonical nonholonomic 3+1 variables, carefully treating
second-class constraints and gauge-fixing conditions (Obj. 2). Particular
attention is given to the nonlinear symmetries of HL structures and their
associated Becchi-Rouet-Stora-Tyutin (BRST) transformations, which
constitute Obj. 3 of this work. 

In section \ref{sec04} (Obj 4), we reformulate the background field method
in N-adapted form and apply it to the BFV quantization of HL structures. We
discuss general renormalization properties of off-diagonal solutions with HL
structures. Additionally, Obj 5 addresses the BFV quantization of
off-diagonal solutions with a (2+1)-dimensional HL configuration and
explores potential applications for quantizing locally anisotropic
cosmological models. 

Finally, in section \ref{sec05} we present conclusions and outline perspectives on nonholonomic BFV quantization for a broad class of MGTs and GR. Appendix \ref{appendixa} provides a concise summary of the AFCDM)for generating off-diagonal, quasi-stationary, and locally anisotropic solutions in GR, including explicit formulas for constructing these classes of solutions. The appendix concludes with technical results concerning N-adapted propagators and the locality of divergences. We emphasize that all quantum expressions systematically encode off-diagonal nonlinear effects into “hat” operators, a feature that is absent in standard BFV-based and quantum gravity approaches.

\section{Nonholonomic 2+2 and 3+1 fibrations on Lorentz manifolds}

\label{sec02} The AFCDM \cite{vv25,vacaru26,vacaru18,partner02} allows the
construction of off-diagonal solutions in GR by employing nonholonomic
dyadic variables with a conventional 2+2 splitting and by distorting the LC
connection to an auxiliary linear connection, which enables the decoupling
and integration of gravitational field equations in rather general forms. In
parallel, the BFV formalism \cite{fv75,bv77,ff78,fr12,barv23,bbd24} for
quantizing GR and Ho\v{r}ava gravity theories has been developed \cite%
{horava1,horava2,lifschitz} using the ADM formalism with a 3+1 splitting 
\cite{misner73}. The aim of this section is to outline the geometry of
Lorentz manifolds endowed with such double nonholonomic fibrations, which
allows both the generation of exact and parametric solutions and the
subsequent application of BFV quantization methods. Appendix \ref{appendixa}
provides the main definitions, formulas, and tables required to implement
the AFCDM in GR and to generate the corresponding classical and quantum HL
deformations. 

\subsection{N-adapted metrics, distortion of connections and geometric
d-objects}

Let us consider a 4-d pseudo-Riemannian manifold $\ V$ of necessary smooth
(differentiability) class enabled with a symmetric metric tensor of
signature $(+++-),$ 
\begin{equation}
\mathbf{g}=g_{\alpha \beta }(u)e^{\alpha }\otimes e^{\beta },  \label{mst}
\end{equation}%
where local coordinates $u=\{u^{\alpha }\}$ for indices $\alpha ,\beta
,...=1,2,3,4,$ and the Einstein convention of summarizing "up-low" repeating
indices is applied. In (\ref{mst}), some co-frames $e^{\alpha }$ are dual to
corresponding frame bases $e_{\alpha }=e_{\ \alpha }^{\underline{\alpha }%
}(u)\partial _{\underline{\alpha }},$ for $\partial _{\underline{\alpha }%
}=\partial /\partial u^{\underline{\alpha }}$ , when underlined indices may
be used to emphasize that we work in a local coordinate basis. We omit such
an underlying if that will not result in ambiguities.\footnote{%
We can also consider transforms to arbitrary frames (tetrads) defined as $%
e_{\alpha ^{\prime }}=e_{\ \alpha ^{\prime }}^{\alpha }(u)e_{\alpha }$ and $%
e^{\alpha ^{\prime }}=e_{\alpha \ }^{\ \alpha ^{\prime }}(u)e^{\alpha }.$ By
definition, such (co) bases are orthonormal if $e_{\alpha \ }^{\ \alpha
^{\prime }}e_{\ \alpha ^{\prime }}^{\beta }=\delta _{\alpha }^{\beta },$
where $\delta _{\alpha }^{\beta }$ is the Kronecker symbol.} 

\subsubsection{Nonholonomic Lorentz manifolds and N-adapted frames}

For a $2+2$ splitting into conventional horizontal (h) and vertical (v)
coordinates (respectively, $x=\{x^{i}\}$ and $y=\{y^{a}\}),$ we can use
indices $j,k,...=1,2$ and $a,b,c,...=3,4;$ and write $u=\{u^{%
\alpha}=(x^{i},y^{a})\}$ as dyadic coordinates on a neighborhood $U\subset
V. $ Such a 2+2 spacetime decomposition can be defined in coordinate-free
form using nonholonomic 2+2 distributions defined by a Whitney sum 
\begin{equation}
\mathbf{N}:\ TV=hV\oplus vV,  \label{ncon}
\end{equation}%
where $TV$ is the tangent bundles and $hV$ and $vV$ are 2-d subspaces. Using
point-by-point decompositions of the tangent bundles, $TV:=\bigcup%
\nolimits_{u}T_{u}V,$ we can say that (\ref{ncon}) defines an N-connection
with as a non-integrable (equivalently, nonholonomic, or anholonomic)
conventional h- and v-splitting. Such a geometric object was used in
coordinate form by E. Cartan \cite{cartan35} (on tangent bundles) and
defined in rigorous mathematical form in \cite{ehresmann55}. On Lorentz
manifolds, N-connections can be considered as nonholonomic fibered
structures.\footnote{%
This different from the Finsler geometry, see details in \cite{vacaru18}
when the N-connections are defined by splitting of type $TTV=hTV\oplus vTV.$
In those geometries with additional velocity and/or momentum-like variables,
it is used also the second tangent bundle $TTV$ and (in some equivalent
forms) the dual constructions for $TT^{\ast }V$), enabled with nonholonomic
distributions defined by a certain splitting with exact sequences.} In local
form, a N-connection (\ref{ncon}) is defined by a set of coefficients $%
N_{i}^{a}(u),$ when 
\begin{equation}
\mathbf{N}=N_{i}^{a}(x,y)dx^{i}\otimes \partial /\partial y^{a}.
\label{nconcoef}
\end{equation}

In our works \cite{vacaru18,partner02,partner06,vv25,vacaru26}, we used the
term \textit{nonholonomic Lorentz manifold} (denoted $\mathbf{V}$) if a
corresponding spacetime in GR, or a MGT, is enabled with a h- and
v-splitting (\ref{ncon}) at least on a neighbourhood $U\subset V.$
Typically, "boldface" symbols are considered to emphasize that certain
spaces or geometric objects are enabled (or adapted) with (to) an
N-connection structure. In brief, we say that such objects and respective
indices are distinguished (d), i.e. adapted to an N-connection and call them
as d-objects, d-vectors, d-tensors, etc.

Using $N_{i}^{a}$ (\ref{nconcoef}), we can define N--elongated
(equivalently, N-adapted) local bases as certain partial derivative
operators, $\mathbf{e}_{\nu },$ and dual (co-) bases, $\mathbf{e}^{\mu },$
used instead of standard differentials. Such local (co) frames are linear on 
$N_{i}^{a},$ 
\begin{eqnarray}
\mathbf{e}_{\nu } &=&(\mathbf{e}_{i},e_{a})=(\mathbf{e}_{i}=\partial
/\partial x^{i}-\ N_{i}^{a}(u)\partial /\partial y^{a},\ e_{a}=\partial
_{a}=\partial /\partial y^{a}),\mbox{ and  }  \label{nader} \\
\mathbf{e}^{\mu } &=&(e^{i},\mathbf{e}^{a})=(e^{i}=dx^{i},\ \mathbf{e}%
^{a}=dy^{a}+\ N_{i}^{a}(u)dx^{i}),  \label{nadif}
\end{eqnarray}%
and, in general, nonholonomic. For instance, a N-elongated basis (\ref{nader}%
) satisfies certain nonholonomic relations 
\begin{equation}
\lbrack \mathbf{e}_{\alpha },\mathbf{e}_{\beta }]=\mathbf{e}_{\alpha }%
\mathbf{e}_{\beta }-\mathbf{e}_{\beta }\mathbf{e}_{\alpha }=W_{\alpha \beta
}^{\gamma }\mathbf{e}_{\gamma },  \label{nonholr}
\end{equation}%
when the (antisymmetric) nontrivial anholonomy coefficients are computed $%
W_{ia}^{b}=\partial _{a}N_{i}^{b},W_{ji}^{a}=\Omega _{ij}^{a}=\mathbf{e}%
_{j}\left( N_{i}^{a}\right) -\mathbf{e}_{i}(N_{j}^{a}).$ In these formulas, $%
\Omega _{ij}^{a}$ define the coefficients of an N-connection curvature $%
\Omega $. Here we note that a $\mathbf{e}_{\alpha }\simeq \partial
_{\alpha}=\partial /\partial u^{\alpha }$ is holonomic if and only if all
anholonomic coefficients $W_{\alpha \beta }^{\gamma }$ are zero. In curved
spacetime coordinates, for holonomic bases, the coefficients $N_{j}^{a}$ may
be non-zero even if all $W_{\alpha \beta }^{\gamma }=0.$ Such nonholonomic
dyadic decompositions are important for formulating the AFCDM.

To elaborate on BFV quantization of gravity theories, we also need a 3+1
splitting and respective ADM\ formalism on Lorentz manifolds \cite{misner73}%
. In this work, we follow such conventions: We consider local frames and
coframes, $e_{\alpha ^{\prime }}=(e_{i^{\prime }},e_{4})$ and $e^{\beta
^{\prime }}=(e^{i^{\prime }},e^{4}),$ were primed indices run values $%
i^{\prime },j^{\prime },...=1,2,3.$ This if we work with some local general
or coordinate frames which are not adapted to a N-connection structure (\ref%
{ncon}). But a 3+1 splitting can be performed for a N-adapted bases (\ref%
{nader}) and (\ref{nadif}). In such cases, we use both boldface symbols and
primed indices with the assumption that we perform a nonholonomic dyadic
decomposition and then (using N-adapted frames) consider a necessary 3+1
splitting. In such cases, we write respectively $\mathbf{e}_{\alpha ^{\prime
}}=(\mathbf{e}_{i^{\prime }},e_{4})$ and $\mathbf{e}^{\beta ^{\prime }}=(%
\mathbf{e}^{i^{\prime }},\mathbf{e}^{4}),$ which emphasize that we use on $%
\mathbf{V}$ a double nonholonomic fibration structure. We can write the
spaces and geometric d-objects as $\mathbf{\acute{V},\acute{T}}$ etc. if
indices are not used. Here we note that the ADM formalism use the so-called
shift and lapse functions (variables), respectively, $\mathbf{N}^{k^{\prime
}}(x^{i^{\prime }},t)$ and $\mathbf{\acute{N}}(x^{i^{\prime }},t),$ when $%
\overrightarrow{x}=\{x^{i^{\prime }\text{ }}\}$ are space-like coordinates
and $u^{4}=y^{4}=t$ is considered as a time-like coordinate. If a 3+1
splitting does not involve a N-connection structure, we write the shift and
lapse coefficients using non-boldface symbols $N^{k^{\prime
}}(x^{i^{\prime}},t)$ and $\acute{N}(x^{i^{\prime }},t).$ Prime labels will
be used if necessary to state that, in principle, we can N-adapt the
geometric objects with 3+1 splitting (such a convention is different from
the notations used in \cite{misner73,fr12,barv23,bbd24}).

\subsubsection{Metric structures adapted to double fibrations}

Using above convention on d-objects, a spacetime metric $\mathbf{g}$ (\ref%
{mst}) can be written equivalently, respectively as a d--metric, $\mathbf{g,}
$ as metric with double fibration, $\mathbf{\acute{g},}$ or as a generic
off-diagonal metric, $\underline{g}_{\alpha \beta },$ using corresponding
formulas 
\begin{eqnarray}
\ \mathbf{g} &=&(hg,vg)=\ g_{ij}(x,y)\ e^{i}\otimes e^{j}+\ g_{ab}(x,y)\ 
\mathbf{e}^{a}\otimes \mathbf{e}^{b}=  \label{dm} \\
\ \mathbf{\acute{g}} &\mathbf{=}&g_{i^{\prime }j^{\prime }}(x,y)\ \mathbf{e}%
^{i^{\prime }}\otimes \mathbf{e}^{j^{\prime }}+g_{44}(x,y)\ \mathbf{e}%
^{4}\otimes \mathbf{e}^{4}  \label{dma} \\
&=&\underline{g}_{\alpha \beta }(u)du^{\alpha }\otimes du^{\beta },
\label{cm}
\end{eqnarray}%
considering that the system of N-adapted coordinates is chosen in a form
when $g_{i^{\prime }4}(x,y)=g_{4i^{\prime }}(x,y)=0.$ We shall use also
notations of type $hg=\{\ g_{ij}\}$ and $\ vg=\{g_{ab}\}.$ The off-diagonal
coefficients in (\ref{dma}) and (\ref{cm}) can be computed respectively if
we introduce the coefficients of (\ref{nadif}) into (\ref{dm}), with a
corresponding regrouping for a coordinate dual basis, and using primed
indices. This way, we obtain 
\begin{equation}
\underline{g}_{\alpha \beta }=\left[ 
\begin{array}{cc}
g_{ij}+N_{i}^{a}N_{j}^{b}g_{ab} & N_{j}^{e}g_{ae} \\ 
N_{i}^{e}g_{be} & g_{ab}%
\end{array}%
\right] =\left[ 
\begin{array}{cc}
g_{i^{\prime }j^{\prime }}+N_{i^{\prime }}N_{j^{\prime }}g_{44} & 
N_{j^{\prime }}g_{44} \\ 
N_{i^{\prime }}g_{44} & g_{44}%
\end{array}%
\right] .  \label{ansatz}
\end{equation}%
We can work with any $\mathbf{g}=\{\underline{g}_{\alpha \beta }\}$ (\ref%
{ansatz}) with coefficients computed with respect to coordinate bases. Such
metrics are generic off-diagonal if the anholonomy coefficients $W_{\alpha
\beta }^{\gamma }$ (\ref{nonholr}) are not identical to zero. On 4-d
(pseudo) Riemannian spaces, such a metric contains, in general, 6
independent coefficients and can't be diagonalized via coordinate transforms
in a local spacetime region. To apply the AFCDM for constructing
off-diagonal solutions is convenient to use parameterizations of type (\ref%
{dm}). Thermodynamic models and canonical approaches to quantization
following the ADM formalism are elaborated for d-metrics of type (\ref{dma}%
). 

\subsubsection{Distinguished connections and fundamental geometric objects}

A general linear (affine) connection structure $D$ on $V$ can be defined in
independent form from $\mathbf{g}$ (\ref{mst}), which defines a
metric-affine structure $(\mathbf{g},V),$ see \cite{misner73} for abstract
geometric constructions and GR. On a nonholonomic Lorentz manifold $\mathbf{%
V,}$ we can consider a class of distinguished (d) connections which are
adapted to a prescribed N-connection structure.

We define \textbf{\ d--connection} $\mathbf{D}=(hD,vD)$ is a linear
connection preserving under parallelism the N--connection splitting (\ref%
{ncon}). It allows us to define, for instance, a covariant N--adapted
derivative $\mathbf{D}_{\mathbf{X}}\mathbf{Y}$ of a d--vector field $\mathbf{%
Y}=hY+vY$ in the direction of a d--vector $\mathbf{X}=hX+vC.$

For N--adapted frames (\ref{nader}) and (\ref{nadif}), any covariant
d-derivative $\mathbf{D}_{\mathbf{X}}\mathbf{Y}$ can be computed in
coefficient form \cite{vacaru18} with h- and v-indices, 
\begin{equation}
\mathbf{D}=\{\mathbf{\Gamma }_{\ \alpha \beta }^{\gamma }=(L_{jk}^{i},\acute{%
L}_{bk}^{a};\acute{C}_{jc}^{i},C_{bc}^{a})\},\mbox{ where }hD=(L_{jk}^{i},%
\acute{L}_{bk}^{a})\mbox{ and }vD=(\acute{C}_{jc}^{i},C_{bc}^{a}),
\label{hvdcon}
\end{equation}%
where $i,j,...=1,2$ and $a,b,...=3,4.$ Such coefficients allow us to perform
a covariant differential and integral calculus in N-adapted form.

Any d--connection $\mathbf{D}$ is characterized by three fundamental
geometric d-objects: 
\begin{eqnarray}
\mathcal{T}(\mathbf{X,Y})&:= &\mathbf{D}_{\mathbf{X}}\mathbf{Y}-\mathbf{D}_{%
\mathbf{Y}}\mathbf{X}-[\mathbf{X,Y}],\mbox{ torsion d-tensor,  d-torsion};
\label{fundgeom} \\
\mathcal{R}(\mathbf{X,Y})&:= &\mathbf{D}_{\mathbf{X}}\mathbf{D}_{\mathbf{Y}}-%
\mathbf{D}_{\mathbf{Y}}\mathbf{D}_{\mathbf{X}}-\mathbf{D}_{\mathbf{[X,Y]}},%
\mbox{ curvature d-tensor, d-curvature};  \notag \\
\mathcal{Q}(\mathbf{X})&:= &\mathbf{D}_{\mathbf{X}}\mathbf{g,}%
\mbox{nonmetricity d-fiels, d-nonmetricity}.  \notag
\end{eqnarray}%
We say that any geometric data $\left( \mathbf{V},\mathbf{N},\mathbf{g,D}%
\right) $ define a nonholonomic, i.e. N-adapted, metric-affine structure on
a Lorentz manifold $V.$\footnote{%
Introducing $\mathbf{X}=\mathbf{e}_{\alpha }$ and $\mathbf{Y}=\mathbf{e}%
_{\beta }$ as in (\ref{nader}) and $\mathbf{D}=\{\mathbf{\Gamma }_{\ \alpha
\beta }^{\gamma }\}$ (\ref{hvdcon}) into (\ref{fundgeom}), we compute in
explicit form the N-adapted coefficients of corresponding fundamental
geometric objects: $\mathcal{T}=\{\mathbf{T}_{\ \alpha \beta }^{\gamma
}=\left( T_{\ jk}^{i},T_{\ ja}^{i},T_{\ ji}^{a},T_{\ bi}^{a},T_{\
bc}^{a}\right) \}$; \newline
$\mathcal{R}=\mathbf{\{R}_{\ \beta \gamma \delta }^{\alpha }=\left( R_{\
hjk}^{i}\mathbf{,}R_{\ bjk}^{a}, R_{\ hja}^{i}, R_{\ bja}^{c}\mathbf{,}R_{\
hba}^{i},R_{\ bea}^{c}\right) \};$ $\mathcal{Q}=\mathbf{\{Q}_{\ \alpha \beta
}^{\gamma }=\mathbf{D}^{\gamma }\mathbf{g}_{\alpha \beta }=(Q_{\
ij}^{k},Q_{\ ij}^{c},Q_{\ ab}^{k},Q_{\ ab}^{c})\}$. Such technical details
can be found in \cite{vacaru18,partner02}.} 

We emphasize that using a metric structure $\mathbf{g,}$ we can always
define the Levi-Civita (LC) connection $\nabla ,$ defined by the metric
compatibility condition, $\mathcal{Q}[\nabla ]=\nabla \mathbf{g=0,}$ and the
zero torsion condition, $\mathcal{T}[\nabla ]=0.$ In such abstract geometric
formulas, $[\nabla ]$ states that corresponding geometry objects consist
certain functionals of $\nabla .$ The abstract and N-adapted coefficient
formulas for the curvature $\mathcal{R}[\nabla ]=\ _{\nabla }\mathcal{R}%
=\{R_{\ \beta \gamma \delta }^{\alpha }=\ _{\nabla }R_{\ \beta \gamma \delta
}^{\alpha }\}$ computed as in (\ref{fundgeom}), see details in \cite%
{misner73,vacaru18,partner02}. N-adapted metric-affine structures on a
nonholonomic Lorentz manifold $\mathbf{V}$ can be characterized by
corresponding distortion relations,%
\begin{equation}
\mathbf{D}=\nabla +\mathbf{Z,}  \label{distr}
\end{equation}%
where $\mathbf{Z}$ is the \textbf{distortion d-tensor} containing
contributions from the nontrivial nonolonomic structure (determined by $%
\mathbf{N}$) and torsion $\mathcal{Q}$ and nonmetricity $\mathcal{T}$
fields. Introducing (\ref{distr}) into (\ref{fundgeom}), we compute
respective distortions relations between the Riemann curvature tensors, 
\begin{equation}
\mathcal{R}[\mathbf{D}]=\ _{\nabla }\mathcal{R+Z}[\nabla ,\mathbf{Z}].
\label{curvdist}
\end{equation}%
We conclude that a large class of N-adapted metric-affine geometries can be
modelled as effective theories on a nonholonomic Lorentz manifold. For
physical applications, we can state a physical causal structure determined
by pseudo-Riemannian data ($\mathbf{g,}\nabla ).$ More general geometric
theories can be constructed by using bi-connection/ distortion structures $%
(\nabla ,\mathbf{D}=\nabla +\mathbf{Z}).$ Classical MGTs are constructed by
postulating respective Lagrangians, or Hamiltonians, as distortions and
nonlinear coupling of respective values in GR. 

Distortions of d-connections $\mathbf{Z}$ (\ref{distr}) and of Riemann d-tensors $\mathcal{Z}$ (\ref{curvdist}) can be formulated in such a way that a classical gravity theory remains within the geometric framework of nonholonomic Lorentz manifolds, provided the $(\mathbf{g,}\nabla ,\ _{\nabla }\mathcal{R})$ reproduce the standard Einstein equations of GR. In an analogous manner, one may employ the data $(\mathbf{g,}\mathcal{R}[\mathbf{D}])$ to generate equivalent Einstein equations expressed in nonholonomic variables, or, alternatively,  to select appropriate distortions d-tensors $\mathbf{Z}$ for the construction of  a classical HL-type gravity theory.

 For the purposes of this work, we can consider distortions when with respect to certain nonholonomic "hatted" variables (introduced in the next subsection), such that  $(\widehat{\mathbf{g}}\mathbf{,}\mathcal{R}[\widehat{\mathbf{D}}])\,\rightarrow (\ ^{HL}\widehat{%
\mathbf{g}}, \ ^{HL}\mathcal{R}[\ ^{HL}\widehat{\mathbf{D}}])$, where the latter encode  quantum deformations of the classical data $(\widehat{\mathbf{g}},\mathcal{R}[\widehat{\mathbf{D}}])\,$\ into certain HL-type quantum models. We shall show that,  in the classical limit,  the resulting theory can be transformed equivalently into the Einstein theory, while its quantum phases exhibit the characteristic features of HL gravity theories.

\subsection{The (modified) Einstein equations in canonical dyadic variables}

The GR theory can be formulated equivalently using different types of
geometric data. Typically $[\mathbf{g},\nabla ]$ are taken in certain
standard forms; or $[\mathbf{g},\mathbf{N},\mathbf{D}],$ when certain (\ref%
{distr}) are used, for instance, to redefine the theory in a teleparallel
form, or to elaborate on new methods of constructing solutions and
formulating certain methods of quantization. In this subsection, we show how
to define an auxiliary d-connection which allows us to decouple and
integrate the (modified) Einstein equations in certain general forms, i.e.
to formulate the AFCDM. 

\subsubsection{The canonical d-connection}

Let us consider a d-metric structure $\mathbf{g}$ (\ref{dm}) and define two
important linear connection structures which satisfy such geometric
properties: 
\begin{equation}
(\mathbf{g,N})\rightarrow \left\{ 
\begin{array}{cc}
\mathbf{\nabla :} & \mathbf{\nabla g}=0;\ _{\nabla }\mathcal{T}=0,\ 
\mbox{\
the LC--connection }; \\ 
\widehat{\mathbf{D}}: & \widehat{\mathbf{Q}}=0;\ h\widehat{\mathcal{T}}=0,v%
\widehat{\mathcal{T}}=0,\ hv\widehat{\mathcal{T}}\neq 0,%
\mbox{ the canonical
d-connection}.%
\end{array}%
\right.  \label{twocon}
\end{equation}
We can consider any parametrization (\ref{dma}) and (\ref{cm}) but formulas
become more simpler if we use nonholonomic dyadic variables. The LC
connection is the standard one in (pseudo) Riemannian geometry, but it is
not suitable for our purposes because it does not allow a general decoupling
of (modified) gravitational equations for generic off-diagonal metrics.

In our works, we prefer to work with the "hat" d-connection $\widehat{%
\mathbf{D}},$ which in GR can be considered as an auxiliary one which allows
to construct off-diagonal solutions. It is related to the LC connection by a
canonical distortion relation of type (\ref{distr}), 
\begin{equation}
\widehat{\mathbf{D}}[\mathbf{g}]=\nabla \lbrack \mathbf{g}]+\widehat{%
\mathcal{Z}}[\mathbf{g,N}].  \label{canondistrel}
\end{equation}%
In these formulas, $\widehat{\mathcal{Z}}=\{\widehat{\mathbf{Z}}_{\ \alpha
\beta }^{\gamma }\}$ is the canonical distortion d-tensor which is
completely defined by the d-metric $\mathbf{g}$ a prescribed $\mathbf{N}$.%
\footnote{%
A canonical d-connection from (\ref{twocon}) is defined by N-adapted
coefficients $\widehat{\mathbf{D}}=\{\widehat{\mathbf{\Gamma }}_{\ \alpha
\beta }^{\gamma }=(\widehat{L}_{jk}^{i},\widehat{L}_{bk}^{a},\widehat{C}%
_{jc}^{i},\widehat{C}_{bc}^{a})\}$ computed for a d--metric $\mathbf{g}%
=[g_{ij},g_{ab}]$ (\ref{dm}) using N--elongated partial derivatives (\ref%
{nader}). In explicit form \cite{vv25,vacaru26,vacaru10,vacaru18,partner02}, 
\begin{eqnarray*}
\widehat{L}_{jk}^{i} &=&\frac{1}{2}g^{ir}(\mathbf{e}_{k}g_{jr}+\mathbf{e}%
_{j}g_{kr}-\mathbf{e}_{r}g_{jk}),\widehat{L}_{bk}^{a}=e_{b}(N_{k}^{a})+\frac{%
1}{2}g^{ac}(\mathbf{e}_{k}g_{bc}-g_{dc}\ e_{b}N_{k}^{d}-g_{db}\
e_{c}N_{k}^{d}), \\
\widehat{C}_{jc}^{i} &=&\frac{1}{2}g^{ik}e_{c}g_{jk},\ \widehat{C}_{bc}^{a}=%
\frac{1}{2}g^{ad}(e_{c}g_{bd}+e_{b}g_{cd}-e_{d}g_{bc}).
\end{eqnarray*}%
In a similar form, we can compute the coefficients of an LC connection $%
\nabla =\{\Gamma _{\ \alpha \beta }^{\gamma }\}$ and the N-adapted
coefficients of the canonical distortion d-tensor from (\ref{canondistrel}), 
$\widehat{\mathbf{Z}}=\{\widehat{\mathbf{Z}}_{\ \alpha \beta }^{\gamma }=%
\widehat{\mathbf{\Gamma }}_{\ \alpha \beta }^{\gamma }-\Gamma _{\ \alpha
\beta }^{\gamma }\}.$}

Using $\widehat{\mathbf{D}}=\{\widehat{\mathbf{\Gamma }}_{\ \alpha
\beta}^{\gamma }\}$\ (\ref{canondistrel}) instead of the coefficients of a
general d-connection $\mathbf{D=\{\Gamma }_{\ \alpha \beta }^{\gamma }\}$,
we can compute the N-adapted coefficients of canonical fundamental
d--objects (\ref{fundgeom}). Such hat symbols are used, for instance, for $%
\widehat{\mathcal{R}}=\{\widehat{\mathbf{R}}_{\ \beta \gamma \delta
}^{\alpha }=(\widehat{R}_{\ hjk}^{i},\widehat{R}_{\ bjk}^{a},...)\};$ for $%
\widehat{\mathcal{T}}= \{\widehat{\mathbf{T}}_{\ \alpha \beta }^{\gamma }=(%
\widehat{T}_{\ jk}^{i},\widehat{T}_{\ ja}^{i},...)\}$ induced
nonholonomically by the d-metric and N-connection coefficients; and for $%
\widehat{\mathcal{Q}}=\{\widehat{\mathbf{Q}}_{\gamma \alpha \beta} =(%
\widehat{Q}_{kij}=0,\widehat{Q}_{kab}=0)=0.$ The canonical distortion
relations for linear connections (\ref{canondistrel}) allow us to compute
respective canonical distortions of the fundamental geometric objects
determined by $\nabla $. Such formulas relate, for instance, two different
curvature tensors, $\ _{\nabla }\mathcal{R}=\{\ _{\nabla }R_{\ \beta \gamma
\delta }^{\alpha }\}$ and $\widehat{\mathcal{R}}=\{\widehat{\mathbf{R}}_{\
\beta \gamma \delta }^{\alpha }\}$, etc., which is similar to (\ref{curvdist}%
) written in hat-variables, $\widehat{\mathcal{R}}[\widehat{\mathbf{D}}]=\
_{\nabla }\mathcal{R}+ \widehat{\mathcal{Z}}[\nabla ,\widehat{\mathbf{Z}} ]$.

\subsubsection{The Einstein d-tensor in canonical dyadic variables}

In this work, the canonical Ricci d-tensor is introduced by contracting the
1st and 4th indices of the canonical curvature d-tensor,%
\begin{equation}
\widehat{\mathbf{R}}ic=\{\widehat{\mathbf{R}}_{\ \beta \gamma }:=\widehat{%
\mathbf{R}}_{\ \beta \gamma \alpha }^{\alpha }\}.  \label{criccidt}
\end{equation}%
In general, this d-tensor is not symmetric, $\widehat{\mathbf{R}}_{\ \beta
\gamma }\neq \widehat{\mathbf{R}}_{\ \gamma \beta }$ (for nonholonomic
geometric objects this is typical). The canonical scalar curvature is
defined in standard form by contracting (\ref{criccidt}) with the inverse
d-metric, $\mathbf{g}^{\alpha \beta },$ when 
\begin{equation}
\widehat{R}sc:=\mathbf{g}^{\alpha \beta }\widehat{\mathbf{R}}_{\ \alpha
\beta }.  \label{criccidsc}
\end{equation}%
The (nonholonomic) canonical Einstein d-tensor can be introduced in abstract
geometric form as in \cite{misner73} but hat-labelled geometric d-objects (%
\ref{criccidt}) and (\ref{criccidsc}), 
\begin{equation}
\widehat{\mathbf{E}}n:=\widehat{\mathbf{R}}ic-\frac{1}{2}\mathbf{g}\widehat{R%
}sc=\{\widehat{\mathbf{R}}_{\ \beta \gamma }-\frac{1}{2}\mathbf{g}_{\ \beta
\gamma }\widehat{R}sc\}.  \label{ceinstdt}
\end{equation}

The d-tensors $\widehat{\mathbf{R}}ic[\widehat{\mathbf{D}}]$ and $\widehat{%
\mathbf{E}}n[\widehat{\mathbf{D}}]$ consist the respective distortions of
the Ricci and Einstein tensors for the LC connection, $Ric[\mathbf{\nabla }]$
and $En[\mathbf{\nabla }],$ when $Rsc[\mathbf{\nabla }]=R.$ Such distortions
can be computed in explicit form using (\ref{canondistrel}).

\subsubsection{Effective sources and generating sources}

We can introduce conventional gravitational and matter fields Lagrange
densities, $\ ^{g}L(\widehat{\mathbf{R}}ic)$ (similarly to GR with $\
^{g}L(R)$) and postulate for matter fields, using a left label $m,$ a $\
^{m}L(\varphi ^{A},\mathbf{g}_{\beta \gamma }).$ The stress-energy d-tensor
of matter fields $\varphi ^{A}$ (labelled by a general index $A$) is defined
and computed as in GR but with N-adapted coefficients defined by respective
dyadic decompositions, 
\begin{equation}
\mathbf{T}_{\alpha \beta }=-\frac{2}{\sqrt{|\mathbf{g}_{\mu \nu }|}}\frac{%
\delta (\ ^{m}L\sqrt{|\mathbf{g}_{\mu \nu }|})}{\delta \mathbf{g}^{\alpha
\beta }}.  \label{emdt}
\end{equation}%
To elaborate on standard physical theories, we should consider more general $%
\ ^{m}\widehat{L}(\varphi ^{A},\mathbf{g}_{\beta \gamma },\widehat{\mathbf{D}%
},...)$ involving on some covariant spinor derivatives, gauge fields, etc.
For simplicity, we do not consider such theories with more sophisticated
energy-momentum d-tensors $\widehat{\mathbf{T}}_{\alpha \beta}. $ Using (\ref%
{emdt}), we define the trace $T:=\mathbf{g}^{\alpha \beta }\mathbf{T}%
_{\alpha \beta }$ and construct certain matter sources of Ricci d-tensors, $%
\widehat{\mathbf{Y}}[\mathbf{g,}\widehat{\mathbf{D}}]\simeq \{\mathbf{T}%
_{\alpha \beta }-\frac{1}{2}\mathbf{g}_{\alpha \beta }T\}.$ If $\widehat{%
\mathbf{D}}$ is used in $\ ^{m}\widehat{L},$ a corresponding source $%
\widehat{\mathbf{Y}}$ includes certain terms determined by the coefficients
of distortion d-tensors.

We can apply the AFCDM and generate exact or parametric solutions for
(effective) sources $\widehat{\mathbf{Y}}[\mathbf{g,}\widehat{\mathbf{D}}%
]=\{\Upsilon _{~\delta }^{\beta }(x,y)\}$ determined by \textbf{two
generating sources} $\ ^{h}\Upsilon (x^{k})$ and $\ ^{v}\Upsilon
(x^{k},y^{a}).$ This means that using certain frame transforms of $\Upsilon
_{\beta \delta }\rightarrow \widehat{\Upsilon }_{\beta \delta }$, we can
write 
\begin{equation}
\widehat{\Upsilon }_{~\delta }^{\beta }=diag[\Upsilon _{\alpha }:\Upsilon
_{~1}^{1}=\Upsilon _{~2}^{2}=~^{h}\Upsilon (x^{k});\Upsilon
_{~3}^{3}=\Upsilon _{~4}^{4}=~^{v}\Upsilon (x^{k},y^{a})],  \label{esourc}
\end{equation}%
where the coefficients are defined with respect to N-adapted frames (\ref%
{nader}) and (\ref{nadif}). The assumption that $\widehat{\mathbf{Y}}$ is of
type (\ref{esourc}) imposes certain nonholonomic constraints on $\mathbf{T}%
_{\alpha \beta }$, a possible cosmological constant $\Lambda $ and more
special splitting of constants into h- and v-components, and the type of
nonholonomic 2+2 splitting. Such constraints involve distortion d-tensors $%
\widehat{\mathbf{Z}}[\mathbf{g}]$ and other values included in $\widehat{%
\mathbf{Y}}.$ The generating sources (\ref{esourc}) allow us to decouple and
integrate in general forms the geometric flows and gravitational and matter
field equations in GR and MGTs as we proved in \cite%
{vv25,vacaru26,vacaru18,partner02,partner06}. In explicit form, physically
important generic off-diagonal solutions can be generated, for instance, if
we $\widehat{\mathbf{Y}}[\mathbf{g,}\widehat{\mathbf{D}},\kappa ]$ contains
a small parameter $\kappa $, or if the gravitational and matter field
dynamics is subjected to certain convenient classes of constraints, trapping
hypersurface conditions, ellipsoid symmetries etc. In such cases, the
solutions can be found in exact form, or generated recurrently using power
decompositions $\kappa ^{0},\kappa ^{1},\kappa ^{2},...$ For various
purposes, $\kappa $ can be related to a gauge field source, anisotropy
parameter, quantum deformations, ..., or another parameter for constructing
ellipsoid deformations, etc. We say that the corresponding classes of
solutions are exact or parametric. 

\subsubsection{Nonholonomic Einstein equations and the LC conditions}

The Einstein equations can be formulated equivalently using the geometric
data $[\mathbf{g},\nabla ]$ or $[\mathbf{g},\mathbf{N},\widehat{\mathbf{D}}%
]. $ We can follow an abstract geometric approach or a variational
formulation as in \cite{misner73} (in general, working with N-adapted
structures). The dyadic hat variables were used in our and co-authors' works
for elaborating the AFCDM \cite{vv25,vacaru26,vacaru18,partner02,partner06}.
That geometric and analytic method allows to decouple and integrate the
Einstein equations in a certain general off-diagonal form using the
corresponding ansatz for d-metric (\ref{dm}). This is not possible if we
work directly off-diagonal ansatz $\mathbf{g}=\{\underline{g}_{\alpha \beta
}\}$ (\ref{ansatz}), or certain 3+1 decompositions (\ref{dma}).

Using the canonical Einstein d-tensor (\ref{ceinstdt}) and the
energy-momentum d-tensor \ (\ref{emdt}), we postulate the Einstein equations
in hat variables, 
\begin{equation}
\widehat{\mathbf{E}}n_{\alpha \beta }=\varkappa \widehat{\mathbf{T}}_{\alpha
\beta }.  \label{einstceq1}
\end{equation}%
In these formulas, the constant $\varkappa $ can be related to the Newton
gravitational constant. Such nonholonomic equations can be equivalent to the
standard Einstein ones in GR if $\widehat{\mathbf{T}}_{\alpha \beta }$ is
constructed in a corresponding way to include as sources the energy-momentum
tensors for matter (in N-adapted bases) but also the distortion terms coming
from $En_{\alpha \beta }[\nabla ]$ and $T_{\alpha \beta }[\nabla ].$
Necessary type distortions are determined by formulas (\ref{canondistrel}).
Alternatively, we obtain an equivalence with the gravitational field
equations in GR if we impose, additionally, certain nonholonomic constraints
for extracting LC configurations: 
\begin{equation}
\widehat{\mathbf{Z}}=0,\mbox{ which is equivalent to }\ \widehat{\mathbf{D}}%
_{\mid \widehat{\mathcal{T}}=0}=\nabla .  \label{lccond}
\end{equation}

The Einstein equations (\ref{einstceq1}) can be written in a form with
canonical Ricci d-tensor on the left, which is more convenient for
decoupling and generating off-diagonal solutions: 
\begin{eqnarray}
\widehat{\mathbf{R}}_{\ \ \beta }^{\alpha } &=&\widehat{\mathbf{\Upsilon }}%
_{\ \ \beta }^{\alpha },  \label{cdeq1} \\
\widehat{\mathbf{T}}_{\ \alpha \beta }^{\gamma } &=&0,%
\mbox{ if we extract
LC configuations with }\nabla .  \label{lccond1}
\end{eqnarray}%
In this system of nonlinear PDEs, the generating sources are of type $%
\widehat{\mathbf{\Upsilon }}_{\ \ \beta }^{\alpha }=[\ ^{h}\Upsilon \delta
_{\ \ j}^{i},\ ^{v}\Upsilon \delta _{\ \ b}^{a}]$ (\ref{esourc}) and the
equations (\ref{lccond1}) are equivalent to (\ref{lccond}). 

Let us discuss the conservation laws for the canonical distorted Einstein
equations (\ref{cdeq1}). We note that, in general, $\widehat{\mathbf{D}}
^{\beta }\widehat{\mathbf{E}}_{\ \ \beta }^{\alpha }\neq 0$ and $\widehat{%
\mathbf{D}}^{\beta }\widehat{\mathbf{\Upsilon }}_{\ \ \beta }^{\alpha }\neq
0.$ This is different from the Einstein and energy-momentum tensors written
in standard form using $\nabla $ and respective Bianchi identities. In
principle, this is not a problem because non-zero covariant divergences are
typical for nonholonomic systems. For instance, in nonholonomic/ continuous
mechanics, the conservation laws are not standard ones but involve Lagrange
multipliers associated with certain classes of nonholonomic constraints.
Solving the constraint equations (if it is possible in an explicit form), we
can redefine the variables and introduce new effective Lagrangians. Then, we
can define standard conservation laws. This can be performed in explicit
form only for some "toy" models with effective and real matter fields in GR.
Finally, we note that using nonholonomic canonical variables and distortions
of connections (\ref{canondistrel}), we can rewrite (\ref{cdeq1})
equivalently in terms of $\nabla ,$ when $\nabla ^{\beta }E_{\ \ \beta
}^{\alpha }=\nabla ^{\beta }T_{\ \ \beta }^{\alpha }=0.$ 

\subsection{General ansatz and generating data for off-diagonal solutions in
GR}

\label{ssdecintei}We proved in our former works (see reviews \cite%
{vv25,vacaru26,vacaru18,partner02,partner06}) that the canonical distorted
Einstein equations (\ref{cdeq1}) can be decoupled and integrated in very
general off-diagonal forms. In Appendix \ref{appendixa}, we provide
necessary formulas and details. In this subsection, we state certain general
ansatz which are important for generating quasi-stationary and locally
anisotropic cosmological solutions. We explain how such solutions may
describe off-diagonal deformations of certain primary d-metrics (in general,
they may not be solutions of the Einstein equations) into certain classes of
solutions of (\ref{cdeq1}) determined by so-called gravitational
polarizations (as generating functions) and generating sources. Respective
nonlinear and dual symmetries and certain conditions of extracting LC
configurations are discussed. All geometric constructions are performed on
nonholonomic Lorentz manifolds when the respective generating functions and
generating sources can be chosen to define target solutions possessing
necessary scaling and anisotropic properties. This allows a self-consistent
and renormalizable BFV quantization of such classes of off-diagonal
solutions. 

\subsubsection{Decoupling properties in canonical nonholonomic variables}

\label{mainoffdans}For the purposes of this work, we consider two general
important ansatz for d-metrics (\ref{dm}) which allow us to construct
off-diagonal solutions in GR\ and MGTs:

\begin{description}
\item[Quasi-stationary solutions ] are generated by off-diagonal ansatz with
Killing symmetry on the time-like coordinate $\partial _{4}=\partial _{t}$
(i.e. not depending on coordinate $y^{4}=t),$%
\begin{eqnarray}
\mathbf{\hat{g}} &=&g_{i}(x^{k})dx^{i}\otimes dx^{i}+h_{3}(x^{k},y^{3})%
\mathbf{e}^{3}\otimes \mathbf{e}^{3}+h_{4}(x^{k},y^{3})\mathbf{e}^{4}\otimes 
\mathbf{e}^{4},  \notag \\
&&\mathbf{e}^{3}=dy^{3}+w_{i}(x^{k},y^{3})dx^{i},\ \mathbf{e}%
^{4}=dy^{4}+n_{i}(x^{k},y^{3})dx^{i}.  \label{dmq}
\end{eqnarray}%
The N-connection coefficients are denoted in a canonical form, $\widehat{N}%
_{i}^{3}=w_{i}(x^{k},y^{3})$ and $\widehat{N}_{i}^{4}=n_{i}(x^{k},y^{3}),$
and the coefficients of a d-metric $\widehat{\mathbf{g}}_{\alpha \beta }=[%
\widehat{g}_{ij}(x^{\kappa }),\widehat{g}_{ab}(x^{\kappa },y^{3})]$ are
considered as functions of the necessary smooth class. Such a
parametrization can be obtained using some frame or coordinate transforms
even, in general, such a $\mathbf{\hat{g}}(u)$ may depend on all spacetime
coordinates, see explanations related to formulas (\ref{qeltorsoffd}) in
Appendix \ref{appendixa}. We use a hat label for $\mathbf{\hat{g}}$ to
emphasize that (\ref{dmq}) is with a Killing symmetry (in this case it
involves a time-like d-vectors). Similarly, we can define ansatz for
d-metrics with other orders of space coordinates than $(x^{k},y^{3}).$ For
instance, we can use $(x^{1},y^{3},x^{2})$, when the $x^{2}$ becomes a
v-coordinate (instead of $y^{3}).$ Such space coordinates can be spherical,
cylindrical, toroidal and other types. The generating function for
quasi-stationary v-distributions is defined in the form $\Psi =\exp (\varpi
),$ where $\varpi =\ln |h_{4}^{\ast}/\sqrt{|h_{3}h_{4}}|,$ see details and
motivation for the quadratic element $\mathbf{\hat{g}}[\Psi ]$ (\ref{qeltors}%
) involving respectively two generating sources $\ ^{h}\Upsilon (x^{k})$ and 
$\ ^{v}\Upsilon (x^{k},y^{3}) $ (\ref{esourc}). In this and our recent works 
\cite{partner02,partner06}, we use such brief notations for partial
derivatives: $\partial _{1}P=P^{\cdot },\partial _{2}P=P^{\prime },\partial
_{3}P=P^{\ast} $ and $\partial _{4}P=\partial _{t}P=P^{\diamond }$ for any
function $P(x^{k},y^{3},t).$ 

\item[Locally anisotropic cosmological solutions] anisotropic cosmological
solutions are generated by such ansatz for d-metrics: 
\begin{eqnarray}
\underline{\mathbf{\hat{g}}} &=&g_{i}(x^{k})dx^{i}\otimes dx^{i}+\underline{h%
}_{3}(x^{k},t)\underline{\mathbf{e}}^{3}\otimes \underline{\mathbf{e}}^{3}+%
\underline{h}_{4}(x^{k},t)\underline{\mathbf{e}}^{4}\otimes \underline{%
\mathbf{e}}^{4},  \notag \\
&&\underline{\mathbf{e}}^{3}=dy^{3}+\underline{n}_{i}(x^{k},t)dx^{i},\ 
\underline{\mathbf{e}}^{4}=dy^{4}+\underline{w}_{i}(x^{k},t)dx^{i}.
\label{dmc}
\end{eqnarray}%
We emphasize that a generic dependence on $y^{4}=t,$ which is important in
cosmology. The ansatz (\ref{dmc}) involves a Killing symmetry on the
space-like coordinate $\partial _{3},$ when the N-connection coefficients
are parameterized $\underline{N}_{i}^{3}=\underline{n}_{i}(x^{k},t)$ and $%
\underline{N}_{i}^{4}=\underline{w}_{i}(x^{k},t)$ and the coefficients of
d-metrics are written $\underline{\mathbf{\hat{g}}}_{\alpha
\beta}=[g_{ij}(x^{\kappa }),\underline{g}_{ab}(x^{\kappa },t)].$ The
generating function for locally anisotropic cosmological v-distributions is
defined in the form $\underline{\Psi }=\exp (\underline{\varpi }),$ where $%
\underline{\varpi }=\ln |\underline{h}_{3}^{\diamond }/\sqrt{|\underline{h}%
_{3}\underline{h}_{4}}|.$ Necessary conventions and formulas are summarized
in Appendix \ref{appendixa}.
\end{description}


We emphasize that the ansatz (\ref{dmq}) and (\ref{dmc}) possess certain
time and space duality properties as explained in Appendix \ref{stduality}.
For instance, we can transform a first type of quasi-stationary solutions
into a locally anisotropic cosmological ones if $h_{3}(x^{k},y^{3})%
\rightarrow \underline{h}_{4}(x^{k},t),$ $h_{4}(x^{k},y^{3})\rightarrow 
\underline{h}_{3}(x^{k},t)$ and $w_{i}(x^{k},y^{3})\rightarrow \underline{n}%
_{i}(x^{k},t), $ $n_{i}(x^{k},y^{3})\rightarrow \underline{w}_{i}(x^{k},t).$
Inverse transforms are also possible. Such "dual" space and time symmetries
can be prescribed only for generic off-diagonal solutions with respective
Killing symmetries on $\partial _{4},$ or $\partial _{3}.$ We can consider
also additional nonholonomic constraints and deformations generating new
classes of solutions of systems of nonlinear PDEs. The coefficients of the
mentioned types d-metrics depend generically on 3 from 4 spacetime
coordinates, and do not reduce the problem to finding solutions of some
"simplified" systems of nonlinear ODEs.

In \cite{vacaru18} (see also references therein) we discussed that certain
decoupling properties of (modified) Einstein equations can be proven for
more general classes d-metrics, for instance, parameterized in the form: 
\begin{eqnarray}
\mathbf{g} &=&g_{i}(x^{k})dx^{i}\otimes dx^{i}+\omega ^{2}(x^{k},y^{a})[%
\underline{h}_{3}(x^{k},y^{4})h_{3}(x^{k},y^{3})\mathbf{e}^{3}\otimes 
\mathbf{e}^{3}+h_{4}(x^{k},y^{3})\underline{h}_{4}(x^{k},y^{4})]\ \mathbf{e}%
^{4}\otimes \mathbf{e}^{4},  \notag \\
&&\mathbf{e}^{3}=dy^{3}+[w_{i}(x^{k},y^{3})+\underline{n}%
_{i}(x^{k},y^{4})]dx^{i},\ \mathbf{e}^{4}=dy^{4}+[n_{i}(x^{k},y^{3})+%
\underline{w}_{i}(x^{k},y^{4})]dx^{i}.  \label{2confgenansatz}
\end{eqnarray}%
Such an ansatz may not have explicit Killing symmetries but involves
vertical co-space conformal transforms with a factor $\omega (x^{k},y^{a}).$
More general ansatz (we can consider other types of frame or conformal
transforms) results in more cumbersome formulas. It implies more complex
nonholonomic structures and results in additional technical difficulties for
generating exact and parametric solutions. We omit such constructions in
this work and consider in explicit form only off-diagonal solutions with at
least one Killing symmetry. Here we note that to apply the BFV methods, is
important to have a nontrivial lapse function $\mathbf{\acute{N}}%
(x^{i^{\prime }},t)=-g_{44}=\omega ^{2}(x^{k},y^{a})h_{4}(x^{k},y^{3}) 
\underline{h}_{4}(x^{k},y^{4})$ as in (\ref{2confgenansatz}) but resulting
in a d-metric (\ref{dmc}). This can be achieved by re-defining the frame and
coordinate structures on $\mathbf{V}$ when for a corresponding $\omega
^{2}h_{a}\underline{h}_{a},$ we transform a chosen quasi-stationary ansatz (%
\ref{dmq}) into a variant of (\ref{dmc}) with $\partial _{4}g_{44}=\partial
_{t}g_{44}\neq 0.$ Such configurations can be modeled as certain effective
projectable or nonprojectable HL theories, see details in section \ref{sec03}%
. 

To construct exact or parametric solutions for ansatz (\ref{dmq}) or (\ref%
{dmc}) is not possible if we work directly with the LC connection $\nabla $
and off-diagonal ansatz (\ref{ansatz}). The AFCDM prescribes to distort
nonholonomically the Einstein equations into (\ref{cdeq1}) using the
canonical d-connection $\widehat{\mathbf{D}}.$ In such cases, the decoupling
and generating solutions in certain general off-diagonal forms is possible
as we explain in Appendix \ref{appendixa}. Then, certain LC configurations
can be extracted by imposing additional nonholonomic constraints (\ref%
{lccond1}) as we explain in \ref{salc}. 

\subsubsection{Off-diagonal transforms and gravitational polarizations of
prime metrics}

The AFCDM allows to construct new classes of off-diagonal solutions in GR in
certain forms describing deformations of a (primary) d-metric $\mathbf{%
\mathring{g}}$ into another type of (target) d-metric $\mathbf{g}$
parameterized in the form(\ref{dmq}) or (\ref{dmc}). In general, a
pseudo-Riemannian prime metric may be or not be a solution of some
gravitational field equations but a target d-metric $\mathbf{g}$ defines an
exact or parametric solution of (\ref{cdeq1}).

Let us denote a \textbf{prime} d-metric as 
\begin{equation}
\mathbf{\mathring{g}=}[\mathring{g}_{\alpha },\mathring{N}_{i}^{a}]
\label{offdiagpm}
\end{equation}%
and consider general off-diagonal transforms it into a \textbf{target}
d-metric $\mathbf{g,}$ 
\begin{eqnarray}
\mathbf{\mathring{g}} &\rightarrow &\mathbf{g}=[g_{\alpha }=\eta _{\alpha }%
\mathring{g}_{\alpha },N_{i}^{a}=\eta _{i}^{a}\ \mathring{N}_{i}^{a}]
\label{offdiagdefr} \\
&\rightarrow &\underline{\mathbf{g}}=[\underline{g}_{\alpha }=\underline{%
\eta }_{\alpha }\mathring{g}_{\alpha },\underline{N}_{i}^{a}=\underline{\eta 
}_{i}^{a}\ \underline{\mathring{N}}_{i}^{a}],%
\mbox{ for locally anisotropic
cosmological configurations}.  \notag
\end{eqnarray}%
The functions $\underline{\eta }_{\alpha }(x^{k},t)$ and $\underline{\eta }
_{i}^{a}(x^{k},t)$ from (\ref{offdiagdefr}) are called gravitational
polarization ($\eta $-polarization) functions. We can also write $\underline{%
\mathbf{\mathring{g}}}=[\underline{\mathring{g}}_{\alpha}, \underline{%
\mathring{N}}_{i}^{a}]$ if such a prime d-metric of cosmological type. For
certain nonholonomic transforms, we can use more general classes of
polarizations functions, $\underline{\eta }_{\alpha }(x^{k},y^{3},t)$ and $%
\underline{\eta }_{i}^{a}(x^{k},y^{3},t),$ or another type ones, for
generating quasi-stationary target solutions, $\eta _{\alpha }(x^{k},y^{3})$
and $\eta _{i}^{a}(x^{k},y^{3}).$ We shall omit underlining of the geometric
and physical objects if that will not result in ambiguities (considering
that we can always redefine the abstract formalism in a necessary form for
generating quasi-stationary or locally anisotropic cosmological
configurations). 

To generate in explicit form target solutions of (\ref{cdeq1}) we can
consider that the nonlinear symmetries (\ref{ntransf2}) are parameterized as 
\begin{eqnarray}
(\Psi ,\ ^{v}\Upsilon ) &\leftrightarrow &(\mathbf{g},\ ^{v}\Upsilon
)\leftrightarrow (\eta _{\alpha }\ \mathring{g}_{\alpha }\sim (\zeta
_{\alpha }(1+\kappa \chi _{\alpha })\mathring{g}_{\alpha },\ ^{v}\Upsilon
)\leftrightarrow  \label{nonlintrsmalp} \\
(\Phi ,\ \Lambda ) &\leftrightarrow &(\mathbf{g},\ \Lambda )\leftrightarrow
(\eta _{\alpha }\ \mathring{g}_{\alpha }\sim (\zeta _{\alpha }(1+\kappa \chi
_{\alpha })\mathring{g}_{\alpha },\ \Lambda ).  \notag
\end{eqnarray}%
In these formulas $\Lambda $ is an effective cosmological and $\kappa $ is a
small parameter $0\leq \kappa <1;$ $\zeta _{\alpha }(x^{k},y^{3})$ and $\chi
_{\alpha }(x^{k},y^{3})$ are respective polarization functions. We can
choose $\kappa \sim \hbar $ (if we study quantum deformations involving the
Planck constant). The nonholonomic transforms (\ref{nonlintrsmalp}) must
result in a target metric $\mathbf{g}$ defined as a solution of type (\ref%
{qeltors}) or, equivalently, (\ref{offdiagcosmcsh}). This is possible if the 
$\eta $- and/or $\chi $-polarizations are subjected to the conditions (\ref%
{ntransf2}), which we write in the form: 
\begin{eqnarray}
\partial _{3}[\Psi ^{2}] &=&-\int dy^{3}\ ^{v}\Upsilon \partial
_{3}h_{4}\simeq -\int dy^{3}\ ^{v}\Upsilon \partial _{3}(\eta _{4}\ 
\mathring{g}_{4})\simeq -\int dy^{3}\ ^{v}\Upsilon \partial _{3}[\zeta
_{4}(1+\kappa \ \chi _{4})\ \mathring{g}_{4}],  \notag \\
\Phi ^{2} &=&-4\ \Lambda h_{4}\simeq -4\ \Lambda \eta _{4}\mathring{g}%
_{4}\simeq -4\ \Lambda \ \zeta _{4}(1+\kappa \chi _{4})\ \mathring{g}_{4}.
\label{nonlinsymrex}
\end{eqnarray}

So, the off-diagonal $\eta $-transforms resulting in d-metrics (\ref%
{offdiagdefr}) can be parameterized to be generated for $\psi $- and $\eta $%
-polarizations, 
\begin{equation}
\psi \simeq \psi (\kappa ;x^{k}),\eta _{4}\ \simeq \eta _{4}(x^{k},y^{3}),
\label{etapolgen}
\end{equation}%
in a form equivalent to (\ref{offdsolgenfgcosmc}). The corresponding
quasi-stationary quadratic element can be written in the form 
\begin{eqnarray}
d\widehat{s}^{2} &=&\widehat{g}_{\alpha \beta }(x^{k},y^{3};\mathring{g}%
_{\alpha };\psi ,\eta _{4};\ \Lambda ,\ ^{v}\Upsilon )du^{\alpha }du^{\beta
}=e^{\psi }[(dx^{1})^{2}+(dx^{2})^{2}]  \label{offdiagpolfr} \\
&&-\frac{[\partial _{3}(\eta _{4}\ \mathring{g}_{4})]^{2}}{|\int dy^{3}\ \
^{v}\Upsilon \partial _{3}(\eta _{4}\ \mathring{g}_{4})|\ \eta _{4}\mathring{%
g}_{4}}\{dy^{3}+\frac{\partial _{i}[\int dy^{3}\ ^{v}\Upsilon \partial
_{3}(\eta _{4}\mathring{g}_{4})]}{\ \ ^{v}\Upsilon \partial _{3}(\eta _{4}%
\mathring{g}_{4})}dx^{i}\}^{2}  \notag \\
&&+\eta _{4}\mathring{g}_{4}\{dt+[\ _{1}n_{k}+\ _{2}n_{k}\int dy^{3}\frac{%
[\partial _{3}(\eta _{4}\mathring{g}_{4})]^{2}}{|\int dy^{3}\ \ ^{v}\Upsilon
\partial _{3}(\eta _{4}\mathring{g}_{4})|\ (\eta _{4}\mathring{g}_{4})^{5/2}}%
]dx^{k}\}^{2}.  \notag
\end{eqnarray}%
We can relate such a solution to another one in the form (\ref%
{offdiagcosmcsh}) if $\Phi ^{2}=-4\ \Lambda h_{4}$ and the $\eta $%
-polarizations are determined by the generating data $(h_{4}=\eta _{4} 
\mathring{g}_{4};\Lambda ,\ ^{v}\Upsilon ).$

The ansatz (\ref{qeltorsc}) from Appendix \ref{appendixa} provides a locally
anisotropic formula for a $t$-transformed d-metric (\ref{offdiagpolfr}) when
the generating data are of type $(\underline{h}_{3}=\underline{\eta }%
_{3}(x^{k},t)\underline{\mathring{g}}_{3}; \underline{\Lambda },\ ^{v}%
\underline{\Upsilon }(x^{k},t)),$ for $\underline{\Phi }^{2}=-4\ \underline{%
\Lambda }\underline{h}_{3},$ and $\eta $-polarizations $\psi \simeq \psi
(\kappa ;x^{k}),\underline{\eta }_{3}\ \simeq \underline{\eta }_{3}(x^{k},t)$
instead of (\ref{etapolgen}). Corresponding time and space duality
properties and related nonholonomic transforms are explained in Appendix \ref%
{stduality}. So, the locally anisotropic cosmological solutions of (\ref%
{cdeq1}) as $t$-duals of quasi-stationary configurations are defined by
quadratic line elements: 
\begin{eqnarray}
d\widehat{s}^{2} &=&\widehat{g}_{\alpha \beta }(x^{k},t;\mathring{g}_{\alpha
};\psi ,\underline{\eta }_{3};\ \underline{\Lambda },\ ^{v}\underline{%
\Upsilon })du^{\alpha }du^{\beta }=e^{\psi }[(dx^{1})^{2}+(dx^{2})^{2}]
\label{offdiagpolcosm} \\
&&+\underline{\eta }_{3}\underline{\mathring{g}}_{3}\{dy^{3}+[\ _{1}n_{k}+\
_{2}n_{k}\int dt\frac{[\partial _{t}(\underline{\eta }_{3}\underline{%
\mathring{g}}_{3})]^{2}}{|\int dt\ \ ^{v}\underline{\Upsilon }\partial _{t}(%
\underline{\eta }_{3}\underline{\mathring{g}}_{3})|\ (\underline{\eta }_{3}%
\underline{\mathring{g}}_{3})^{5/2}}]dx^{k}\}^{2}  \notag \\
&&-\frac{[\partial _{t}(\underline{\eta }_{3}\underline{\mathring{g}}%
_{3})]^{2}}{|\int dt\ \ ^{v}\underline{\Upsilon }\partial _{t}(\underline{%
\eta }_{3}\underline{\mathring{g}}_{3})|\ \underline{\eta }_{3}\underline{%
\mathring{g}}_{3}}\{dt+\frac{\partial _{i}[\int dt\ ^{v}\underline{\Upsilon }%
\partial _{t}(\underline{\eta }_{3}\underline{\mathring{g}}_{3})]}{\ \ ^{v}%
\underline{\Upsilon }\partial _{t}(\underline{\eta }_{3}\underline{\mathring{%
g}}_{3})}dx^{i}\}^{2}.  \notag
\end{eqnarray}%
Such off-diagonal solutions are important for elaborating on (in general,
no-perturbative) quantum cosmological models. In another turn, the d-metrics
(\ref{offdiagpolfr}) can be used for study quantum deformations of black
hole, wormhole and other types of quasi-stationary solutions in GR. 

\subsection{Modelling HL-configurations on noholonomic Lorentz manifolds}

The AFCDM allows us to generate various classes of off-diagonal solutions of
the Einstein equations encoding, for instance, nonassociative and
noncommutative structures, supersymmetric configurations, etc. \cite%
{vv25,vacaru26,partner02,partner06,vacaru18}. Such solutions can be
quantized following different approaches related, for instance, to
deformation quantization, DQ, or generalized BFV methods \cite%
{fv75,bv77,ff78,fr12,vacaru25}. Certain subclasses of off-diagonal solutions
in mentioned type MGTs possess asymptotic safety properties \cite%
{vapny24,weinberg79,niedermaier06}. Nevertheless, the issue of selecting
renormalizable (in a nonlinear BFV sense) solutions in GR has not been
investigated in classical gravity theories and QG. 

Let us explain how, for instance, the generating function $\underline{\eta }%
_{3}\ \simeq \underline{\eta }_{3}(x^{k},t)$ can be chosen to generate
off-diagonal cosmological solutions (\ref{offdiagpolcosm}) with anisotropies
between the time and space coordinates like the Lifshitz works on phase
space transitions in condensed matter physics \cite%
{lifschitz,horava1,horava2}. We emphasize that the geometric data for a
primary d-metric $\left( \underline{\mathring{g}}_{i},\underline{\mathring{g}%
}_{a},\underline{\mathring{n}}_{i},\underline{\mathring{w}}_{i}\right) $ can
be chosen define a nonholonomic Lorentz manifold when the gravitational
polarizations determined by an anisotropic $\underline{\eta }_{3}(x^{k},t)$
determines a class of off-diagonal solutions modelling effective
anisotropies in the framework of classical GR. For such target d-metrics $%
\underline{\mathbf{g}}=[\underline{g}_{\alpha }=\underline{\eta }_{\alpha }%
\mathring{g}_{\alpha },\underline{N}_{i}^{a}=\underline{\eta }_{i}^{a}\ 
\underline{\mathring{N}}_{i}^{a}]$ (\ref{offdiagdefr}) defined in the form (%
\ref{offdiagpolcosm}), certain cosmological evolution scenarios can involve
an (effective) anisotropy between time and space. We can elaborate on
explicit examples when the generating function 
\begin{eqnarray}
\underline{\eta }_{3}(x^{k},t) &\simeq &\underline{\eta }_{3}(x^{k},t)+\
^{HL}\underline{\eta }_{3}(x^{k},t),\mbox{ where }  \label{hlqdef} \\
\ ^{HL}\underline{\eta }_{3}(x^{k},t) &=&\sigma _{0}\phi (x^{k},t),\sigma
_{0}=const,\sigma _{0}\simeq \hbar ,\mbox{ or }\sigma _{0}\simeq \kappa %
\mbox{ as in }(\ref{offdncelepsilon}),  \notag
\end{eqnarray}
is modelled as a scalar field also in $\widetilde{d}$ spacetime dimensions. An integration measure is determined by $\underline{\mathbf{g}}$ and an effective action can be defined:%
\begin{equation}
\ ^{\phi }S=-\int \delta ^{\widetilde{d}+1}u\left( \mathbf{\hat{e}}_{\alpha
}\phi \right) \left( \mathbf{\hat{e}}^{\alpha }\phi \right) \Longrightarrow
\ ^{\phi }\tilde{S}=\int dt\delta ^{\widetilde{d}}u\left[ (\partial _{t}\phi
)^{2}+\phi \ ^{z}\mathbf{D}\phi \right] .  \label{anisotract}
\end{equation}%
In these formulas, we consider a higher order differential operator in
N-elongated spacial derivatives%
\begin{equation*}
\ ^{z}\mathbf{D}=-\breve{M}^{-2(z-1)}(-\widehat{\triangle })^{z}+...,%
\widehat{\triangle }=\mathbf{\hat{e}}_{i^{\prime }}\mathbf{\hat{e}}%
^{i^{\prime }},
\end{equation*}%
and the parameters $\sigma _{0}$ and $\breve{M}$ are introduced to keep the
physical dimensionality correct and the lower derivative. The symbol $\delta
^{\widetilde{d}}u$ is used to emphasize that we use N-elongated
differentials as in (\ref{nadif}) but for respective space dimension $%
\widetilde{d}.$ 

We argue that generating functions subjected to conditions of type (\ref%
{anisotract}) effectively broke the local Lorentz symmetry even at the
classical level. For small parametric deformations (\ref{nonlinsymrex}), we
can chose $\underline{\chi }_{3}(x^{k},t)\simeq \sigma _{0}\phi (x^{k},t).$
The off-diagonal solutions of type (\ref{offdiagpolcosm}), (\ref{qeltorsc})
and $t$-duals of (\ref{offdncelepsilon}) are generated to model Lorentz
spacetime phase transitions between two types of configurations of the
target d-metrics $\underline{\mathbf{g}}:$ One phase is with local Lorentz
symmetry as in GR, another phase is gravitationally polarized to effective
configurations with space and time anisotropy. 

In the phase with local Lorentz invariance violations (for generating
well-defined quantum deformations of GR, we can consider only quantum such
phase), an effective action (\ref{anisotract}) can be constructed, which is
invariant under special anisotropic scaling: 
\begin{equation}
\left\{ 
\begin{array}{c}
u^{\alpha }\longrightarrow b^{-1}u^{\alpha },b=const; \\ 
\phi \longrightarrow b^{(\widetilde{d}-1)/2}\phi ,[\phi ]=\frac{\widetilde{d}%
-1}{2}%
\end{array}%
\right. \implies \left\{ 
\begin{array}{c}
u^{i^{\prime }}\longrightarrow b^{-1}u^{i^{\prime }},t\longrightarrow b^{-z}t
\\ 
\phi \longrightarrow b^{(\widetilde{d}-z)/2}\phi ,[\phi ]=\frac{\widetilde{d}%
-z}{2}%
\end{array}%
\right. .  \label{localscaltransf}
\end{equation}%
Such transforms define the so-called anisotropic scaling dimension of the
field $\phi $ when its dimension $[\phi ]$ transform under anisotropic
scaling as $\phi \longrightarrow b^{r}\phi .$ So, we assign to the spacial
coordinates $u^{i^{\prime }}$ the dimension $-1$ and to the time coordinate $%
t$ the dimension $-z.$ For $z\neq 1,$ a scaling dimension is not equal to a
physical dimension. Such scaling properties are considered in Ho\v{r}ava
gravity theories \cite{horava1,horava2} (which are different from GR) and
related QG theories involving anisotropic scaling \cite{barv23,bbd24}. For
certain critical values $z_{crit},$ corresponding theories become
renormalizable, and superrenormalizable below such critical values. In this
work, BFV quantization is considered for quantum configurations with locally
anisotropic scaling (\ref{localscaltransf}). In other types of HL-theories,
such properties were assummed for certain classical configurations which
resulted in gravity models which are different from GR. If we consider $%
\sigma _{0}\simeq \hbar ,$ or $\sigma _{0}\simeq \kappa ,$ the quantum
HL-models reduse to GR. 

The main Hypothesis of this work (see the end of the Introduction), stating
that HL-like configurations can emerge via quantum gravitational nonlinear
polarizations of classical configurations and respective physical constants
within GR, is supported by the new classes of exact and parametric
off-diagonal solutions constructed above. Phases with anisotropic scaling of 
$\underline{\mathbf{g}}$ can exist even when the generating sources $\ ^{v}%
\underline{\Upsilon }(x^{k},t)$ in (\ref{offdiagpolcosm}) are not invariant
under the transformations (\ref{localscaltransf}). Notably, quasi-stationary
solutions (\ref{offdiagpolfr}) may lack anisotropic scaling for certain
generating functions $h_{3}(x^{k},y^{3})$. Models can be quantized without
considering anisotropic scaling if the generating functions and effective
sources are time-independent. In the following sections, we demonstrate that
BFV quantization of such nonholonomic and locally anisotropic configurations
leads to renormalizable QG models. 

\section{BFV quantization of nonholonomic HL-structures}

\label{sec03} In this section, we define effective HL models based on
off-diagonal solutions of nonholonomically distorted Einstein equations and
their associated nonholonomic ADM structures. We then develop the BFV
quantization of these models in canonical nonholonomic variables,
incorporating second-class constraints and corresponding gauge-fixing
conditions.

\subsection{Classical nonholonomic ADM structures and effective models}

In our approach, we can choose some prime metrics as classical solutions of
the standard Einstein equations and then to consider parametric deformations
into target metrics as solutions of effective systems of nonlinear PDEs
characterized by nonlinear symmetries and anisotropic scaling properties. 

\subsubsection{Nonholonomic foliations and N-adapted preserving
diffeomorphysms}

We can define on a nonholonomic Lorentz manifold $\mathbf{V}$ enabled with
N-connection dyadic structure an additional N-adapted ADM splitting of the
gravitational configuration space. For such geometric constructions, we
introduce a spatial d-metric $\mathbf{g}_{i^{\prime }j^{\prime }}$ and
respective lapse $\mathbf{\acute{N}}(x^{i^{\prime }},t)$ and shift $\mathbf{N%
}^{k^{\prime }}(x^{i^{\prime }},t)$ functions, when the quadratic element (%
\ref{dma}) is re-defined for a space dimension $\widetilde{d}$: 
\begin{eqnarray}
ds^{2} &=&\mathbf{g}_{i^{\prime }j^{\prime }}(\mathbf{e}^{i^{\prime }}+%
\mathbf{N}^{i^{\prime }}dt)(\mathbf{e}^{j^{\prime }}+\mathbf{N}^{j^{\prime
}}dt)-\mathbf{\acute{N}}^{2}(\mathbf{e}^{4})^{2}  \label{dma1} \\
&=&g_{i^{\prime }j^{\prime }}(du^{i^{\prime }}+N^{i^{\prime
}}dt)(du^{j^{\prime }}+N^{j^{\prime }}dt)-\acute{N}^{2}(dt)^{2},
\label{dmab}
\end{eqnarray}%
for $i^{\prime },j^{\prime },...=1,2,...,\widetilde{d}.$ In terms of
generating functions for (\ref{offdiagpolcosm}), $\mathbf{\acute{N}}%
^{2}=[\partial _{t}(\underline{\eta }_{3}\underline{\mathring{g}}%
_{3})]^{2}/|\int dt\ \ ^{v}\underline{\Upsilon }\partial _{t}(\underline{%
\eta }_{3}\underline{\mathring{g}}_{3})|\ \underline{\eta }_{3}\underline{%
\mathring{g}}_{3},$ when $\mathbf{g}_{i^{\prime }j^{\prime }},$ $\mathbf{e}%
^{\alpha ^{\prime }}=(\mathbf{e}^{i^{\prime }},\mathbf{e}%
^{4}=dt+N_{i}^{4}dx^{i})$, and $\mathbf{N}^{i^{\prime }}$ can be expressed
via N-adapted coefficients of a class of off-diagonal solutions by a
corresponding re-grouping of terms in (\ref{dma1}). Such a split may be used
for HL-quantum deformations (\ref{hlqdef}) and does not violate the
diffeomorphism invariance at the classical level because we can consider any
N-adapted double 2+2 and 3+1 splitting and then arbitrary frame/coordinate
transforms. We can always define (with respect to some fixed nonholonomic
(co) frames (\ref{nader}) and (\ref{nadif})) certain reparametrization of
coordinates and coefficients of (\ref{dma1}),%
\begin{eqnarray}
u^{i^{\prime }} &\rightarrow &\tilde{u}^{i^{\prime }}=\tilde{u}^{i^{\prime
}}(u^{i^{\prime }},t)\mbox{ and }t\rightarrow \tilde{t}=\tilde{t}(t);
\label{coordtr} \\
\mathbf{N}^{i^{\prime }} &\rightarrow &\widetilde{\mathbf{N}}^{i^{\prime }}=(%
\widetilde{\mathbf{N}}^{j^{\prime }}\frac{\partial \tilde{u}^{i^{\prime }}}{%
\partial u^{j^{\prime }}}-\frac{\partial \tilde{u}^{i^{\prime }}}{\partial
u^{j^{\prime }}})\frac{dt}{d\tilde{t}}\mbox{ and }\mathbf{\acute{N}%
\rightarrow }\widetilde{\mathbf{\acute{N}}}=\mathbf{\acute{N}}\frac{dt}{d%
\tilde{t}};  \notag \\
\mathbf{g}_{i^{\prime }j^{\prime }} &\rightarrow &\mathbf{\tilde{g}}%
_{i^{\prime }j^{\prime }}=\mathbf{g}_{k^{\prime }l^{\prime }}\frac{\partial
u^{k^{\prime }}}{\partial \tilde{u}^{i^{\prime }}}\frac{\partial
u^{l^{\prime }}}{\partial \tilde{u}^{j^{\prime }}}\mbox{ and }\mathbf{K}%
_{i^{\prime }j^{\prime }}\rightarrow \widetilde{\mathbf{K}}_{k^{\prime
}l^{\prime }}\frac{\partial u^{k^{\prime }}}{\partial \tilde{u}^{i^{\prime }}%
}\frac{\partial u^{l^{\prime }}}{\partial \tilde{u}^{j^{\prime }}},  \notag
\end{eqnarray}%
where the extrinsic curvature of a d-connection $\mathbf{D}_{\alpha }=(%
\mathbf{D}_{i^{\prime }},\mathbf{D}_{4})$ is defined as%
\begin{equation*}
\mathbf{K}_{i^{\prime }j^{\prime }}=\frac{1}{2\mathbf{\acute{N}}}(\partial
_{t}\mathbf{g}_{i^{\prime }j^{\prime }}-\mathbf{D}_{i^{\prime }}\mathbf{N}%
_{j^{\prime }}-\mathbf{D}_{j^{\prime }}\mathbf{N}_{i^{\prime }}).
\end{equation*}%
We can use hat variables and write $\widehat{\mathbf{K}}_{i^{\prime
}j^{\prime }}$ for a canonical d-connection (\ref{canondistrel}) and define $%
\mathbf{\hat{g}=(\mathbf{\hat{g}}}_{i^{\prime }j^{\prime }},\widehat{\mathbf{%
\mathbf{N}}}^{i^{\prime }},\widehat{\mathbf{\acute{N}}}\mathbf{)}$ for a 3+1
decomposition of a quasi-stationary solution (\ref{dmq}) of (\ref{cdeq1}).
For locally anisotropic cosmological d-metrics (\ref{dmc}), we can underline
the symbols, $\underline{\mathbf{\hat{g}}}=(\underline{\mathbf{\mathbf{\hat{g%
}}}}_{i^{\prime }j^{\prime }},\underline{\widehat{\mathbf{\mathbf{N}}}}%
^{i^{\prime }},\underline{\widehat{\mathbf{\acute{N}}}}),$ and construct $%
\underline{\widehat{\mathbf{K}}}_{i^{\prime }j^{\prime }}$ (typically, we
shall omit in the next such underlining if that will not result in
ambiguities). Here we note that even an N-connection structure $\mathbf{N}$ (%
\ref{ncon}) can be prescribed in coordinate-free form, the local
coefficients for such an N-connection also transform $N_{i}^{a}\rightarrow 
\tilde{N}_{i}^{a}$ under coordinate changing (\ref{coordtr}). Such
coefficients must satisfy the conditions: 
\begin{equation}
\mathbf{N}=N_{i}^{a}(x,y)dx^{i}\otimes \partial /\partial y^{a}=\tilde{N}%
_{i}^{a}(\tilde{x},\tilde{y})d\tilde{x}^{i}\otimes \partial /\partial \tilde{%
y}^{a},  \label{nconcoordtr}
\end{equation}%
when $\tilde{u}=\{\tilde{x}(x,y),\tilde{y}(x,y)\}=\{\tilde{x}^{i}(x,y),%
\tilde{y}^{a}(x,y)\}=\{\tilde{x}^{i}(x^{k},y^{b}),\tilde{y}%
^{a}(x^{k},y^{b})\}=\{\tilde{x}^{i}(u^{i^{\prime }},t),\tilde{y}%
^{a}(u^{i^{\prime }},t)\}$ etc. Transforms to "tilde" coordinates also
result in "tilde" modifications of the coefficients in nonholonomic (co)
frames (\ref{nader}) and (\ref{nadif}). Such transforms of nonholonomic
N-adapted 3+1 splitting are different from those considered, for instance,
in \cite{misner73,barv23,bbd24}.

In this work, we present a proof of renormalization of general classes of
off-diagonal solutions in GR, which can be defined as nonholonomic double
fibrations with 2+2 and 3+1 conventional splitting. Typical d-metric ansatz
defining such solutions can be represented in certain forms (\ref{dma}), or (%
\ref{dmc}) (or, in more general forms, as (\ref{2confgenansatz})). For
nonholonomic 3+1 splitting, such d-metrics can be represented in the form (%
\ref{dma1}). In our approach, we work on nonholonomic Lorentz manifolds when
HL type structures can be generated effectively by $\underline{\eta }%
_{3}(x^{k},t)\simeq \sigma _{0}\phi (x^{k},t),$ or $\underline{\chi }%
_{3}(x^{k},t)\simeq \sigma _{0}\phi (x^{k},t),$ as we explain for the
effective action (\ref{anisotract}) involving anisotropic scaling of space
and time coordinates (\ref{localscaltransf}) and HL-generating functions of
type (\ref{hlqdef}). Such classes of off-diagonal solutions and
nonholonomically deformed gravity models are described by nonlinear
symmetries (\ref{nonlintrsmalp}) and re-parametrization of coordinates (\ref%
{coordtr}) adapted to the N-connection structure transforms (\ref%
{nconcoordtr}). 

In our approach, we work with effective theories modeled on nonhlonomic
Lorentz manifolds, when nonlinear gauge groups are stated by N-adapted
foliation preserving diffeomorphisms (FDiffN). We note that FDiffN
transforms into FDiff for holonomic structures. Such effective theories are
unitary and renormalized (being quantized under the BFV formalism \cite%
{barv23,bbd24,vacaru25}). We can always extract LC configurations (\ref%
{lccond1}), which might yield the classical dynamics in GR for respective
classes of generating functions and generating source, and even for locally
anisotropic scaling of quantum fluctuations at the large-distance limit.

Let us analyze the corresponding nonlinear gauge symmetries given by FDiffN.
So for any class of off-diagonal solutions (\ref{dma}), or \ref{dmc}) we can
compute the ADM nonholonomic variables as in (\ref{dma}) \ and (\ref{dma1}).
As we explained for (\ref{2confgenansatz}), we can deal only with the
nonprojectable versions of HL models when the lapse function $\mathbf{\acute{%
N}}$ can depend generically on time and space coordinates. The FDiffN acts
infinitesimaly as $\mathbf{e}^{i^{\prime }}=-\zeta ^{i^{\prime }}$ and $%
\mathbf{e}^{4}=f(t)$ (for holonomic configurations with $N_{i}^{a}=0$, we
obtain respectively $\mathbf{e}^{i^{\prime }}\rightarrow \delta u^{i^{\prime
}}$ and $\mathbf{e}^{4}\rightarrow \delta t$ as in \cite{bbd24}). This
results in N-adapted FDiffN transforms of the ADM variables: 
\begin{eqnarray}
\delta \mathbf{g}_{i^{\prime }j^{\prime }} &=&\zeta ^{k^{\prime }}\mathbf{e}%
_{k^{\prime }}\mathbf{g}_{i^{\prime }j^{\prime }}+2\mathbf{g}_{k^{\prime
}(i^{\prime }}\mathbf{e}_{j^{\prime })}\zeta ^{k^{\prime }}+f\partial _{t}%
\mathbf{g}_{i^{\prime }j^{\prime }}  \label{nagaugetr} \\
\delta \mathbf{N}^{i^{\prime }} &=&\zeta ^{k^{\prime }}\mathbf{e}_{k^{\prime
}}\mathbf{N}^{i^{\prime }}-\mathbf{N}^{k^{\prime }}\mathbf{e}_{k^{\prime
}}\zeta ^{i^{\prime }}+\partial _{t}\zeta ^{i^{\prime }}+\partial _{t}(f%
\mathbf{N}^{i^{\prime }}),\delta \mathbf{\acute{N}=}\zeta ^{k^{\prime }}%
\mathbf{e}_{k^{\prime }}\mathbf{\acute{N}+}\partial _{t}(f\mathbf{\acute{N}}%
).  \notag
\end{eqnarray}%
Such transforms were studied in \cite{brs12} in the context of
asymptotically flat conditions in the context of Ho\v{r}ava gravity, where a
definition of the spacetime metric does not exist. For such a theory, the
restriction $f=0$ is equivalent to a partial gauge-fixing condition for
chosen slices, when the Fdiff gauge symmetric of quantum models is
characterized by a time-dependent spacial vector $\zeta ^{i^{\prime
}}(x^{k^{\prime }},t).$ 

In the case of nonholonomic Lorentz manifolds with nontrivial N-connection
structure and solutions of type (\ref{qeltors}), we can always select a
subclass of generating and integration functions when the condition $f=0$ is
encoded for FDiffN. This can considered for generating functions in GR or
for quantum extensions of type (\ref{hlqdef}). Denoting the nonlinear
N-adapted transforms by $\ _{\zeta }\delta $ (we put a left label because
write ones can be confused with coefficient indices for d-objects), we write
the corresponding nonlinear gauge transforms (\ref{nagaugetr}),%
\begin{equation}
\ _{\zeta }\delta \mathbf{g}_{i^{\prime }j^{\prime }}=\zeta ^{k^{\prime }}%
\mathbf{e}_{k^{\prime }}\mathbf{g}_{i^{\prime }j^{\prime }}+2\mathbf{g}%
_{k^{\prime }(i^{\prime }}\mathbf{e}_{j^{\prime })}\zeta ^{k^{\prime }},\
_{\zeta }\delta \mathbf{N}^{i^{\prime }}=\zeta ^{k^{\prime }}\mathbf{e}%
_{k^{\prime }}\mathbf{N}^{i^{\prime }}-\mathbf{N}^{k^{\prime }}\mathbf{e}%
_{k^{\prime }}\zeta ^{i^{\prime }}+\partial _{t}\zeta ^{i^{\prime }},\delta 
\mathbf{\acute{N}=}\zeta ^{k^{\prime }}\mathbf{e}_{k^{\prime }}\mathbf{%
\acute{N}}.  \label{nagaugetrgf}
\end{equation}%
For instance, the d-operator $\ _{\zeta }\delta $ \ acts respectively on a
time-dependent spacial d-tenor, $\mathbf{P}^{i^{\prime }j^{\prime }},$ or a
d-tensor density, $\mathbf{p}^{i^{\prime }j^{\prime }},$ in such forms: 
\begin{eqnarray}
\ _{\zeta }\delta \mathbf{P}^{i^{\prime }j^{\prime }} &=&\zeta ^{k^{\prime }}%
\mathbf{e}_{k^{\prime }}\mathbf{P}^{i^{\prime }j^{\prime }}-\mathbf{P}%
^{k^{\prime }j^{\prime }}\mathbf{e}_{k^{\prime }}\zeta ^{i^{\prime }}-%
\mathbf{P}^{i^{\prime }k^{\prime }}\mathbf{e}_{k^{\prime }}\zeta ^{j^{\prime
}},\mbox{ or }  \label{nagaugetrgftens} \\
\ _{\zeta }\delta \mathbf{p}^{i^{\prime }j^{\prime }} &=&\zeta ^{k^{\prime }}%
\mathbf{e}_{k^{\prime }}\mathbf{p}^{i^{\prime }j^{\prime }}+\mathbf{p}%
^{i^{\prime }j^{\prime }}\mathbf{e}_{k^{\prime }}\zeta ^{k^{\prime }}-%
\mathbf{p}^{k^{\prime }j^{\prime }}\mathbf{e}_{k^{\prime }}\zeta ^{i^{\prime
}}-\mathbf{p}^{i^{\prime }k^{\prime }}\mathbf{e}_{k^{\prime }}\zeta
^{j^{\prime }}.  \notag
\end{eqnarray}%
%

In this work, the term (N-adapted, nonlinear) FDiffN gauge symmetry of
nonholonomic Einstein equations (\ref{cdeq1}) and their off-diagonal
solutions refers to transformations (\ref{nagaugetrgf}) and (\ref%
{nagaugetrgftens}). For generating functions with anisotropic space and time
scaling, such FDiffN transforms encode nonholonomic HL phases of with
quantum off-diagonal deformations (\ref{hlqdef}) of GR. We can generate two
types of HL configurations in this approach. The first ones correspond to
the so-called \textit{projectable} Ho\v{r}ava models where the lapse
function depends only on time, $\mathbf{\acute{N}=\acute{N}}(t).$ In such
cases, we can re-parameterize the time and set $\mathbf{\acute{N}}(t)=1$ and
consider that the time-dependent spatial N-adapted diffeomorphisms remain as
certain nonlinear gauge transforms. We can generalize in nonholonomic form,
for instance, the constructions from section 3 of \cite{barv23}, and perform
the BFV quantization of quasi-stationary d-metric configurations (\ref{dmq}%
). We omit such "projectable" considerations in this paper. We address
non-projectable HL models which can be off-diagonally generated by locally
anisotropic d-metrics (\ref{dmc}) (in more general cases, by (\ref%
{2confgenansatz})). In such cases, we can chose such nonholonomic structures
when $\mathbf{\acute{N}=\acute{N}}(x^{k},t).$ In certain sense, this work
consists an (associative and commutative) nonholonomic N-adapted deformation
(\ref{hlqdef}) of the "non-projectable results \cite{bbd24,vacaru25}, see
also references therein and discussion from section 4 of \cite{barv23}. 

\subsubsection{Effective actions with nonholonomic 3+1 splitting encoding HL
structures}

According to our main Hypothesis (formulated in Introduction), generic
off-diagonal gravitational and matter field interactions in certain phases
classical locally isotropic and homogeneous properties, but another type
phase spacetime transitions result in HL configurations. The last type of
ones can be quantized following certain forms to undertake the
renormalizatoin of gauge theories, which is based on a synthesis of the
AFCDM and the background formalism. Here, we emphasize that in this work we
do not substitute GR by a HL gravity theory but try to keep (at least in the
classical level and possibly anisotropic scaling deformations) the Einstein
theory paradigm. We study the properties of generic off-diagonal solutions
in GR and MGTs, which for respective classes of nonlinear configurations
transform into effective models with "good" quantum renormalization
properties. 

In the next sections, the classical and quantum constructions will be
elaborated for cosmological d-metrics $\underline{\mathbf{\hat{g}}}\mathbf{=(%
\mathbf{\hat{g}}}_{i^{\prime }j^{\prime }},\widehat{\mathbf{\mathbf{N}}}%
^{i^{\prime }},\widehat{\mathbf{\acute{N}}})$ \ (\ref{dmc}) and
corresponding canonical d-connections $\widehat{\mathbf{D}}=(\widehat{%
\mathbf{D}}_{i^{\prime }},\widehat{\mathbf{D}}_{4})$ (\ref{canondistrel})
constructed as solutions of (\ref{cdeq1}) with N-adapted 3+1 decomposition.
For simplicity, we shall omit underlining of symbols, considering that the
corresponding geometric d-objects may involve non-projectable HL
configurations for a $\mathbf{\acute{N}=\acute{N}}(x^{k},t)$ with generic
dependence on time and space coordinates. Such geometric data allow us to
construct an effective action 
\begin{equation}
\widehat{S}=\int \delta ^{3}\acute{u}\ \delta t\sqrt{\mathbf{\mathbf{\hat{g}}%
}_{i^{\prime }j^{\prime }}}\widehat{\mathbf{\acute{N}}}(\widehat{\mathbf{K}}%
_{i^{\prime }j^{\prime }}\widehat{\mathbf{K}}^{i^{\prime }j^{\prime }}-%
\widehat{\lambda }\widehat{\mathbf{K}}^{2}-\widehat{\mathcal{V}}),
\label{effects}
\end{equation}%
where $\widehat{\mathbf{K}}=\widehat{\mathbf{g}}^{i^{\prime }j^{\prime }}%
\widehat{\mathbf{K}}_{i^{\prime }j^{\prime }}$ and $\widehat{\lambda }$ is
the coupling constant (we put a hat symbol to distinguish such a notation
from a cosmological constant $\lambda $ in standard Einstein equations for $%
\nabla $). The term $\widehat{\mathcal{V}}=\widehat{\mathcal{V}}(\mathbf{%
\mathbf{\hat{g}}}_{i^{\prime }j^{\prime }},\mathbf{a}_{i^{\prime }}=\mathbf{e%
}_{i^{\prime }}\widehat{\mathbf{\acute{N}}}/\widehat{\mathbf{\acute{N}}})$
is called the potential. Such a complete potential is huge and it results in
very sophisticated types of nonlinear symmetries (\ref{nonlintrsmalp})
between the generating functions and generating sources. Such symmetries and
geometric data are encoded in (\ref{effects}) using geometric d-objects with
hat variables  For a general class of quasi-stationary of cosmological 
off-diagonal solutions, we can always transform effective sources in an
effective cosmological constant $\Lambda ,$ which facilitate the procedures
of constructiong exact/parametric solutions and performing BFV quantization.

In this work, we consider N-adapted nonholonomic generalizations of the
potential constructed and studied in \cite{cgs14,bbd22,bbd23,bbd24},%
\begin{equation}
\ ^{z=3}\widehat{\mathcal{V}}=-\alpha _{3}(\mathbf{\acute{D}}^{2}\acute{R})(%
\widehat{\mathbf{D}}_{i^{\prime }}\mathbf{a}^{i^{\prime }})-\alpha _{4}(%
\mathbf{\acute{D}}^{2}\mathbf{a}_{i^{\prime }})(\mathbf{\acute{D}}^{2}%
\mathbf{a}^{i^{\prime }})-\beta _{3}(\widehat{\mathbf{D}}_{i^{\prime }}%
\mathbf{\acute{R}}_{j^{\prime }k^{\prime }})(\widehat{\mathbf{D}}^{i^{\prime
}}\mathbf{\acute{R}}^{j^{\prime }k^{\prime }})-\beta _{4}(\widehat{\mathbf{D}%
}_{i^{\prime }}\acute{R})(\widehat{\mathbf{D}}^{i^{\prime }}\acute{R}),
\label{potz3}
\end{equation}%
where $\mathbf{\acute{D}}^{2}:=\widehat{\mathbf{D}}_{i^{\prime }}\widehat{%
\mathbf{D}}^{i^{\prime }},\mathbf{\acute{R}}_{i^{\prime }j^{\prime }}$ is
the Ricci d-tensor constructed from the $\mathbf{\mathbf{\hat{g}}}%
_{i^{\prime }j^{\prime }},\acute{R}=\mathbf{\mathbf{\hat{g}}}^{i^{\prime
}j^{\prime }}\mathbf{\acute{R}}_{i^{\prime }j;};$ and $\widehat{\lambda }%
,\alpha _{3},\alpha _{4},\beta _{3}$ and $\beta _{4}$ are coupling constants
of our HL deformation of the nonholonomic Einstein gravity. This potential
is constructed to satisfy the power-counting criterion for renormalizability
which requires terms of order $z=3$ for the theories in 3 spatial
dimensions. In the next section, we shall provide the formulas how the
propagators are defined in explicit forms. For the vertices, we have to know
the possible orders that they contributed in the spatial derivatives. The
nonholonomic structure of a Lorentz manifold enabled with a prime d-metric $%
\underline{\mathbf{\mathring{g}}}=[\underline{\mathring{g}}_{\alpha },%
\underline{\mathring{N}}_{i}^{a}]$ (\ref{offdiagpm}) can be organized in a
form to define any physical important solution in GR (for instance, a black
hole, a wormhole, or a cosmological metric etc.). The target off-diagonal
d-metric $\underline{\mathbf{g}}=[\underline{g}_{\alpha }=\underline{\eta }%
_{\alpha }\mathring{g}_{\alpha },\underline{N}_{i}^{a}=\underline{\eta }%
_{i}^{a}\ \underline{\mathring{N}}_{i}^{a}]$ (\ref{offdiagdefr}) may involve
arbitrary generating functions and generating sources, which model
configurations with effective $\ ^{z=3}\widehat{\mathcal{V}}$ (\ref{potz3})
for certain $\underline{\eta }_{\alpha }(x^{k},t)$ and $\underline{\eta }%
_{i}^{a}(x^{k},t)$ determined by $\underline{\eta }_{3}(x^{k},t)+\ ^{HL}%
\underline{\eta }_{3}(x^{k},t)$ (\ref{hlqdef}) defining HL configurations
(for classical condigurations, $\ ^{HL}\underline{\eta }_{3}\rightarrow 0$). 

Using any $\widehat{\mathcal{V}},$ we can introduce respective effective
classical Hamiltonian 
\begin{equation}
\widehat{H}_{0}=\int \delta ^{3}u\ \widehat{\mathcal{H}}_{0},\mbox{ for }\ 
\widehat{\mathcal{H}}_{0}\equiv \sqrt{\mathbf{\mathbf{\hat{g}}}_{i^{\prime
}j^{\prime }}}\widehat{\mathbf{\acute{N}}}\left[ |\mathbf{\mathbf{\hat{g}}}%
_{i^{\prime }j^{\prime }}|^{-1}(\widehat{\mathbf{\pi }}^{i^{\prime
}j^{\prime }}\widehat{\mathbf{\pi }}_{i^{\prime }j^{\prime }}+\frac{\widehat{%
\lambda }}{1-3\widehat{\lambda }}\widehat{\mathbf{\pi }}^{2}+\widehat{%
\mathcal{V}})\right] .  \label{effecthlham}
\end{equation}%
The nonholonomic structure on $\mathbf{V}$ and respective generating and
integration data are supposed to be prescribed for the conditions $1-3%
\widehat{\lambda }\neq 0$ when velocity type variables $\mathbf{e}_{4}%
\mathbf{\mathbf{\hat{g}}}_{i^{\prime }j^{\prime }}$ can be solved in the
favour of $\widehat{\mathbf{\pi }}^{i^{\prime }j^{\prime }}.$ This allows us
to determine conjugate nonholonomic pairs $(\mathbf{\mathbf{\hat{g}}}%
_{i^{\prime }j^{\prime }},\widehat{\mathbf{\pi }}^{i^{\prime }j^{\prime }}),$
where $\widehat{\mathbf{\pi }}\equiv \mathbf{\hat{g}}^{i^{\prime }j^{\prime
}}\widehat{\mathbf{\pi }}_{i^{\prime }j^{\prime }},$ and the canonical
momentum of $\widehat{\mathbf{\acute{N}}}$ can be fixed to be zero for
corresponding holonomic data of off-diagonal solutions determining such HL
type model. In this N-adapted ADM formalism (for holonomic Lorentz
manifolds, we obtain similar formulas as in \cite{br11,bbd24}). Here we note
that the momentum variables can be defined in the form 
\begin{equation*}
\widehat{\mathbf{\pi }}^{i^{\prime }j^{\prime }}=\sqrt{\mathbf{\mathbf{\hat{g%
}}}_{i^{\prime }j^{\prime }}}\mathbf{G}^{i^{\prime }j^{\prime }k^{\prime
}l^{\prime }}\widehat{\mathbf{K}}_{k^{\prime }l^{\prime }}\mbox{ for }%
\mathbf{G}^{i^{\prime }j^{\prime }k^{\prime }l^{\prime }}=\frac{1}{2}(%
\mathbf{\mathbf{\hat{g}}}^{i^{\prime }k^{\prime }}\mathbf{\mathbf{\hat{g}}}%
^{j^{\prime }l^{\prime }}+\mathbf{\mathbf{\hat{g}}}^{i^{\prime }l^{\prime }}%
\mathbf{\mathbf{\hat{g}}}^{j^{\prime }k^{\prime }})-\widehat{\lambda }%
\mathbf{\mathbf{\hat{g}}}^{i^{\prime }j^{\prime }}\mathbf{\mathbf{\hat{g}}}%
^{k^{\prime }l^{\prime }}.
\end{equation*}

We shall be able to apply the BFV procedure \cite{fv75,bv77,ff78} if we
identify the constraints that are involutive under certain Dirac brackets.
For the effective HL model (\ref{effecthlham}), such momentum constraints
can be defined for the generators of N-adapted spacial diffeomorphisms on
the canonical pair $(\mathbf{\hat{g}}_{i^{\prime }j^{\prime }},\widehat{%
\mathbf{\pi }}^{i^{\prime }j^{\prime }}),$ 
\begin{equation}
\widehat{\mathcal{H}}_{i^{\prime }}=-2\mathbf{\mathbf{\hat{g}}}_{i^{\prime
}j^{\prime }}\widehat{\mathbf{D}}_{k^{\prime }}\widehat{\mathbf{\pi }}%
^{k^{\prime }j^{\prime }}=0.  \label{momconst1}
\end{equation}%
We emphasize that the effective classical Lagrangiand (\ref{effecthlham})
and constraints (\ref{momconst1}) do not define a gravity which would be
equivanlent to certrain nonholonomic deformations of GR if they involve a
nontrivial generating HL-component $\ ^{HL}\underline{\eta }_{3}(x^{k},t)$ (%
\ref{hlqdef}). They define nonholonomic deformations of GR to  respective
HL-configurations which allows us to construct self-consistent quantum
models. In the classical limits, we have to impose  $\ ^{HL}\underline{\eta }%
_{3}(x^{k},t)\rightarrow 0.$

In this work, the term canonical has a double meaning: the variables are
defined for certain canonical distortion relation (\ref{canondistrel}) and
respective canonical off-diagonal solutions, (\ref{dmq}) or (\ref{dmc}), and
when canonical ADM formalism is defined in respective nonholonomic variables
(\ref{dma1}). The Dirack bracket $\{,\}_{D}$ between two $\widehat{\mathcal{H%
}}_{i^{\prime }}$ (\ref{momconst1}) coincides with the Poisson bracket,
which defines a d-algebra of spatial N-adapted diffeomorphism. For two
spatial points $^{1}u^{i^{\prime }}$ and $^{2}u^{i^{\prime }}$ and denoting
the N-adapted derivatives (\ref{nader}) in respective forms $\ ^{1}\mathbf{e}%
_{i^{\prime }}$ and $^{2}\mathbf{e}_{i^{\prime }}$, we define such a
canonical d-algebra 
\begin{equation*}
\{\widehat{\mathcal{H}}_{i^{\prime }}(^{1}u^{i^{\prime }}),\widehat{\mathcal{%
H}}_{j^{\prime }}(^{1}u^{i^{\prime }})\}_{D}=(\ ^{1}\mathbf{e}_{i^{\prime
}}\delta (\ ^{1}u-\ ^{2}u))\widehat{\mathcal{H}}_{j^{\prime }}(\ ^{1}u)-(\
^{2}\mathbf{e}_{j^{\prime }}\delta (\ ^{1}u-\ ^{2}u))\widehat{\mathcal{H}}%
_{i^{\prime }}(\ ^{2}u).
\end{equation*}%
We shall use the coefficients of this d-algebra for defining the BRST charge
and the gauge-fixed Hamiltonian (see next section).

For the above nonholonomic HL system, we can impose second-class constraints
given by the conditions of vanishing of the momentum conjugate to $\widehat{ 
\mathbf{\acute{N}}}$ (we considered them as solved): 
\begin{equation}
\ ^{1}\widehat{\theta }\equiv \widehat{\mathbf{\acute{N}}}\frac{\delta 
\widehat{H}_{0}}{\delta \widehat{\mathbf{\acute{N}}}}=\frac{\widehat{\mathbf{%
\acute{N}}}}{\sqrt{\mathbf{\mathbf{\hat{g}}}_{i^{\prime }j^{\prime }}}}(%
\widehat{\mathbf{\pi }}^{i^{\prime }j^{\prime }}\widehat{\mathbf{\pi }}%
_{i^{\prime }j^{\prime }}+\frac{\widehat{\lambda }}{1-3\widehat{\lambda }}%
\widehat{\mathbf{\pi }}^{2})+\sqrt{\mathbf{\mathbf{\hat{g}}}_{i^{\prime
}j^{\prime }}}\widehat{\mathbf{\acute{N}}}\widehat{\mathcal{V}}+\widehat{%
\mathcal{B}}=0.  \label{secclasscon}
\end{equation}%
In this formula, $\widehat{\mathcal{B}}$ is used for total N-adapted
derivatives, when the variations of $\mathbf{a}_{i^{\prime }}$ with respect
to $\widehat{\mathbf{\acute{N}}}$ produce such a total d-derivative $\delta 
\mathbf{a}_{i^{\prime }}=\mathbf{e}_{i^{\prime }}(\delta \widehat{\mathbf{%
\acute{N}}}/ \widehat{\mathbf{\acute{N}}}).$ This term contribute to the
propagators if we elaborate on a respective nonholonomic BFV quantum model
and come from effective $\ ^{z=3}\widehat{\mathcal{V}}$ (\ref{potz3}) such
that 
\begin{equation*}
\ ^{z=3}\widehat{\mathcal{B}}=-\alpha _{3}\sqrt{\mathbf{\mathbf{\hat{g}}}%
_{i^{\prime }j^{\prime }}}(\mathbf{\acute{D}}^{2}\ \widehat{\mathbf{\acute{N}%
}}\ \mathbf{\acute{D}}^{2}\acute{R})+2\alpha _{4}\sqrt{\mathbf{\mathbf{\hat{g%
}}}_{i^{\prime }j^{\prime }}}\widehat{\mathbf{D}}^{i^{\prime }}(\mathbf{%
\acute{D}}^{2}(\widehat{\mathbf{\acute{N}}}(\mathbf{\acute{D}}^{2}\mathbf{a}%
_{i^{\prime }}))).
\end{equation*}

We note that the integral of second class constraint (\ref{secclasscon}), $%
\widehat{H}_{0}=\int \delta ^{3}u\ \ ^{1}\widehat{\theta },$ are equivalent
to the effective Hamiltonian (\ref{effecthlham}) encoding the phase sector
with HL configurations. If the generating functions of off-diagonal
solutions are fixed to not determine such phases with anisotropic scaling,
we can consider the variant of nonholonomic BV quantization elaborated in 
\cite{vacaru25} (for associative and commutative geometric and quantum
d-objects). In both cases (as with nonholonomic HL induced off-diagonal
solutions from this work, or for respective phases with non-violated local
invariance), we can prove that asymptotic safe configurations exist \cite%
{vapny24}. 

\subsection{Quantization with canonical noholonomic variables and
second-class constraints}

Using the canonical nonholonomic pair $(\mathbf{\mathbf{\hat{g}}}_{i^{\prime
}j^{\prime }},\widehat{\mathbf{\pi }}^{i^{\prime }j^{\prime }}),$ the BFV
quantization can be performed by introducing (in our case) nonholonomic
ghost pairs, $(\mathcal{C}^{i^{\prime }},\overline{\mathcal{P}}_{i^{\prime
}}).$ and their duals involving Hermitian conjugation operations, $(%
\overline{\mathcal{C}}_{i^{\prime }},\mathcal{P}^{i^{\prime }}).$\footnote{%
To not exaggerate with abstract geometric notations, we do not write, for
instance, $(\widehat{\mathcal{C}}^{i^{\prime }},\overline{\widehat{\mathcal{P%
}}}_{i^{\prime }})$ considering that the observable canonical variables are
defined by hat ones as $(\mathbf{\mathbf{\hat{g}}}_{i^{\prime }j^{\prime }},%
\widehat{\mathbf{\pi }}^{i^{\prime }j^{\prime }})$.} We extend the geometric
constructions from a nonholonomic Lorentz manifold $\mathbf{V}$ to a
relativistic phase space $\widehat{\mathcal{V}}$ enabled with a measure $%
\delta (\ ^{1}\widehat{\theta })\det [\delta (\ ^{1}\widehat{\theta }%
)/\delta \widehat{\mathbf{\acute{N}}}]$ associated \ to the second-class
constraints (\ref{secclasscon}), see details and motivations in \cite%
{fv75,bbd24} and references therein. To elaborate on self-consistent quantum
models the phase space $\widehat{\mathcal{V}}$ is additionally HL-deformed
by using generating functions of type $\ ^{HL}\underline{\eta }_{3}(x^{k},t)$
(\ref{hlqdef}), which correspodingly modify the pairs  $(\mathbf{\mathbf{%
\hat{g}}}_{i^{\prime }j^{\prime }},\widehat{\mathbf{\pi }}^{i^{\prime
}j^{\prime }}),\widehat{H}_{0}$ (\ref{effecthlham}), $\{,\}_{D}$ between two 
$\widehat{\mathcal{H}}_{i^{\prime }}$ (\ref{momconst1}) etc.   So, we define
the nonholonomic BFV path integral $\widehat{\mathcal{Z}}$ (with a
respective fixed gauge condition),%
\begin{eqnarray}
\widehat{\mathcal{Z}} &\mathcal{=}&\int \mathcal{D}\widehat{\mathcal{V}}e^{i%
\widehat{\mathcal{S}}}\mbox{ for }\widehat{\mathcal{S}}\mathcal{=}\int \sqrt{%
\mathbf{\mathbf{\hat{g}}}_{i^{\prime }j^{\prime }}}\widehat{\mathbf{\acute{N}%
}}\delta ^{3}u\delta t(\widehat{\mathbf{\pi }}^{i^{\prime }j^{\prime }}%
\mathbf{\partial }_{t}\mathbf{\mathbf{\hat{g}}}_{i^{\prime }j^{\prime }}+%
\mathbf{\pi }_{i^{\prime }}\mathbf{\partial }_{t}(\widehat{\mathbf{\mathbf{N}%
}}^{i^{\prime }})+\mathcal{P}^{i^{\prime }}\mathbf{\partial }_{t}(\overline{%
\mathcal{C}}_{i^{\prime }})+\overline{\mathcal{P}}^{i^{\prime }}\mathbf{%
\partial }_{t}(\mathcal{C}_{i^{\prime }})-\ _{\psi }\widehat{\mathcal{H}});
\label{bfvpathint} \\
\mathcal{D}\widehat{\mathcal{V}} &\mathcal{=}&\mathcal{D}\mathbf{\mathbf{%
\hat{g}}}_{i^{\prime }j^{\prime }}\mathcal{D}\widehat{\mathbf{\pi }}%
^{i^{\prime }j^{\prime }}\mathcal{D}\widehat{\mathbf{\mathbf{N}}}^{k^{\prime
}}\mathcal{D}\mathbf{\pi }_{k^{\prime }}\mathcal{DC}^{i^{\prime }}\mathcal{D}%
\overline{\mathcal{P}}_{i^{\prime }}\mathcal{D}\overline{\mathcal{C}}%
_{i^{\prime }}\mathcal{DP}^{i^{\prime }}\times \delta (\ ^{1}\widehat{\theta 
})\det [\frac{\delta (\ ^{1}\widehat{\theta })}{\delta \widehat{\mathbf{%
\acute{N}}}}]\mbox{ is the integration measure }.  \label{integrmeas}
\end{eqnarray}%
In (\ref{bfvpathint}), the gauge-fixed Hamiltonian density $\ _{\psi }%
\widehat{\mathcal{H}}$ is defined for a gauge fermion $\psi ,$\footnote{%
we use a small Greek letter instead of a capital one \cite{bbd24} because in
this paper $\Psi $ is used as a generating function for $\mathbf{\hat{g}}%
[\Psi ]$ (\ref{qeltors})} 
\begin{eqnarray}
\ _{\psi }\widehat{\mathcal{H}} &=&\widehat{\mathcal{H}}_{0}+\{\psi ,\acute{%
\Omega}\}_{D},\mbox{ see }(\ref{effecthlham}),\mbox{ for }  \label{generbrst}
\\
\acute{\Omega} &=&\int \sqrt{\mathbf{\mathbf{\hat{g}}}_{i^{\prime }j^{\prime
}}}\delta ^{3}u(\widehat{\mathcal{H}}_{k^{\prime }}\mathcal{C}^{k^{\prime }}+%
\mathbf{\pi }_{i^{\prime }}\mathcal{P}^{i^{\prime }}-\mathcal{C}^{k^{\prime
}}\mathbf{e}_{k^{\prime }}\mathcal{C}^{i^{\prime }}\overline{\mathcal{P}}%
_{i^{\prime }})\mbox{ is the generator of the BRST symmetry}.  \notag
\end{eqnarray}

The integration mesure (\ref{integrmeas}) is incorporated in the quantum
action $\widehat{\mathcal{S}}$ with the help of new auxiliary variables (a
bosonic scalar field, $\mathcal{A},$ and a pair of scalar ghosts, $\acute{%
\eta},\overline{\acute{\eta}}$),%
\begin{equation*}
\delta (\ ^{1}\widehat{\theta })\det [\frac{\delta (\ ^{1}\widehat{\theta })%
}{\delta \widehat{\mathbf{\acute{N}}}}]=\int \mathcal{DA\ D}\acute{\eta}\ 
\mathcal{D}\overline{\acute{\eta}}\exp [i\int \sqrt{\mathbf{\mathbf{\hat{g}}}%
_{i^{\prime }j^{\prime }}}\delta ^{3}u\delta t(\mathcal{A}\ ^{1}\widehat{%
\theta }-\overline{\acute{\eta}}\frac{\delta (\ ^{1}\widehat{\theta })}{%
\delta \widehat{\mathbf{\acute{N}}}}\acute{\eta})].
\end{equation*}%
This also involves a gauge-fixing condition and its associated gauge fermion
defined by the equations%
\begin{equation*}
\mathbf{\partial }_{t}(\widehat{\mathbf{\mathbf{N}}}^{i^{\prime }})-\acute{%
\chi}^{i^{\prime }}=0\mbox{ and }\psi =\overline{\mathcal{P}}_{i^{\prime }}%
\widehat{\mathbf{\mathbf{N}}}^{i^{\prime }}+\overline{\mathcal{C}}%
_{i^{\prime }}\acute{\chi}^{i^{\prime }},
\end{equation*}%
where the coefficients $\acute{\chi}^{i^{\prime }}$ will be chosen below (we
use symbols with "prime" because such coefficients are different from the $\
\eta $- and $\chi $-polarizations in nonlinear symmetries (\ref%
{nonlintrsmalp})). With these setting, we write the path integral (\ref%
{bfvpathint}) in the form%
\begin{eqnarray}
\widehat{\mathcal{Z}} &\mathcal{=}&\int \mathcal{D}\mathbf{\mathbf{\hat{g}}}%
_{i^{\prime }j^{\prime }}\mathcal{D}\widehat{\mathbf{\pi }}^{i^{\prime
}j^{\prime }}\mathcal{D}\widehat{\mathbf{\mathbf{N}}}^{k^{\prime }}\mathcal{D%
}\mathbf{\pi }_{k^{\prime }}\mathcal{DC}^{i^{\prime }}\mathcal{D}\overline{%
\mathcal{P}}_{i^{\prime }}\mathcal{D}\overline{\mathcal{C}}_{i^{\prime }}%
\mathcal{DP}^{i^{\prime }}\mathcal{DA\ D}\acute{\eta}\ \mathcal{D}\overline{%
\acute{\eta}}  \label{bfvpathint1} \\
&&\exp [\int \sqrt{\mathbf{\mathbf{\hat{g}}}_{i^{\prime }j^{\prime }}}%
\widehat{\mathbf{\acute{N}}}\delta ^{3}u\delta t(\widehat{\mathbf{\pi }}%
^{i^{\prime }j^{\prime }}\mathbf{\partial }_{t}\mathbf{\mathbf{\hat{g}}}%
_{i^{\prime }j^{\prime }}+\mathbf{\pi }_{i^{\prime }}\mathbf{\partial }_{t}(%
\widehat{\mathbf{\mathbf{N}}}^{i^{\prime }})+\mathcal{P}^{i^{\prime }}%
\mathbf{\partial }_{t}(\overline{\mathcal{C}}_{i^{\prime }})+\overline{%
\mathcal{P}}^{i^{\prime }}\mathbf{\partial }_{t}(\mathcal{C}_{i^{\prime }})-%
\widehat{\mathcal{H}}_{0}-\widehat{\mathcal{H}}_{k^{\prime }}\widehat{%
\mathbf{\mathbf{N}}}^{k^{\prime }}  \notag \\
&&-\overline{\mathcal{P}}_{i^{\prime }}\mathcal{P}^{i^{\prime }}+\overline{%
\mathcal{P}}_{i^{\prime }}(\mathcal{C}^{k^{\prime }}\mathbf{e}_{k^{\prime }}%
\widehat{\mathbf{\mathbf{N}}}^{i^{\prime }}-\widehat{\mathbf{\mathbf{N}}}%
^{k^{\prime }}\mathbf{e}_{k^{\prime }}\mathcal{C}^{i^{\prime }})-\mathbf{\pi 
}_{i^{\prime }}\acute{\chi}^{i^{\prime }}-\overline{\mathcal{C}}_{i^{\prime
}}\{\acute{\chi}^{i^{\prime }},\widehat{\mathcal{H}}_{k^{\prime }}\}\mathcal{%
C}^{k^{\prime }})+\mathcal{A}\ ^{1}\widehat{\theta }-\overline{\acute{\eta}}%
\frac{\delta (\ ^{1}\widehat{\theta })}{\delta \widehat{\mathbf{\acute{N}}}}%
\acute{\eta}].  \notag
\end{eqnarray}%
In this formula, $\acute{\chi}^{i^{\prime }}$ do not depend on $\widehat{%
\mathbf{\mathbf{N}}}^{i^{\prime }}$ nor any of ghosts $\acute{\eta},%
\overline{\acute{\eta}}.$

Finally, we speculate how $\acute{\chi}^{i^{\prime }}$ can be defined in
explicit form. A typical example is to consider perturbative variables
around a flat background as in formulas (2.28) - (2.29) for Ho\v{r}ava
gravity in \cite{bbd24}. For nonholonomic Lorentz manifolds, we can
introduce N-adapted perturbative values around a primary off-diagonal
background $\mathbf{\mathring{g}=}[\mathring{g}_{\alpha },\mathring{N}%
_{i}^{a}]$ (without scaling anisotropies), or using general off-diagonal
transforms to a target d-metric $\mathbf{g,}$ $\mathbf{\mathring{g}}%
\rightarrow \underline{\mathbf{g}}=[\underline{g}_{\alpha }=\underline{\eta }%
_{\alpha }\mathring{g}_{\alpha },\underline{N}_{i}^{a}=\underline{\eta }%
_{i}^{a}\ \underline{\mathring{N}}_{i}^{a}]$ (\ref{offdiagdefr}). Target
d-metrics can encode HL configurations if we use generating functions $\
^{HL}\underline{\eta }_{3}(x^{k},t)$ (\ref{hlqdef}), and, in general, we can
consider both N-adapted and 3+1 decomposed perturbations:%
\begin{equation*}
\mathbf{\acute{g}}_{i^{\prime }j^{\prime }}=\underline{\mathbf{g}}%
_{i^{\prime }j^{\prime }}+\mathbf{h}_{i^{\prime }j^{\prime }},\widehat{%
\mathbf{\pi }}^{i^{\prime }j^{\prime }}=\mathbf{p}^{i^{\prime }j^{\prime }},%
\widehat{\mathbf{\mathbf{N}}}^{i^{\prime }}=\mathbf{\acute{n}}^{i^{\prime }}%
\mathbf{,}\widehat{\mathbf{\acute{N}}}=\mathbf{\acute{n}.}
\end{equation*}%
We can choose 
\begin{eqnarray}
\acute{\chi}^{i^{\prime }}[\underline{\mathbf{g}}_{i^{\prime }j^{\prime }}]
&=&\rho \left( \mathfrak{D}^{i^{\prime }j^{\prime }}\mathbf{\pi }_{j^{\prime
}}-2\triangle ^{2}\mathbf{e}_{j^{\prime }}\mathbf{h}^{i^{\prime }j^{\prime
}}+2\widehat{\lambda }(\kappa +1)\triangle ^{2}\mathbf{e}^{i^{\prime }}(%
\mathbf{h}_{k^{\prime }j^{\prime }}\underline{\mathbf{g}}^{k^{\prime
}j^{\prime }})-2\kappa \triangle \mathbf{e}^{i^{\prime }}(\mathbf{e}%
_{j^{\prime }}\mathbf{e}_{k^{\prime }}\mathbf{h}_{j^{\prime }k^{\prime
}})\right) ,\mbox{
where }  \notag \\
\mathfrak{D}^{i^{\prime }j^{\prime }} &=&\underline{\mathbf{g}}^{i^{\prime
}j^{\prime }}\triangle ^{2}+\kappa \triangle \mathbf{e}^{i^{\prime }}\mathbf{%
e}^{j^{\prime }},\triangle =\mathbf{e}_{k^{\prime }}\mathbf{e}_{k^{\prime }},
\label{chicoeff}
\end{eqnarray}%
for some independent constants $\rho $ and $\kappa .$ The procedure of gauge
fixing depends on the explicit form of $\acute{\chi}^{i^{\prime }}[%
\underline{\mathbf{g}}_{i^{\prime }j^{\prime }}].$ We can consider similar
perturbation schemes using $\acute{\chi}^{i^{\prime }}[\underline{\mathbf{%
\mathring{g}}}_{i^{\prime }j^{\prime }}].$ The coefficients in above
formulas are chosen to get certain analogy (for holonomic configurations)
and simplify the formulas for propagators as in \cite{bbd24}.

\subsection{The nonholonomic BRST transforms}

We elaborate on nonholonomic BVF quantization of the classes of off-diagonal
solutions \ defined by ansatz (\ref{dmc}) involving additional 3+1
decompositions and a respective nonholonomic ADM formalism. Let us begin
with nonlinear BRST symmetry transformations on nonholonomic canonical
fields $\Theta $ determined by $\acute{\Omega}$ from (\ref{generbrst}): $%
\delta _{\acute{\Omega}}\Theta =\{\Theta ,\acute{\Omega}\}_{D}\epsilon $,
where $\epsilon $ is the fermionic parameter of such transforms. To
constrain in N-adapted form the QG model on phase spaces as we defined in
previous section we consider such actions of the d-operator $\delta _{\acute{%
\Omega}}:$%
\begin{eqnarray*}
\delta _{\acute{\Omega}}\mathbf{\mathbf{\hat{g}}}_{i^{\prime }j^{\prime }}
&=&\delta _{\mathcal{C}\epsilon }\mathbf{\mathbf{\hat{g}}}_{i^{\prime
}j^{\prime }},\delta _{\acute{\Omega}}\widehat{\mathbf{\mathbf{N}}}%
^{i^{\prime }}=\mathcal{P}^{i^{\prime }}\epsilon ,\delta _{\acute{\Omega}}%
\widehat{\mathbf{\acute{N}}}=\delta _{\mathcal{C}\epsilon }\widehat{\mathbf{%
\acute{N}}};\delta _{\acute{\Omega}}\widehat{\mathbf{\pi }}^{i^{\prime
}j^{\prime }}=\delta _{\mathcal{C}\epsilon }\widehat{\mathbf{\pi }}%
^{i^{\prime }j^{\prime }},\delta _{\acute{\Omega}}\mathbf{\pi }_{j^{\prime
}}=0; \\
\delta _{\acute{\Omega}}\mathcal{C}^{i^{\prime }} &=&(\mathbf{e}_{j^{\prime
}}\mathcal{C}^{i^{\prime }})\mathcal{C}^{j^{\prime }}\epsilon ,\delta _{%
\acute{\Omega}}\mathcal{P}^{i^{\prime }}=0,\delta _{\acute{\Omega}}\overline{%
\mathcal{P}}^{i^{\prime }}=\delta _{\mathcal{C}\epsilon }\overline{\mathcal{P%
}}^{i^{\prime }},\delta _{\acute{\Omega}}\overline{\mathcal{C}}^{i^{\prime
}}=\mathbf{\pi }_{j^{\prime }}\epsilon .
\end{eqnarray*}%
In these formulas, $\delta _{\mathcal{C}\epsilon }\in $FDiffN determined by
a time-dependent d-vector parameter $\mathcal{C}^{i^{\prime }}\epsilon ;$
and $\widehat{\mathbf{\pi }}^{i^{\prime }j^{\prime }}$ and $\overline{%
\mathcal{P}}_{i^{\prime }}$ are time-dependent spatial d-tensor densities.
The d-operator $\delta _{\acute{\Omega}}$ also acts on auxiliary fields
(they are not canonical) when their BRST transformations are defined by a
left-invariant measure:%
\begin{equation*}
\delta _{\acute{\Omega}}\mathcal{A}=\delta _{\mathcal{C}\epsilon }\mathcal{A}%
,\delta _{\acute{\Omega}}\acute{\eta}=\delta _{\mathcal{C}\epsilon }\acute{%
\eta},\delta _{\acute{\Omega}}\overline{\acute{\eta}}=\delta _{\mathcal{C}%
\epsilon }\overline{\acute{\eta}}.
\end{equation*}%
In these formulas, the fields $\mathcal{A},\acute{\eta}$ and $\overline{%
\acute{\eta}}$ transform as time-dependent spatial scalar fields; the
product $\mathcal{A}\ ^{1}\widehat{\theta },$ see (\ref{secclasscon}), is
invariant on the constrained surface.

Let us denote in brief such a set of N-adapted fields%
\begin{equation}
\ ^{A}\varphi =\{\mathbf{\mathbf{\hat{g}}}_{i^{\prime }j^{\prime }},\widehat{%
\mathbf{\pi }}^{i^{\prime }j^{\prime }},\widehat{\mathbf{\mathbf{N}}}%
^{i^{\prime }},\widehat{\mathbf{\acute{N}}},\mathcal{A},\acute{\eta},%
\overline{\acute{\eta}}\}.  \label{setfields}
\end{equation}
Performing the integration on the ghost fields $\mathcal{P}^{i^{\prime }}$
and $\overline{\mathcal{P}}_{i^{\prime }}$ in the path integral (\ref%
{bfvpathint1}) and regrouping the terms for respective two sectors, we
obtain the canonical action:%
\begin{eqnarray}
\widehat{\mathcal{S}} &=&\ _{0}\widehat{\mathcal{S}}[\ ^{A}\varphi ]\mathcal{%
+}\ _{\acute{\Omega}}\widehat{\mathcal{S}},\mbox{ where }
\label{bfvpathint2} \\
\ _{0}\widehat{\mathcal{S}}[\ ^{A}\varphi ] &\mathcal{=}&\int \sqrt{\mathbf{%
\mathbf{\hat{g}}}_{i^{\prime }j^{\prime }}}\widehat{\mathbf{\acute{N}}}%
\delta ^{3}u\delta t((\widehat{\mathbf{\pi }}^{i^{\prime }j^{\prime }}%
\mathbf{\partial }_{t}\mathbf{\mathbf{\hat{g}}}_{i^{\prime }j^{\prime }}-%
\widehat{\mathcal{H}}_{0}-\widehat{\mathcal{H}}_{k^{\prime }}\widehat{%
\mathbf{\mathbf{N}}}^{k^{\prime }}\mathcal{A}\ ^{1}\widehat{\theta }-%
\overline{\acute{\eta}}\frac{\delta (\ ^{1}\widehat{\theta })}{\delta 
\widehat{\mathbf{\acute{N}}}}\acute{\eta})  \notag \\
\ _{\acute{\Omega}}\widehat{\mathcal{S}} &=&\int \sqrt{\mathbf{\mathbf{\hat{g%
}}}_{i^{\prime }j^{\prime }}}\widehat{\mathbf{\acute{N}}}\delta ^{3}u\delta
t[\mathbf{\pi }_{i^{\prime }}(\mathbf{\partial }_{t}(\widehat{\mathbf{%
\mathbf{N}}}^{i^{\prime }})-\acute{\chi}^{i^{\prime }})-\mathbf{\partial }%
_{t}(\overline{\mathcal{C}}_{i^{\prime }})(\mathbf{\partial }_{t}\mathcal{C}%
^{i^{\prime }}+\mathcal{C}^{k^{\prime }}\mathbf{e}_{k^{\prime }}\widehat{%
\mathbf{\mathbf{N}}}^{i^{\prime }}-\widehat{\mathbf{\mathbf{N}}}^{k^{\prime
}}\mathbf{e}_{k^{\prime }}\mathcal{C}^{i^{\prime }})-\overline{\mathcal{C}}%
_{i^{\prime }}\{\acute{\chi}^{i^{\prime }},\widehat{\mathcal{H}}_{k^{\prime
}}\}\mathcal{C}^{k^{\prime }}].  \notag
\end{eqnarray}%
As a consequence of the integration of ghost fields, we have to revise the
nonholonomic BRST symmetry transforms:%
\begin{eqnarray*}
\delta _{\acute{\Omega}}\mathbf{\mathbf{\hat{g}}}_{i^{\prime }j^{\prime }}
&=&\delta _{\mathcal{C}\epsilon }\mathbf{\mathbf{\hat{g}}}_{i^{\prime
}j^{\prime }},\delta _{\acute{\Omega}}\widehat{\mathbf{\mathbf{N}}}%
^{i^{\prime }}=\delta _{\mathcal{C}\epsilon }\widehat{\mathbf{\mathbf{N}}}%
^{i^{\prime }},\delta _{\acute{\Omega}}\widehat{\mathbf{\acute{N}}}=\delta _{%
\mathcal{C}\epsilon }\widehat{\mathbf{\acute{N}}};\delta _{\acute{\Omega}}%
\widehat{\mathbf{\pi }}^{i^{\prime }j^{\prime }}=\delta _{\mathcal{C}%
\epsilon }\widehat{\mathbf{\pi }}^{i^{\prime }j^{\prime }},\delta _{\acute{%
\Omega}}\mathbf{\pi }_{j^{\prime }}=0; \\
\delta _{\acute{\Omega}}\mathcal{C}^{i^{\prime }} &=&(\mathbf{e}_{j^{\prime
}}\mathcal{C}^{i^{\prime }})\mathcal{C}^{j^{\prime }}\epsilon ,\delta _{%
\acute{\Omega}}\overline{\mathcal{C}}^{i^{\prime }}=\mathbf{\pi }_{j^{\prime
}}\epsilon ;\delta _{\acute{\Omega}}\mathcal{A}=\delta _{\mathcal{C}\epsilon
}\mathcal{A},\delta _{\acute{\Omega}}\acute{\eta}=\delta _{\mathcal{C}%
\epsilon }\acute{\eta},\delta _{\acute{\Omega}}\overline{\acute{\eta}}%
=\delta _{\mathcal{C}\epsilon }\overline{\acute{\eta}}.
\end{eqnarray*}%
For such transforms, the d-operator $\delta _{\acute{\Omega}}$ defines
FDiffN along $\mathcal{C}^{i^{\prime }}\epsilon $ and the nonholonomic BRST
invariance of $\ _{0}\widehat{\mathcal{S}}[\ ^{A}\varphi ]$ become obvious.

Now, we can write the Q action (\ref{bfvpathint2}) in a standard form for
the BRST symmetry:%
\begin{eqnarray}
\widehat{\mathcal{S}} &=&\ _{0}\widehat{\mathcal{S}}[\ ^{A}\varphi ]\mathcal{%
+}\int \sqrt{\mathbf{\mathbf{\hat{g}}}_{i^{\prime }j^{\prime }}}\widehat{%
\mathbf{\acute{N}}}\delta ^{3}u\delta t\ \widehat{s}\tilde{\psi},%
\mbox{
where }  \label{bfvpathint3} \\
\ _{\acute{\Omega}}\widehat{\mathcal{S}} &=&\int \sqrt{\mathbf{\mathbf{\hat{g%
}}}_{i^{\prime }j^{\prime }}}\widehat{\mathbf{\acute{N}}}\delta ^{3}u\delta
t\ \widehat{s}\tilde{\psi},\mbox{ for }\tilde{\psi}=\overline{\mathcal{C}}%
_{i^{\prime }}(\mathbf{\partial }_{t}(\widehat{\mathbf{\mathbf{N}}}%
^{i^{\prime }})-\acute{\chi}^{i^{\prime }}),  \notag
\end{eqnarray}%
for $\acute{\chi}^{i^{\prime }}$ defined by formula (\ref{chicoeff}). In
these formulas, we denote by $\ \widehat{s}$ the N-adapted canonical BRST
d-operator defined by actions%
\begin{equation}
\widehat{s}\ ^{A}\varphi =\delta _{\mathcal{C}}^{A}\varphi ,\widehat{s}%
\mathcal{C}^{i^{\prime }}=-\mathcal{C}^{j^{\prime }}(\mathbf{e}_{j^{\prime }}%
\mathcal{C}^{i^{\prime }})\ ,\widehat{s}\overline{\mathcal{C}}_{i^{\prime }}=%
\mathbf{\pi }_{i^{\prime }},\widehat{s}\mathbf{\pi }_{i^{\prime }}=0.
\label{actbrst}
\end{equation}

The effective Lagrangian in (\ref{bfvpathint3}) is completely local. We can
extend this action at second order in perturbation and calculate propagators
as in section 4 of \cite{bbd24}. The results (using N-adapted indices) are
provided in Appendix \ref{apendpropag}. All propagators have a regular
structure under integration on the spatial momentum $k^{i^{\prime }}.$ The
locality of divergences produced by such integration is ensured by the
analysis of Lorentz-violating theories \cite{anselmi07,anselmi08}. In a
similar form, the superficial degree of divergence $D_{div}$ is computed
using the scaling of propagators and the maximal number of spatial
derivatives in the vertices. For an arbitrary (sub) diagram, the formula is $%
D_{div}=6-3\ _{p}E-2\ _{n}E-\ _{\pi }E-X,$ where $\ _{p}E $ is the number of
external $p_{i^{\prime }j^{\prime }}$ legs, $\ _{n}E$ is the number of
external $n^{i^{\prime }}$-legs, $\ _{\pi }E$ is the number of external $%
\mathbf{\pi }_{i^{\prime }}$-legs, and $X$ is the total number of spatial
derivatives on external legs. So, we obtain that $D_{div}=6$ is the highest
order of divergence.

\section{The N-adapted background field method and renormalization}

\label{sec04} In this section, we re-formulate the background-field approach
in N-adapted form to include canonical nonholonomic variables. The main goal
is to prove the renormalization of off-diagonal solutions in GR involving HL
structures.

\subsection{Canonical gravitational backgrounds}

Background fields are introduced in QFT with the aim to get a
background-gauge symmetry in the gauge-fixed quantum action. In this work,
we consider the set of quantum fields $\ ^{A}\varphi $ (\ref{setfields}) and
respective background filed $\ _{\flat }^{A}\varphi $ subjected to FDiffN
gauge transforms of nonholonomic Einstein equations (\ref{cdeq1}) and their
off-diagonal solutions refers to transformations (\ref{nagaugetrgf}) and (%
\ref{nagaugetrgftens}). In our approach, we consider only N-adapted
background fields only for $\mathbf{\mathbf{\hat{g}}}_{i^{\prime }j^{\prime
}}$ and $\widehat{\mathbf{\mathbf{N}}}^{i^{\prime }}$ with $\ _{\flat
}^{A}\varphi =\{\ _{\flat }\mathbf{\mathbf{\hat{g}}}_{i^{\prime }j^{\prime
}},\ _{\flat }\widehat{\mathbf{\mathbf{N}}}^{i^{\prime }}\}.$ Hereafter, we
shall use such notations for the differences of N-adapted fields:%
\begin{equation*}
\mathbf{h}_{i^{\prime }j^{\prime }}=\mathbf{\mathbf{\hat{g}}}%
_{i^{\prime}j^{\prime }}- \ _{\flat }\mathbf{\mathbf{\hat{g}}}_{i^{\prime
}j^{\prime }}\mbox{ and }\mathbf{n}^{i^{\prime }}=\widehat{\mathbf{\mathbf{N}%
}}^{i^{\prime }}-\ _{\flat }\widehat{\mathbf{\mathbf{N}}}^{i^{\prime }}.
\end{equation*}%
Not to confuse such notations with similar ones in the previous sections (in
those formulas, $i,j=1,2$ for dyadic splitting but $i^{\prime },j^{\prime
}=1,2,3$ for space nonholonomic ADM structures. Linearizing on the above
deformations, we get N-adapted gauge transforms as time-dependent spatial
d-vectors under background-gauge transformations acting as%
\begin{equation}
\ _{\zeta }\delta \mathbf{h}_{i^{\prime }j^{\prime }}=\zeta ^{k^{\prime }}%
\mathbf{e}_{k^{\prime }}\mathbf{h}_{i^{\prime }j^{\prime }}+2\mathbf{h}%
_{k^{\prime }(i^{\prime }}\mathbf{e}_{j^{\prime })}\zeta ^{k^{\prime }}%
\mbox{ and }\ _{\zeta }\delta \mathbf{n}^{i^{\prime }}=\zeta ^{k^{\prime }}%
\mathbf{e}_{k^{\prime }}\mathbf{n}^{i^{\prime }}-\mathbf{n}^{k^{\prime }}%
\mathbf{e}_{k^{\prime }}\zeta ^{i^{\prime }}.  \label{lingauge}
\end{equation}

The term $\ _{0}\widehat{\mathcal{S}}[\ ^{A}\varphi ]$ from (\ref%
{bfvpathint3}) is automatically invariant under the above linear gauge
transforms (\ref{lingauge}), but we have to replace the gauge fermion by an
N-adapted background one, 
\begin{eqnarray}
\tilde{\psi} &=&\overline{\mathcal{C}}_{i^{\prime }}(\mathbf{\partial }_{t}(%
\widehat{\mathbf{\mathbf{N}}}^{i^{\prime }})-\acute{\chi}^{i^{\prime
}})\rightarrow \ _{\flat }\tilde{\psi}=\overline{\mathcal{C}}_{i^{\prime
}}(\ _{\flat }\mathbf{D}_{t}\mathbf{n}^{i^{\prime }}-\rho \ _{\flat }\Theta
^{i^{\prime }j^{\prime }k^{\prime }}\mathbf{h}_{i^{\prime }j^{\prime }}-\
_{\flat }\mathcal{D}^{i^{\prime }j^{\prime }}(\mathbf{\pi }_{i^{\prime }}/%
\sqrt{\ _{\flat }\mathbf{\mathbf{\hat{g}}}_{i^{\prime }j^{\prime }}})) 
\notag \\
&&-\mathbb{T}^{i^{\prime }j^{\prime }}\mathbf{h}_{i^{\prime }j^{\prime }}-%
\mathbb{K}_{i^{\prime }j^{\prime }}\widehat{\mathbf{\pi }}^{i^{\prime
}j^{\prime }}-\mathbb{T}_{i^{\prime }}\mathbf{n}^{i^{\prime }}-\mathbb{T}%
\widehat{\mathbf{\acute{N}}}-\mathbb{S}\mathcal{A}-\overline{\mathbb{N}}%
\acute{\eta}-\overline{\acute{\eta}}\mathbb{N+}\ _{\flat }\overline{\mathbf{J%
}}_{i^{\prime }}\mathcal{C}^{i^{\prime }}.  \label{fermgaugetr}
\end{eqnarray}%
In this formula, we use such a collective d-source, $\ _{\flat }^{A}\gamma
=\{\mathbb{T}^{i^{\prime }j^{\prime }}, \mathbb{K}_{i^{\prime }j^{\prime }}, 
\mathbb{T}_{i^{\prime }},\mathbb{T},\mathbb{S},\overline{\mathbb{N}},\mathbb{%
N}\},$ when $_{\flat }\overline{\mathbf{J}}_{i^{\prime }}$ is defined as the
source for $\widehat{s}\overline{\mathcal{C}}_{i^{\prime }}$ in (\ref%
{actbrst}) (we do not consider sources for the auxiliary fields $\overline{%
\mathcal{C}}_{i^{\prime }}$ and $\mathbf{\pi }_{i^{\prime }}$). Raising and
lowering indices are defined by background data $\ _{\flat }\mathbf{\mathbf{%
\hat{g}}}_{i^{\prime }j^{\prime }}$ and $\ _{\flat }\widehat{\mathbf{D}}%
_{i^{\prime }}$ and used to define the canonical d-operators%
\begin{eqnarray*}
\ _{\flat }\mathbf{D}_{t}\mathbf{n}^{i^{\prime }} &=&\mathbf{\partial }_{t}%
\mathbf{n}^{i^{\prime }}-\ _{\flat }\widehat{\mathbf{\mathbf{N}}}^{k^{\prime
}}\ _{\flat }\widehat{\mathbf{D}}_{k^{\prime }}\mathbf{n}^{i^{\prime }}+%
\mathbf{n}^{k^{\prime }}\ _{\flat }\widehat{\mathbf{D}}_{k^{\prime }}\
_{\flat }\widehat{\mathbf{\mathbf{N}}}^{i^{\prime }}\ ,\ _{\flat }\mathcal{D}%
^{i^{\prime }j^{\prime }}=\ _{\flat }\mathbf{\mathbf{\hat{g}}}^{i^{\prime
}j^{\prime }}(\ _{\flat }\widehat{\mathbf{D}})^{4}+(\ _{\flat }\widehat{%
\mathbf{D}})^{2}\ _{\flat }\widehat{\mathbf{D}}^{i^{\prime }}\ _{\flat }%
\widehat{\mathbf{D}}^{j^{\prime }}, \\
\Theta ^{i^{\prime }j^{\prime }k^{\prime }} &=&-2\ _{\flat }\mathbf{\mathbf{%
\hat{g}}}^{i^{\prime }j^{\prime }}(\ _{\flat }\widehat{\mathbf{D}})^{4}\
_{\flat }\widehat{\mathbf{D}}^{k^{\prime }}+2\widehat{\lambda }\overline{%
\kappa }\mathbf{\mathbf{\hat{g}}}^{j^{\prime }k^{\prime }}(\ _{\flat }%
\widehat{\mathbf{D}})^{4}\ _{\flat }\widehat{\mathbf{D}}^{i^{\prime
}}-2\kappa (\ _{\flat }\widehat{\mathbf{D}})^{2}\ _{\flat }\widehat{\mathbf{D%
}}^{i^{\prime }}\ _{\flat }\widehat{\mathbf{D}}^{j^{\prime }}\ _{\flat }%
\widehat{\mathbf{D}}^{k^{\prime }},
\end{eqnarray*}%
for $(\ _{\flat }\widehat{\mathbf{D}})^{2}=\ _{\flat }\widehat{\mathbf{D}}%
^{k^{\prime }}\ _{\flat }\widehat{\mathbf{D}}_{k^{\prime }}.$ The above
values are defined in such a way that $\tilde{\psi}$ (\ref{fermgaugetr}) is
invariant under N-adapted background-gauge transformations.

Using above formulas, we can write the quantum (nonlinear) gauge-fixed
action in the presence of N-adapted background fields, 
\begin{eqnarray}
\ _{0}\widehat{\mathbf{\Sigma }} &=&\ _{0}\widehat{\mathcal{S}}[\
^{A}\varphi ]\mathcal{+}\int \sqrt{\mathbf{\mathbf{\hat{g}}}_{i^{\prime
}j^{\prime }}}\widehat{\mathbf{\acute{N}}}\delta ^{3}u\delta t[\mathbf{\pi }%
_{i^{\prime }}(\ _{\flat }\mathbf{D}_{t}\mathbf{n}^{i^{\prime }}-\rho \
_{\flat }\Theta ^{i^{\prime }j^{\prime }k^{\prime }}\mathbf{h}_{i^{\prime
}j^{\prime }}-\ _{\flat }\mathcal{D}^{i^{\prime }j^{\prime }}(\mathbf{\pi }%
_{i^{\prime }}/\sqrt{\ _{\flat }\mathbf{\mathbf{\hat{g}}}_{i^{\prime
}j^{\prime }}}))  \notag \\
&&+\overline{\mathcal{C}}_{i^{\prime }}\widehat{s}(\ _{\flat }\mathbf{D}_{t}%
\mathbf{n}^{i^{\prime }}-\rho \ _{\flat }\Theta ^{i^{\prime }j^{\prime
}k^{\prime }}\mathbf{h}_{j^{\prime }k^{\prime }})-\ _{\flat }^{A}\gamma 
\widehat{s}\ ^{A}\varphi +\ _{\flat }\overline{\mathbf{J}}_{i^{\prime }}%
\widehat{s}\mathcal{C}^{i^{\prime }}  \notag \\
&&+\ ^{A}\Omega \overline{\mathcal{C}}_{i^{\prime }}\frac{\delta }{\delta \
_{\flat }^{A}\varphi }(\ _{\flat }\mathbf{D}_{t}\mathbf{n}^{i^{\prime
}}-\rho \ _{\flat }\Theta ^{i^{\prime }j^{\prime }k^{\prime }}\mathbf{h}%
_{i^{\prime }k^{\prime }}-\ _{\flat }\mathcal{D}^{i^{\prime }j^{\prime }}(%
\mathbf{\pi }_{j^{\prime }}/\sqrt{\ _{\flat }\mathbf{\mathbf{\hat{g}}}%
_{i^{\prime }j^{\prime }}}))+\ ^{A}\Omega \ _{\flat }^{A}\gamma ]=\ _{0}%
\widehat{\mathcal{S}}[\ ^{A}\varphi ]\mathcal{+}\int \sqrt{\mathbf{\mathbf{%
\hat{g}}}_{i^{\prime }j^{\prime }}}\mathbf{Q}\ _{\flat }\tilde{\psi}.  \notag
\end{eqnarray}%
In these formulas, we use a nilpotent d-operator $\mathbf{Q=}\widehat{%
\mathbf{s}}+\ ^{A}\Omega \ \delta /\delta \ _{\flat }^{A}\varphi ,$ where $\
^{A}\Omega =\{\Omega _{i^{\prime }j^{\prime }},\Omega ^{i^{\prime }}\}$ are
external Grassman fields. We emphasize that such objects are different from
the curvature of N-connection involved in the computation of anholonomy
coefficients (\ref{nonholr}). 

\subsection{Renormalization of solutions with HL-structure}

On the basis of actions (\ref{bfvpathint3}), in an N-adapted BRST invariant
form, a corresponding gauge fixed action, the renormalization of effective
models defined by off-diagonal solutions of (\ref{cdeq1}) can be proven
following the procedure from \cite{barv17,bbd24}. We present a summary of
such renormalization.

The canonical action involving variables with double fibration at the $l$-th
order in loops (in the inductive approach, we assume that at order $(l-1) $
the divergences have been subtracted) can be expressed in the form 
\begin{equation*}
\ _{l}\widehat{\mathbf{\Sigma }}=\ _{l-1}\widehat{\mathbf{\Sigma }}-\hbar
^{l}\Gamma _{l,\infty }+\mathcal{O}(\hbar ^{l+1}).
\end{equation*}%
In this formula, the effective action $\Gamma $ is used as a functional of
the quantum fields ($\infty $ stands for its divergent part). Important
identities (the Slavnov-Taylor and the Ward identities, and the field
equations for the auxiliary d-fields $\mathbf{\pi }_{i^{\prime }}$) can be
established as in \cite{barv17}. They imply respective conditions of
annihilation:%
\begin{eqnarray}
\mathbf{Q}_{+}\check{\Gamma}_{l,\infty } &=&0,%
\mbox{ for a nilptent
operator, where }  \label{annih} \\
\check{\Gamma} &=&\Gamma -(\mathbf{\pi }_{i^{\prime }}+\ ^{A}\Omega 
\overline{\mathcal{C}}_{i^{\prime }}\frac{\delta }{\delta \ ^{A}\varphi }(%
\mathbf{D}_{t}\mathbf{n}^{i^{\prime }}-\rho \ \Theta ^{i^{\prime }j^{\prime
}k^{\prime }}\mathbf{h}_{i^{\prime }k^{\prime }}-\mathcal{D}^{i^{\prime
}j^{\prime }}(\mathbf{\pi }_{j^{\prime }}/\sqrt{\ \mathbf{\mathbf{\hat{g}}}%
_{i^{\prime }j^{\prime }}}))),  \notag \\
\mathbf{Q}_{+} &=&\frac{\delta \ _{0}\widehat{\mathbf{\Sigma }}}{\delta \
^{A}\widehat{\gamma }}\frac{\delta }{\delta \ ^{A}\varphi }+\frac{\delta \
_{0}\widehat{\mathbf{\Sigma }}}{\delta \ ^{A}\varphi }\frac{\delta }{\delta
\ ^{A}\widehat{\gamma }}+\frac{\delta \ _{0}\widehat{\mathbf{\Sigma }}}{%
\delta \mathbf{J}_{i^{\prime }}}\frac{\delta }{\delta \mathcal{C}^{i^{\prime
}}}+\frac{\delta \ _{0}\widehat{\mathbf{\Sigma }}}{\delta \mathcal{C}%
^{i^{\prime }}}\frac{\delta }{\delta \mathbf{J}_{i^{\prime }}}+\ ^{A}\Omega 
\frac{\delta }{\delta \ ^{A}\varphi },  \notag \\
\widehat{\gamma }_{i^{\prime }} &=&\mathbb{T}_{i^{\prime }}-\overline{%
\mathcal{C}}_{i^{\prime }}\ \mathbf{D}_{t},\ \widehat{\gamma }^{i^{\prime
}j^{\prime }}=\mathbb{T}^{i^{\prime }j^{\prime }}+\rho \ \overline{\mathcal{C%
}}_{k^{\prime }}\Theta ^{k^{\prime }i^{\prime }j^{\prime }},\ ^{A}\widehat{%
\gamma }=\ \ ^{A}\gamma \mbox{ for other sets of indices }.  \notag
\end{eqnarray}%
In the above formulas, $\ _{0}\widehat{\mathbf{\Sigma }}$ is the tree-level
reduced action which does not depend explicitly on the N-adapted background
fields,%
\begin{equation*}
\ _{0}\widehat{\mathbf{\Sigma }}=\ _{0}\widehat{\mathcal{S}}[\ ^{A}\varphi ]%
\mathcal{+}\int \sqrt{\mathbf{\mathbf{\hat{g}}}_{i^{\prime }j^{\prime }}}%
\widehat{\mathbf{\acute{N}}}\delta ^{3}u\delta t\mathbf{Q}(\tilde{\psi}-\
^{A}\widehat{\gamma }\ ^{A}\varphi +\overline{\mathbf{J}}_{i^{\prime }}%
\mathcal{C}^{i^{\prime }})
\end{equation*}

The solution of (\ref{annih}) at $l$-order can be written in the form:%
\begin{equation}
\check{\Gamma}_{l,\infty }=\widehat{\mathbf{S}}_{l}[\ ^{A}\varphi ]+\mathbf{Q%
}_{+}\yen _{l}.  \label{qsol}
\end{equation}%
This formula is obtained by a procedure of expanding $\check{\Gamma}%
_{l,\infty }$ on the ghost fields $\mathcal{C}^{i^{\prime }}$ and $\
^{A}\Omega ,$ the cohomology of $\mathbf{Q}_{+}$ and other d-operators which
guarantee the existence of $\widehat{\mathbf{S}}_{l}[\ ^{A}\varphi ]$ and $%
\yen _{l}.$ Introducing the solution (\ref{qsol}) in (\ref{annih}), we find
such recurrent formulas for the $l$-th order \ gauge fermion $\tilde{\psi}%
_{l} $ and respective counter-term, $\widehat{\mathbf{\Sigma }}^{C}$:%
\begin{equation*}
\tilde{\psi}_{l}=\tilde{\psi}_{l-1}-\hbar ^{l}\yen _{l}\mbox{ and }\widehat{%
\mathbf{\Sigma }}^{C}=-\widehat{\mathbf{S}}_{l}[\ ^{A}\varphi ]-\mathbf{Q}%
_{+}\yen _{l}.
\end{equation*}%
Then, re-defining the fields as in \cite{barv17} but for N-adapted
configurations, $(\ ^{A}\varphi ,\mathcal{C}^{i^{\prime }})\rightarrow $ $(\
^{A}\check{\varphi},\mathcal{\check{C}}^{i^{\prime }}),$ when%
\begin{equation*}
\ ^{A}\varphi =\ ^{A}\check{\varphi}+\hbar ^{l}\frac{\delta \yen _{l}}{%
\delta \ ^{A}\widehat{\gamma }}(\ ^{A}\check{\varphi},\mathcal{\check{C}}%
^{i^{\prime }},...)+\mathcal{O(}\hbar ^{l+1}),\mathcal{C}^{i^{\prime }}=%
\mathcal{\check{C}}^{i^{\prime }}-\hbar ^{l}\frac{\delta \yen _{l}}{\delta \
^{A}\widehat{\gamma }}(\ ^{A}\check{\varphi},\mathcal{\check{C}}^{i^{\prime
}},...)+\mathcal{O(}\hbar ^{l+1}),
\end{equation*}%
we obtain the $l$-th order quantum action%
\begin{eqnarray}
\ _{l}\widehat{\mathbf{\Sigma }} &=&\ _{l}\widehat{\mathcal{S}}[\
^{A}\varphi ]\mathcal{+}\int \sqrt{\mathbf{\mathbf{\hat{g}}}_{i^{\prime
}j^{\prime }}}\widehat{\mathbf{\acute{N}}}\delta ^{3}u\delta t\mathbf{Q}%
\tilde{\psi}_{l},\mbox{ where }  \label{actq4} \\
\tilde{\psi}_{l} &=&(\mathbf{D}_{t}\mathbf{n}^{i^{\prime }}-\rho \ \Theta
^{i^{\prime }j^{\prime }k^{\prime }}\mathbf{h}_{i^{\prime }k^{\prime }}-%
\mathcal{D}^{i^{\prime }j^{\prime }}(\mathbf{\pi }_{j^{\prime }}/\sqrt{\ 
\mathbf{\mathbf{\hat{g}}}_{i^{\prime }j^{\prime }}})))-\ _{\flat }^{A}\gamma
(\ ^{A}\varphi -\ _{\flat }^{A}\varphi )+\overline{\mathbf{J}}_{i^{\prime }}%
\mathcal{C}^{i^{\prime }}+\mathcal{O(}\hbar ).  \notag
\end{eqnarray}%
The term $\ _{l}\widehat{\mathcal{S}}[\ ^{A}\varphi ]$ in action (\ref{actq4}%
) is a FDiffN gauge invariant functional and possess a N-adapted BRST
structure.

Finally, we note that the nonholonomic generalized BFV quantization
technique can be applied directly to quantize in general form any class of
off-diagonal solutions determined by an ansatz of type (\ref{dmq}), i.e. for
quasi-stationary configurations; and (\ref{2confgenansatz}), involving an
additional time-depending 2-d conformal factor. In particular, we can
quantize various types of black hole, wormhole, and toroid solutions which
are subjected to off-diagonal deformations involving time and space scaling
with respective nonholonomic HL sectors. How to perform such geometric
quantum constructions we shall study in our future works. In the next
section, we apply above formulated nonholonomic geometric quantum formalism
to a "simpler" case when the BFV formalism is applied to HL deformed locally
anisotropic configurations in GR. 

\subsection{HL-configurations and locally anisotropic cosmology}

The class of off-diagonal locally anisotropic solutions determined by ansatz
of type (\ref{dmc}) consists a special example because it can be related in
effective form to certain (2+1)-d Ho\v{r}ava type models. This is possible
if we consider classes of solutions of type (\ref{offdiagpolcosm}), or (\ref%
{qeltorsc}), when the generating function is of type $\underline{\eta }%
_{3}(x^{k},t)\simeq \sigma _{0}\phi (x^{k},t),$ $\sigma _{0}=const,$
subjected to HL-conditions (\ref{anisotract}) and (\ref{localscaltransf}).

The ansatz (\ref{dmc}) involve two spacial coordinates, $x^{i},$ and the
time-like one, $t,$ when the nonholonomic structure can be chosen in a form
that the geometric d-objects a labeled by indices $i,j,...=1,2$ instead of $%
i^{\prime },j^{\prime },...=1,2,3.$ In certain N-adapted frame and systems
of coordinates, we eliminate the dependence on space coordinate $u^{3}=x^{3}$%
, $N_{i}^{3}(x^{k},t)=0,$ and the off-diagonal cosmological d-metric is
trivially extended to 4-d configurations via a term $+(dx^{3})^{2},$ when 
\begin{equation}
ds^{2}=g_{ij}(x^{k})+(dx^{3})^{2}+g_{4}(x^{k},t)(dt+N_{i}^{4}(x^{k},t)dx^{i})^{2}.
\label{dmef3d}
\end{equation}%
Such d-metrics define exact or parametric solutions of (\ref{cdeq1}) in 4-d
nonholonomic Einstein gravity, but encode via nonlinear gauge and 2-d
vertical conformal transforms as in (\ref{2confgenansatz}) a nontrivial HL
(2+1)-d phase sector. We can write equivalently (\ref{dmef3d}) in any form (%
\ref{dma1}) or (\ref{dmab}) with a nontrivial $\mathbf{\acute{N}}(x^{k},t).$

We can quantize locally anisotropic cosmological systems (\ref{dmef3d}) by
generalizing in nonholonomic canonical forms the results and methods from 
\cite{svw11,bbd22,bbd23,bbd24}. Such approaches require a potential of $z=2 $
order of power-counting renormalizability when all inequivalent terms $z=1,2 
$ are compatible, with the FDiffN symmetry. The (2+1)-d and $z=2$ version of
(\ref{potz3}) can written in the form 
\begin{eqnarray}
\ ^{z=2}\widehat{\mathcal{V}} &=&-\check{\beta}\acute{R}-\check{\alpha}%
\mathbf{a}_{i}\mathbf{a}^{i}+\check{\alpha}_{1}\acute{R}^{2}+\check{\alpha}%
_{2}(\mathbf{a}_{i}\mathbf{a}^{i})^{2}+\check{\alpha}_{3}\acute{R}\mathbf{a}%
_{i}\mathbf{a}^{i}+\check{\alpha}_{4}\mathbf{a}_{i}\mathbf{a}^{i}\widehat{%
\mathbf{D}}_{k}\mathbf{a}^{k}  \notag \\
&&+\check{\alpha}_{5}\acute{R}(\widehat{\mathbf{D}}_{i}\mathbf{a}^{i})+%
\check{\alpha}_{6}(\widehat{\mathbf{D}}^{k}\mathbf{a}^{j})(\widehat{\mathbf{D%
}}_{k}\mathbf{a}_{j})+\check{\alpha}_{7}(\widehat{\mathbf{D}}_{k}\mathbf{a}%
^{k})^{2},  \label{21theory}
\end{eqnarray}%
where $\mathbf{a}_{i}=\mathbf{e}_{i}\mathbf{\acute{N}}/\mathbf{\acute{N}})$
and other d-objects $\widehat{\mathbf{D}}_{k},\acute{R},$... are computed
for hat variables for (\ref{dmef3d}) involving nonholonomic HL
(2+1)-configurations. The constants $\check{\beta},\check{\alpha},\check{%
\alpha}_{1},...,\check{\alpha}_{7}$ defines the effective quantum HL model (%
\ref{21theory}). We put "inverse hats" to distinguish such configurations
from the (3+1) above considered in previous sections. 

The primary classical Hamiltonian resulting in $\ ^{z=2}\widehat{\mathcal{V}}
$ (\ref{21theory}) has a structure which is similar to the N-adapted
functional form (\ref{effecthlham}), with the same combinations of constants 
$\widehat{\sigma }$ $=\widehat{\lambda }/(1-3\widehat{\lambda })$ and $%
\overline{\sigma }=(1-\widehat{\lambda })/(1-3\widehat{\lambda })$ used in
formulas (\ref{potz3}) and (\ref{chicoeff}) but with respective $\check{\beta%
},\check{\alpha},\check{\alpha}_{1},..$. The analogues of second-class
constraints (\ref{secclasscon}) are defined from the condition of vanishing
of the momentum conjugate to $\mathbf{\acute{N}}$ for a respective class of
locally anisotropic cosmological solutions. In N-adapted form, for $\widehat{%
\mathbf{\pi }}=\widehat{\mathbf{\pi }}^{ij}\widehat{g}_{ij}$ ($\widehat{g}%
_{ij}=g_{ij}(x^{k})$ is a canonical d-metric (\ref{dmc}) redefined as (\ref%
{dmef3d}); here $\widehat{\mathbf{D}}^{2}=\widehat{\mathbf{D}}_{k}\widehat{%
\mathbf{D}}^{k}$) we write 
\begin{eqnarray}
\ ^{1}\widehat{\theta } &\equiv &\sqrt{|\widehat{g}_{ij}|}\mathbf{\acute{N}(}%
\frac{\widehat{\mathbf{\pi }}^{ij}\widehat{\mathbf{\pi }}_{ij}}{|\widehat{g}%
_{ij}|}+\overline{\sigma }\frac{\widehat{\mathbf{\pi }}^{2}}{|\widehat{g}%
_{ij}|}+\widehat{\mathcal{V}}\ \mathbf{)+}\sqrt{|\widehat{g}_{ij}|}[2\check{%
\alpha}\widehat{\mathbf{D}}_{k}(\mathbf{\acute{N}a}^{k})-4\check{\alpha}_{2}%
\widehat{\mathbf{D}}_{k}(\mathbf{\acute{N}\mathbf{a}}^{i}\mathbf{\mathbf{a}}%
_{k}\mathbf{a}^{k})-2\check{\alpha}_{3}\widehat{\mathbf{D}}_{k}(\mathbf{%
\acute{N}\acute{R}a}^{k})  \label{secclconst2} \\
&&+\check{\alpha}_{4}\left( \widehat{\mathbf{D}}^{2}(\mathbf{\acute{N}%
\mathbf{a}}_{k}\mathbf{a}^{k})-2\widehat{\mathbf{D}}_{k}(\mathbf{\acute{N}%
\mathbf{a}}^{k}\ \widehat{\mathbf{D}}_{i}\mathbf{a}^{i})\right) +\check{%
\alpha}_{5}\widehat{\mathbf{D}}^{2}(\mathbf{\acute{N}\acute{R}})+2\check{%
\alpha}_{6}\widehat{\mathbf{D}}^{i}\widehat{\mathbf{D}}^{j}(\mathbf{\acute{N}%
}\widehat{\mathbf{D}}_{j}\mathbf{a}_{k})+2\check{\alpha}_{7}\widehat{\mathbf{%
D}}^{2}(\mathbf{\acute{N}}\widehat{\mathbf{D}}_{j}\mathbf{a}^{j})]=0  \notag
\end{eqnarray}%
The primary Hamiltonian for such locally anisotropic cosmological HL
configurations is equivalent to the integral of (\ref{secclconst2}), $%
\widehat{H}_{0}=\int d^{2}x\ ^{1}\widehat{\theta }.$ This is similar to the
3+1 case (\ref{secclasscon}) with $\widehat{H}_{0}=\int \delta ^{3}u\ \ ^{1}%
\widehat{\theta },$ but for a different integration measures.

There are two ways to perform the BFV quantization of the above effective HL
cosmological theory. The first one is to repeat in N-adapted form the
procedures from \cite{bbd22,bbd24}. We can also consider all formulas for
the above nonholonomic 3+1 cofigurations, by redefining the propagators \ref%
{apendpropag} by a respective changing of the constants and indices as in (%
\ref{21theory}). We omit such technical results in this work.

Let us explain how to include in the constraints (\ref{secclconst2})
classical cosmological solutions in GR with possible nonholonomic
deformations of the Einstein equations as off-diagonal solutions of (\ref%
{cdeq1}). We can consider a prime d-metric $\underline{\mathbf{\mathring{g}}}%
=[\underline{\mathring{g}}_{\alpha },\underline{\mathring{N}}_{i}^{a}]$ (\ref%
{offdiagdefr}) for a cosmological model in GR, or an off-diagonal
deformation, $\mathbf{\mathring{g}}\rightarrow \underline{\mathbf{g}},$ to a
target locally anisotroic cosmological configuration $\underline{\mathbf{g}}$
(\ref{offdiagpolcosm}) in GR or related MGT \cite{v16,v18}. We can consider
a chain of such nonholonomic transforms $\mathbf{\mathring{g}}\rightarrow 
\underline{\mathbf{g}}\rightarrow \ ^{HL}\underline{\mathbf{g}}$ for
respective gravitational $\eta $-polarization functions defining 
\begin{equation*}
\underline{\mathbf{g}}=[\underline{g}_{\alpha }=\underline{\eta }_{\alpha }%
\mathring{g}_{\alpha },\underline{N}_{i}^{a}=\underline{\eta }_{i}^{a}\ 
\underline{\mathring{N}}_{i}^{a}]\mbox{ and }\ ^{HL}\underline{\mathbf{g}}%
=[\ ^{HL}\underline{g}_{\alpha }=(\underline{\eta }_{\alpha }+\ ^{HL}%
\underline{\eta }_{\alpha })g_{\alpha },\ ^{HL}\underline{N}_{i}^{a}=(%
\underline{\eta }_{i}^{a}+\ ^{HL}\underline{\eta }_{i}^{a})\ \underline{N}%
_{i}^{a}].
\end{equation*}%
From an infinite number of generating functions we can consider subsets with 
$\ ^{HL}\underline{\eta }_{\alpha }=$ $\underline{\eta }_{a}(x^{k},t)$ (\ref%
{hlqdef}) which define HL\ structures with time- and space- anisotropic
scaling (\ref{anisotract}) and (\ref{localscaltransf}). Corresponding
classes of off-diagonal cosmological solutions are defined by nonlinear
cosmological evolution equations with nonlinear symmetries resulting in
effective cosmological constants. The prime configurations can be chosen  as
some physically important classical solutions of the (modified) Einstein
equations when the target solutions can be selected/ generated to be of HL
type. These off-diagonal cosmological models in GR are self-consistently
defined in the classical limit and can be renormalized using generating
functions $\ ^{HL}\underline{\eta }_{\alpha }$ allowing to work the
framework of the nonholonomic BFV formalism. Locally anisotropic
cosmological models of this type have been extensively studied in the
context of generalized Finsler and HL gravity theories formulated at the
classical level; for reviews, see \cite{v16,v18}. Further research on
accelerating cosmology scenarios, as well as dark energy and dark matter
physics, can be developed by encoding quantum models based on (3+1)- and
(2+1)-dimensional renormalizable theories with effective Hamiltonians,
incorporating nonholonomic constraints of the type (\ref{secclasscon}) and (%
\ref{secclconst2}). This approach opens new and promising perspectives for
the study of quantum cosmology and anisotropic gravitational dynamics. 

\section{Conclusions and perspectives}

\label{sec05} UV renormalization of Ho\v{r}av-Lifshitz (HL) gravity has been
established for specific projectable and non-projectable models, with
asymptotic freedom demonstrated in several cases \cite%
{horava1,horava2,barv23,bbd24}. BRST and BFV quantization techniques provide
efficient tools for handling modified gravity theories with broken Lorentz
symmetry \cite{fv75,bv77,ff78,br11,svw11,bbd22,bbd23}. However, Ho\v{r}ava
gravity generally does not recover Einstein's general relativity in the
quasi-classical limit without additional, theoretically opaque assumptions.
Formulating classical and quantum models on Lorentz manifolds and
(co)tangent bundles offers well-defined causal structures and nonlinear QFT
constructions that can capture renormalizable configurations more
consistently. 

\vskip4pt Even though various black hole thermodynamic models and
cosmological scenarios have been developed within the HL framework \cite%
{mukohyama10,du19}, these constructions do not correspond to solutions of
the Einstein equations in GR. Certain diagonal ansatzes encoding HL
space-time anisotropic scaling could potentially be quantized following the
formalism of \cite{bbd22,bbd23,bbd24}, yet such models generally fail to
produce quasi-classical limits that match exact or parametric solutions in
GR. These observations are often invoked as serious arguments for
reconsidering or extending the standard paradigms of GR and QG. 

\vskip4pt 
We have shown that the orthodox paradigm of general relativity can be consistently extended to generate effective, self-consistent, and HL – renormalizable models via nonholonomic BFV quantization. This extension is realized through broad classes of off-diagonal solutions to canonically distorted Einstein equations, constructed using advanced geometric methods with double nonholonomic 2+2 and  3+ 1 splittings. The 2+2 dyadic formalism enables the explicit application of the AFCDM to generate exact and parametric solutions in GR and modified gravity theories. A key advantage of this geometric–analytic approach is that very general and physically relevant solution classes in GR can be obtained in fully off-diagonal form, determined by generating functions $\eta _{a}\rightarrow \eta _{a}+\ ^{HL}\eta _{a}$  (see, for example, (\ref{hlqdef})). The 3+1 ADM-type splitting, in turn, facilitates the construction of effective Hamiltonians encoding HL-type structures and their BFV quantization. By imposing suitable nonholonomic second-class constraints and N-adapted gauge-fixing conditions, together with nonholonomic BRST transformations, one achieves a consistent treatment of nonlinear symmetries and HL-type deformations. This demonstrates the feasibility of unifying GR-based geometric methods with HL-inspired quantum constructions. It should be emphasized that within our framework one may choose (or, conversely, dynamically generate) classical off-diagonal quasi-stationary or cosmological configurations in GR, while HL-type phases are introduced specifically for the elaboration of quantum gravity models.


\vskip4pt 
The AFCDM enables the explicit construction of off-diagonal solutions of the (modified) Einstein equations that depend on spacetime and phase-space variables through generating functions and effective sources associated with physical constants or boundary data. Specific subclasses of these generating functions define solutions that either remain within standard GR or effectively model HL – type spacetime anisotropies. The resulting classical gravity configurations are determined by the chosen classes of generating functions and effective sources, corresponding either to locally Lorentz-invariant dynamics or to HL-type deformations. Certain terms in the generating functions can be selected such that all classical GR properties are preserved, while additional contributions encode dynamical quantum modeling of nonlinear interactions and their associated symmetries. Within this framework, locally anisotropic cosmological solutions can be quantized using N-adapted BFV methods, leading to renormalizable models and suggesting a viable pathway toward a nonholonomic BFV quantization of (2+1)-dimensional HL cosmological configurations.

\vskip4pt 
We conclude that the results of this work strongly support the central hypothesis stated in the Introduction: generic off-diagonal solutions of the Einstein equations in GR, formulated in canonical dyadic variables with equivalent nonholonomic ADM data, admit a consistent quantization framework. Such solutions can be quantized via suitable nonholonomic generalizations of the BFV formalism, incorporating additional distortions encoded by generating functions with Ho\v{r}ava–Lifshitz–type deformations. The resulting classes of quantized configurations are perturbatively renormalizable for arbitrary primary d-metrics in GR that are off-diagonally deformed into effective HL-type geometries.

We conjecture that alternative classes of solutions, which do not admit such nonholonomic and off-diagonal structures, exhibit poor quantum behavior and fail to define consistent quantum gravity models. In this framework, locally anisotropic spacetime scaling arises dynamically from nonlinear off-diagonal gravitational and matter-field interactions, leading to phases characterized by effective local Lorentz symmetry breaking. In the (quasi-)classical limit, the theory reproduces physically relevant quasi-stationary and locally anisotropic configurations in GR, as well as in certain modified gravity theories.

\vskip4pt 
Finally, we conclude that the methods and results developed in this work can be applied to the construction of quasi-stationary and cosmological off-diagonal solutions for coupled Einstein–Yang–Mills–Higgs–Dirac systems. The BFV quantization of such nonlinear configurations provides a consistent framework with potential relevance for effective modeling of dark energy and dark matter sectors. Moreover, nonholonomic BFV methods play a significant role in classical and quantum information flow theories, as well as in quantum gravity. Closely related developments in (nonassociative) gauge gravity and BV quantization have been presented recently in \cite{partner06,vapny24,vacaru25}.

\vskip5pt \textbf{Acknowledgement:} 

This work was carried out within the framework of a visiting fellowship of SV at Kocaeli University, T\"{u}rkiye, and builds upon earlier volunteer research programs conducted at California State University, Fresno, USA, and Taras Shevchenko National University of Kyiv, Ukraine. The authors are grateful to the editors and referees for their insightful and constructive comments, which enabled a substantial extension of the manuscript and contributed to making the geometric and quantum methods -- specifically the synthesis of the AFCDM and BFV formalism -- more accessible to researchers in modern cosmology and astrophysics.

\appendix\setcounter{equation}{0} 
\renewcommand{\theequation}
{A.\arabic{equation}} \setcounter{subsection}{0} 
\renewcommand{\thesubsection}
{A.\arabic{subsection}}

\section{The AFCDM for off-diagonal integration of the Einstein equations}

\label{appendixa}

In this appendix, we provide necessary formulas summarizing the AFCDM for
generating exact and parametric off-diagonal solutions in GR, see reviews of
results and proofs in \cite{vv25,vacaru26,vacaru18,partner02,partner06}.

\subsection{Parametrization of off-diagonal solutions and their nonlinear
symmetries}

Let us introduce parameterizations for a d-metric (\ref{dm}) and respective
off-diagonal ansatz (\ref{ansatz}) which allow to generate quasi-stationary
solutions (\ref{dmq}) for (\ref{cdeq1}). We consider a $\mathbf{\hat{g}}=$ $%
\underline{\widehat{g}}_{\alpha \beta }(u)du^{\alpha }\otimes du^{\beta }$,
when the respective coefficients are parameterized in the form: 
\begin{equation}
\widehat{\underline{g}}_{\alpha \beta } = \left[ 
\begin{array}{cccc}
e^{\psi }+(w_{1})^{2}h_{3}+(n_{1})^{2}h_{4} & w_{1}w_{2}h_{3}+n_{1}n_{2}h_{4}
& w_{1}h_{3} & n_{1}h_{4} \\ 
w_{1}w_{2}h_{3}+n_{1}n_{2}h_{4} & e^{\psi }+(w_{2})^{2}h_{3}+(n_{2})^{2}h_{4}
& w_{2}h_{3} & n_{2}h_{4} \\ 
w_{1}h_{3} & w_{2}h_{3} & h_{3} & 0 \\ 
n_{1}h_{4} & n_{2}h_{4} & 0 & h_{4}%
\end{array}%
\right] ,  \label{qeltorsoffd}
\end{equation}%
where $g_{i}=g_{i}(x^{k})$ and $g_{a}=g_{a}(x^{k},y^{a});$ $%
N_{i}^{3}=w_{i}(x^{k},y^{3})$ and $N_{i}^{a}=n_{i}(x^{k},y^{3}).$

We consider $g_{1}=g_{2}=e^{\psi (x^{k})},$ were $\psi (x^{k})$ is a
solution of 2-d Poisson equations $\partial _{11}^{2}\psi +\partial
_{22}^{2}\psi =2\ ^{v}\Upsilon (x^{k})$. Then we introduce as a generating
function $\Psi (x^{k},y^{3})=\exp (\varpi ),$ where $\varpi =\ln
|h_{4}^{\ast }/\sqrt{|h_{3}h_{4}}|,$ for $h_{4}^{\ast }=\partial _{3}h_{4},$
and use generating sources $\ ^{h}\Upsilon (x^{k})$ and $\ ^{v}\Upsilon
(x^{k},y^{3})\,$ as in (\ref{esourc}). Tedious proofs \cite%
{vv25,vacaru26,vacaru18,partner02,partner06} show that quadratic elements 
\begin{eqnarray}
d\widehat{s}^{2} &=&e^{\psi (x^{k},\ ^{h}\Upsilon
)}[(dx^{1})^{2}+(dx^{2})^{2}]+\frac{[\Psi ^{\ast }]^{2}}{4(\ ^{v}\Upsilon
)^{2}\{g_{4}^{[0]}-\int dy^{3}[\Psi ^{2}]^{\ast }/4(\ ^{v}\Upsilon )\}}%
(dy^{3}+\frac{\partial _{i}\Psi }{\Psi ^{\ast }}dx^{i})^{2}+  \label{qeltors}
\\
&&\{g_{4}^{[0]}-\int dy^{3}\frac{[\Psi ^{2}]^{\ast }}{4(\ ^{v}\Upsilon )}%
\}\{dt+[\ _{1}n_{k}+\ _{2}n_{k}\int dy^{3}\frac{[(\Psi )^{2}]^{\ast }}{4(\
^{v}\Upsilon )^{2}|g_{4}^{[0]}-\int dy^{3}[\Psi ^{2}]^{\ast }/4(\
^{v}\Upsilon )|^{5/2}}]dx^{k}\},  \notag
\end{eqnarray}%
define off-diagonal quasi-stationary solutions of the Einstein equations (%
\ref{cdeq1}) written in canonical nonholonomic variables. In (\ref{qeltors}%
), $g_{4}^{[0]}(x^{k}),\ _{1}n_{k}(x^{k})$ and $\ _{2}n_{k}(x^{k})$ are
integration functions. 

A very important and new property of the quasi-stationary solutions (\ref%
{qeltors}) is that they are described by some nonlinear symmetries which
allow use different types of generating functions and generating. Such
nonlinear symmetries also allow introducing certain effective cosmological
constants which are important, for instance, for computing G. Perelman's
variables \cite{partner06,vacaru18}. By tedious computations, we can prove
that off-diagonal solutions (\ref{qeltors}) admit a changing of the
generating data, $(\Psi ,\ \ ^{v}\Upsilon )\leftrightarrow (\Phi ,\
^{v}\Lambda =const\neq 0).$ In general, a v-cosmological constant $\
^{v}\Lambda $ may be different from a h-cosmological constant $\ ^{h}\Lambda
,$ but in GR we can consider an effective $\ \ ^{h}\Lambda =\ ^{v}\Lambda =$ 
$\Lambda .$ So, the quasi-stationary solutions $\mathbf{\hat{g}}[\Psi ]$ (%
\ref{qeltors}) of $\widehat{\mathbf{R}}_{\ \ \beta }^{\alpha }[\Psi ]=%
\widehat{\mathbf{\Upsilon }}_{\ \ \beta }^{\alpha }$ (\ref{cdeq1}) can be
expressed in equivalent forms as solutions of 
\begin{equation}
\widehat{\mathbf{R}}_{\ \ \beta }^{\alpha }[\Phi ]=\Lambda \mathbf{\delta }%
_{\ \ \beta }^{\alpha }.  \label{cdeq1a}
\end{equation}

The conditions that quasi-stationary configurations (\ref{qeltors}) are
transformed into solutions of (\ref{cdeq1a}) are defined by such
differential or integral equations (we can check this by straightforward
computations): 
\begin{equation}
\frac{\lbrack \Psi ^{2}]^{\ast }}{\ ^{v}\Upsilon } =\frac{[\Phi ^{2}]^{\ast }%
}{\Lambda },\mbox{ or integrating,  } \Phi ^{2} =\ \Lambda \int dy^{3}(\
^{v}\Upsilon )^{-1}[\Psi ^{2}]^{\ast }\mbox{ and/or }\Psi ^{2}=\Lambda
^{-1}\int dy^{3}(\ ^{v}\Upsilon )[\Phi ^{2}]^{\ast }.  \label{ntransf2}
\end{equation}%
Using (\ref{ntransf2}), we obtain $h_{4}=h_{4}^{[0]}-\frac{\ \Phi ^{2}}{4\
\Lambda },$ which allows us to express the coefficients of d-metrics in
terms of new generating data. First, we write $(\Psi )^{\ast }/\
^{v}\Upsilon $ in terms of such $(\Phi ,\Lambda )$ and then write (\ref%
{ntransf2}) as $\frac{\Psi (\ \Psi )^{\ast }}{\ ^{v}\Upsilon }=\frac{(\Phi
^{2})^{\ast }}{2\Lambda }$ and $\ \Psi =|\Lambda |^{-1/2}\sqrt{|\int dy^{3}\
^{v}\Upsilon \ (\Phi ^{2})^{\ast }|}.$ Second, we introduce $\Psi $ from the
above second equation in the first above equation and re-define $\Psi ^{\ast
}$ in terms of generating data $(\ ^{v}\Upsilon ,\Phi ,\Lambda ),$ when $%
\frac{\Psi ^{\ast }}{\ ^{v}\Upsilon }=\frac{[\Phi ^{2}]^{\ast }}{2\sqrt{|\
\Lambda \int dy^{3}(\ ^{v}\Upsilon )[\Phi ^{2}]^{\ast }|}}.$ This allows us
to conclude that any quasi-stationary solution (\ref{qeltors}) possesses
important nonlinear symmetries of type (\ref{ntransf2}) which are trivial or
do not exist for the diagonal ansatz. 

The quasi-stationary solutions (\ref{qeltors}) can be written in an
equivalent form using generating data $(\ ^{v}\Upsilon ,\Phi ,\Lambda )$ 
\begin{eqnarray}
d\widehat{s}^{2} &=&\widehat{g}_{\alpha \beta }(x^{k},y^{3},\Phi ,\Lambda
)du^{\alpha }du^{\beta }=e^{\psi (x^{k})}[(dx^{1})^{2}+(dx^{2})^{2}]-\frac{%
\Phi ^{2}[\Phi ^{\ast }]^{2}}{|\Lambda \int dy^{3}\ ^{v}\Upsilon \lbrack
\Phi ^{2}]^{\ast }|[h_{4}^{[0]}-\Phi ^{2}/4\Lambda ]}\{dy^{3}
\label{offdiagcosmcsh} \\
&&+\frac{\partial _{i}\ \int dy^{3}\ ^{v}\Upsilon \ [\Phi ^{2}]^{\ast }}{\
^{v}\Upsilon \ [(\ \Phi )^{2}]^{\ast }}dx^{i}\}^{2}-\{h_{4}^{[0]}-\frac{\Phi
^{2}}{4\Lambda }\}\{dt+[\ _{1}n_{k}+\ _{2}n_{k}\int dy^{3}\frac{\Phi
^{2}[\Phi ^{\ast }]^{2}}{|\Lambda \int dy^{3}\ ^{v}\Upsilon \lbrack \Phi
^{2}]^{\ast }|[h_{4}^{[0]}-\Phi ^{2}/4\Lambda ]^{5/2}}]\}.  \notag
\end{eqnarray}%
In this formula, the generating functions are parameterized $\psi (x^{k})$
and $\Phi (x^{k},y^{3}),$ when the effective cosmological constant is $\
\Lambda .$ The quasi-stationary solutions represented in the form (\ref%
{offdiagcosmcsh}) "re-distribute" into respective off-diagonal forms the
prescribed generating data $(\Psi ,\ ^{v}\Upsilon )$ into another type of
ones, $(\Phi ,\ \Lambda )$. Here we note that the contributions of a
generating source $\ ^{v}\Upsilon $ are not completely transformed into a
cosmological constant $\Lambda .$

Taking the partial derivative on $y^{3}$ of the coefficient $h_{4}$ from (%
\ref{qeltors}) allows us to write $h_{4}^{\ast }=-[\Psi ^{2}]^{\ast }/4\
^{v}\Upsilon .$ So, prescribing $h_{4}(x^{i},y^{3})$ and $\ ^{v}\Upsilon
(x^{i},y^{3})$, we can compute (up to an integration function) a generating
function $\ \Psi $ \ defined by $[\Psi ^{2}]^{\ast }=\int dy^{3}\
^{v}\Upsilon h_{4}^{\ast }.$ This allows us to work in equivalent form with
generating data $(h_{4},\ ^{v}\Upsilon )$ and rewrite the quadratic element (%
\ref{qeltors}) as 
\begin{eqnarray}
d\widehat{s}^{2} &=&\widehat{g}_{\alpha \beta }(x^{k},y^{3};h_{4},\
^{v}\Upsilon )du^{\alpha }du^{\beta }=e^{\psi
(x^{k})}[(dx^{1})^{2}+(dx^{2})^{2}]-\frac{(h_{4}^{\ast })^{2}}{|\int
dy^{3}[\ \ ^{v}\Upsilon h_{4}]^{\ast }|\ h_{4}}\times
\label{offdsolgenfgcosmc} \\
&&\{dy^{3}+\frac{\partial _{i}[\int dy^{3}(\ ^{v}\Upsilon )\ h_{4}^{\ast }]}{%
\ ^{v}\Upsilon \ h_{4}^{\ast }}dx^{i}\}^{2}+h_{4}\{dt+[\ _{1}n_{k}+\
_{2}n_{k}\int dy^{3}\frac{(h_{4}^{\ast })^{2}}{|\int dy^{3}[\ ^{v}\Upsilon
h_{4}]^{\ast }|\ (h_{4})^{5/2}}]dx^{k}\}.  \notag
\end{eqnarray}

The nonlinear symmetries (\ref{ntransf2}) allow us to perform similar
computations related to (\ref{offdiagcosmcsh}). Expressing $\Phi ^{2}=-4\
\Lambda h_{4},$ we can eliminate $\Phi $ from the nonlinear element and
generate a solution of type (\ref{offdsolgenfgcosmc}) which are determined
by the generating data $(h_{4};\Lambda ,\ ^{v}\Upsilon ).$ 

\subsection{Decomposing off-diagonal solutions on a small parameter}

We can consider nonlinear transforms (\ref{nonlintrsmalp}) for (\ref%
{offdiagpolfr}) involving $\kappa $-linear decompositions with generating
functions as $\chi $-polarizations. This way, we define small nonholonomic
deformations of a prime d-metric $\mathbf{\mathring{g}}$ into so-called $%
\kappa $-parametric solutions with $\zeta $- and $\chi $-coefficients: 
\begin{eqnarray}
\psi &\simeq &\psi (x^{k})\simeq \psi _{0}(x^{k})(1+\kappa \ _{\psi }\chi
(x^{k})),\mbox{ for }\   \label{epsilongenfdecomp} \\
\ \eta _{2} &\simeq &\eta _{2}(x^{k_{1}})\simeq \zeta _{2}(x^{k})(1+\kappa
\chi _{2}(x^{k})),\mbox{ we consider }\ \eta _{2}=\ \eta _{1};\eta
_{4}\simeq \eta _{4}(x^{k},y^{3})\simeq \zeta _{4}(x^{k},y^{3})(1+\kappa
\chi _{4}(x^{k},y^{3})).  \notag
\end{eqnarray}%
In these formulas, $\psi $ and $\eta _{2}=\ \eta _{1}$ are such way chosen
to relate $g_{1}$ and $g_{2}$ to the solutions of the 2-d Poisson equation $%
\partial _{11}^{2}\psi +\partial _{22}^{2}\psi =2\ ^{v}\Upsilon (x^{k}).$

Introducing formulas (\ref{epsilongenfdecomp}) for respective coefficients
of a d-metric, we compute $\kappa $-parametric deformations with $\chi $%
-generating functions, 
\begin{equation*}
d\ \widehat{s}^{2}=\widehat{g}_{\alpha \beta }(x^{k},y^{3};\psi
,g_{4};^{v}\Upsilon )du^{\alpha }du^{\beta }=e^{\psi _{0}}(1+\kappa \ ^{\psi
}\chi )[(dx^{1})^{2}+(dx^{2})^{2}]
\end{equation*}%
\begin{eqnarray*}
&&-\{\frac{4[\partial _{3}(|\zeta _{4}\mathring{g}_{4}|^{1/2})]^{2}}{%
\mathring{g}_{3}|\int dy^{3}\{\ \ ^{v}\Upsilon \partial _{3}(\zeta _{4}%
\mathring{g}_{4})\}|}-\kappa \lbrack \frac{\partial _{3}(\chi _{4}|\zeta _{4}%
\mathring{g}_{4}|^{1/2})}{4\partial _{3}(|\zeta _{4}\mathring{g}_{4}|^{1/2})}%
-\frac{\int dy^{3}\{\ ^{v}\Upsilon \partial _{3}[(\zeta _{4}\mathring{g}%
_{4})\chi _{4}]\}}{\int dy^{3}\{\ ^{v}\Upsilon \partial _{3}(\zeta _{4}%
\mathring{g}_{4})\}}]\}\mathring{g}_{3} \\
&&\{dy^{3}+[\frac{\partial _{i}\ \int dy^{3}\ ^{v}\Upsilon \ \partial
_{3}\zeta _{4}}{(\mathring{N}_{i}^{3})\ ^{v}\Upsilon \partial _{3}\zeta _{4}}%
+\kappa (\frac{\partial _{i}[\int dy^{3}\ ^{v}\Upsilon \ \partial _{3}(\zeta
_{4}\chi _{4})]}{\partial _{i}\ [\int dy^{3}\ ^{v}\Upsilon \partial
_{3}\zeta _{4}]}-\frac{\partial _{3}(\zeta _{4}\chi _{4})}{\partial
_{3}\zeta _{4}})]\mathring{N}_{i}^{3}dx^{i}\}^{2}
\end{eqnarray*}%
\begin{eqnarray}
&&+\zeta _{4}(1+\kappa \ \chi _{4})\ \mathring{g}_{4}\{dt+[(\mathring{N}%
_{k}^{4})^{-1}[\ _{1}n_{k}+16\ _{2}n_{k}[\int dy^{3}\frac{\left( \partial
_{3}[(\zeta _{4}\mathring{g}_{4})^{-1/4}]\right) ^{2}}{|\int dy^{3}\partial
_{3}[\ ^{v}\Upsilon (\zeta _{4}\mathring{g}_{4})]|}]  \label{offdncelepsilon}
\\
&&+\kappa \frac{16\ _{2}n_{k}\int dy^{3}\frac{\left( \partial _{3}[(\zeta
_{4}\mathring{g}_{4})^{-1/4}]\right) ^{2}}{|\int dy^{3}\partial _{3}[\
^{v}\Upsilon (\zeta _{4}\mathring{g}_{4})]|}(\frac{\partial _{3}[(\zeta _{4}%
\mathring{g}_{4})^{-1/4}\chi _{4})]}{2\partial _{3}[(\zeta _{4}\mathring{g}%
_{4})^{-1/4}]}+\frac{\int dy^{3}\partial _{3}[\ ^{v}\Upsilon (\zeta _{4}\chi
_{4}\mathring{g}_{4})]}{\int dy^{3}\partial _{3}[\ ^{v}\Upsilon (\zeta _{4}%
\mathring{g}_{4})]})}{\ _{1}n_{k}+16\ _{2}n_{k}[\int dy^{3}\frac{\left(
\partial _{3}[(\zeta _{4}\mathring{g}_{4})^{-1/4}]\right) ^{2}}{|\int
dy^{3}\partial _{3}[\ ^{v}\Upsilon (\zeta _{4}\mathring{g}_{4})]|}]}]%
\mathring{N}_{k}^{4}dx^{k}\}^{2}.  \notag
\end{eqnarray}%
We can redefine this formula using $t$-transforms to describe small
off-diagonal transforms of cosmological solutions (with underlined symbols).
The parametric solutions (\ref{offdncelepsilon}) allow us to define, for
instance, ellipsoidal deformations of BH metrics into BE ones, or to
describe an off-diagonal quasi-stationary vacuum structure. 

\subsubsection{Space and time duality of quasi-stationary and cosmological
solutions}

\label{stduality}We considered such a specific space and time duality
between ansatz (\ref{dmq}) and (\ref{dmc}). A corresponding duality
principle can be formulated in a general form for generic off-diagonal
solutions. It allows to use of abstract geometric formulas and does not
repeat all computations presented for quasi-stationary metrics with
nontrivial partial derivatives $\partial _{3}$ for generating locally
anisotropic cosmological solutions with nontrivial partial derivatives $%
\partial _{4}=\partial _{t}$.

So, the \textbf{principle of space and time duality } of generic
off-diagonal configurations with one Killing symmetry on a space-like $%
\partial _{3}$ or time-like $\partial _{t}$ can formulated: 
\begin{eqnarray*}
y^{3} &\longleftrightarrow &y^{4}=t,h_{3}(x^{k},y^{3})\longleftrightarrow 
\underline{h}_{4}(x^{k},t),h_{4}(x^{k},y^{3})\longleftrightarrow \underline{h%
}_{3}(x^{k},t), \\
N_{i}^{3} &=&w_{i}(x^{k},y^{3})\longleftrightarrow N_{i}^{4}=\underline{n}%
_{i}(x^{k},t),N_{i}^{4}=n_{i}(x^{k},y^{3})\longleftrightarrow N_{i}^{3}=%
\underline{w}_{i}(x^{k},t)
\end{eqnarray*}%
allows to transform (\ref{dmq}) into (\ref{dmc}), and inversely. In explicit
form, we can change respectively 
\begin{equation*}
\Upsilon _{~3}^{3}=\Upsilon _{~4}^{4}=~^{v}\Upsilon
(x^{k},y^{3})\longleftrightarrow \underline{\Upsilon }_{~4}^{4}=\underline{%
\Upsilon }_{~3}^{3}=~^{v}\underline{\Upsilon }(x^{k},t),\mbox{ see }(\ref%
{esourc});
\end{equation*}%
\begin{equation}
\begin{array}{ccc}
\begin{array}{c}
(\Psi ,~^{v}\Upsilon )\leftrightarrow (\mathbf{g},\ ~^{v}\Upsilon
)\leftrightarrow \\ 
(\eta _{\alpha }\ \mathring{g}_{\alpha }\sim (\zeta _{\alpha }(1+\kappa \chi
_{\alpha })\mathring{g}_{\alpha },~^{v}\Upsilon )\leftrightarrow%
\end{array}
& \Longleftrightarrow & 
\begin{array}{c}
(\underline{\Psi },\ ~^{v}\underline{\Upsilon })\leftrightarrow (\underline{%
\mathbf{g}},\ ~^{v}\underline{\Upsilon })\leftrightarrow \\ 
(\underline{\eta }_{\alpha }\ \underline{\mathring{g}}_{\alpha }\sim (%
\underline{\zeta }_{\alpha }(1+\kappa \underline{\chi }_{\alpha })\underline{%
\mathring{g}}_{\alpha },\ ~^{v}\underline{\Upsilon })\leftrightarrow%
\end{array}
\\ 
\begin{array}{c}
(\Phi ,\ \Lambda )\leftrightarrow (\mathbf{g},\ \Lambda )\leftrightarrow \\ 
(\eta _{\alpha }\ \mathring{g}_{\alpha }\sim (\zeta _{\alpha }(1+\kappa \chi
_{\alpha })\mathring{g}_{\alpha },\ \Lambda ),%
\end{array}
& \Longleftrightarrow & 
\begin{array}{c}
(\underline{\Phi },\ \underline{\Lambda })\leftrightarrow (\underline{%
\mathbf{g}},\ \underline{\Lambda })\leftrightarrow \\ 
(\underline{\eta }_{\alpha }\ \underline{\mathring{g}}_{\alpha }\sim (%
\underline{\zeta }_{\alpha }(1+\kappa \underline{\chi }_{\alpha })\underline{%
\mathring{g}}_{\alpha },\ \underline{\Lambda }).%
\end{array}%
\end{array}
\label{dualnonltr}
\end{equation}%
The duality conditions are also extended to the corresponding systems of
nonlinear PDE (to which the nonholonomic Einstein equations (\ref{cdeq1})
reduce for the corresponding off-diagonal ansatz): 
\begin{equation}
\begin{array}{ccc}
\Psi ^{\ast }h_{4}^{\ast }=2h_{3}h_{4}\ ~^{v}\Upsilon \Psi , & 
\longleftrightarrow & \sqrt{|\underline{h}_{3}\underline{h}_{4}|}\underline{%
\Psi }=\underline{h}_{3}^{\diamond }, \\ 
\sqrt{|h_{3}h_{4}|}\Psi =h_{4}^{\ast }, & \longleftrightarrow & \underline{%
\Psi }^{\diamond }\underline{h}_{3}^{\diamond }=2\underline{h}_{3}\underline{%
h}_{4}\ \ ~^{v}\underline{\Upsilon }\underline{\Psi }, \\ 
\Psi ^{\ast }w_{i}-\partial _{i}\Psi =\ 0, & \longleftrightarrow & 
\underline{n}_{i}^{\diamond \diamond }+\left( \ln \frac{|\underline{h}%
_{3}|^{3/2}}{|\underline{h}_{4}|}\right) ^{\diamond }\underline{n}%
_{i}^{\diamond }=0, \\ 
\ n_{i}^{\ast \ast }+\left( \ln \frac{|h_{4}|^{3/2}}{|h_{3}|}\right) ^{\ast
}n_{i}^{\ast }=0 & \longleftrightarrow & \underline{\Psi }^{\diamond }%
\underline{w}_{i}-\partial _{i}\underline{\Psi }=\ 0.%
\end{array}%
.  \label{dualcosm}
\end{equation}

Considering (\ref{dualcosm}), the nonlinear symmetries (\ref{ntransf2}) can
be written in respective dual forms (for locally anisotropic cosmological
solutions): 
\begin{equation*}
\frac{\lbrack \underline{\Psi }^{2}]^{\diamond }}{~^{v}\underline{\Upsilon }}%
=\frac{[\underline{\Phi }^{2}]^{\diamond }}{\underline{\Lambda }},%
\mbox{ which can be
integrated as  }\underline{\Phi }^{2}=\ \underline{\Lambda }\int dt(~^{v}%
\underline{\Upsilon })^{-1}[\underline{\Psi }^{2}]^{\diamond }%
\mbox{ and/or
}\underline{\Psi }^{2}=(\underline{\Lambda })^{-1}\int dt(~^{v}\underline{%
\Upsilon })[\underline{\Phi }^{2}]^{\diamond }.
\end{equation*}%
Using such nonlinear symmetries, we can redefine for different types of
cosmological models the quasi-stationary d-metrics (\ref{qeltors}), (\ref%
{offdiagcosmcsh}), (\ref{offdsolgenfgcosmc}), (\ref{offdiagpolfr}) and (\ref%
{offdncelepsilon}). As an example of applications of such an abstract
symbolic calculus, we provide the formula for the dualized d-metric (\ref%
{qeltors}): 
\begin{eqnarray}
d\underline{s}^{2} &=&e^{\psi
(x^{k})}[(dx^{1})^{2}+(dx^{2})^{2}]+\{g_{3}^{[0]}-\int dt\frac{[\underline{%
\Psi }^{2}]^{\diamond }}{4~^{v}\underline{\Upsilon }}\}\{dy^{3}+[\ _{1}n_{k}
\label{qeltorsc} \\
&&+\ _{2}n_{k}\int dt\frac{[(\underline{\Psi })^{2}]^{\diamond }}{4(\ ~^{v}%
\underline{\Upsilon })^{2}|g_{3}^{[0]}-\int dt[\underline{\Psi }%
^{2}]^{\diamond }/4\ ~^{v}\underline{\Upsilon }|^{5/2}}]dx^{k}\}+\frac{[%
\underline{\Psi }^{\diamond }]^{2}}{4(\ ~^{v}\underline{\Upsilon }%
)^{2}\{g_{3}^{[0]}-\int dt[\underline{\Psi }^{2}]^{\diamond }/4\ ~^{v}%
\underline{\Upsilon }\}}(dt+\frac{\partial _{i}\underline{\Psi }}{\underline{%
\Psi }^{\diamond }}dx^{i})^{2}.  \notag
\end{eqnarray}

\subsubsection{Constraints on generating functions and sources for
extracting LC configurations}

\label{salc}The generic off--diagonal solutions of (\ref{cdeq1}) from the
previous subsections are constructed for an auxiliary canonical
d--connection $\widehat{\mathbf{D}}.$ In general, such solutions are
characterized by nonholonomically induced d--torsion coefficients $\widehat{%
\mathbf{T}}_{\ \alpha \beta }^{\gamma }.$ To generate exact and parametric
solutions in GR, we have to solve additional anholonomic constraints of type
(\ref{lccond}) or (\ref{lccond1}). By straightforward computations, we can
check that $\widehat{\mathbf{T}}_{\ \alpha \beta }^{\gamma }=0$ if (for
quasi-stationary configurations) 
\begin{eqnarray}
\ w_{i}^{\ast }(x^{i},y^{3}) &=&\mathbf{e}_{i}\ln \sqrt{|\
h_{3}(x^{i},y^{3})|},\mathbf{e}_{i}\ln \sqrt{|\ h_{4}(x^{i},y^{3})|}%
=0,\partial _{i}w_{j}=\partial _{j}w_{i}\mbox{ and }n_{i}^{\ast }=0;  \notag
\\
n_{k}(x^{i}) &=&0\mbox{ and }\partial _{i}n_{j}(x^{k})=\partial
_{j}n_{i}(x^{k}).  \label{zerot1}
\end{eqnarray}%
The solutions for such $w$- and $n$-functions depend on the class of vacuum
or non--vacuum metrics we are generating.

We consider two steps: If we prescribe a generating function $\Psi =\check{%
\Psi}(x^{i},y^{3})$ for which $[\partial _{i}(\ _{2}\check{\Psi})]^{\ast
}=\partial _{i}(\ _{2}\check{\Psi})^{\ast },$ we can solve the equations for 
$w_{j}$ (from (\ref{zerot1})). This is possible in explicit form if $\
^{v}\Upsilon =const,$ or if the effective source is expressed as a
functional $\ ^{v}\Upsilon (x^{i},y^{3})=\ \ ^{v}\Upsilon \lbrack \ _{2}%
\check{\Psi}].$ Non-explicit solutions for (\ref{zerot1}) can be also
physically important. At the next step, we can solve the third conditions $%
\partial _{i}w_{j}=\partial _{j}w_{i}$ if we chose a generating function $\ 
\check{A}=\check{A}(x^{k},y^{3})$ and define such N-connection coefficients
in the form 
\begin{equation*}
w_{i}(x^{i},y^{3})=\check{w}_{i}(x^{i},y^{3})=\partial _{i}\ \check{\Psi}/(%
\check{\Psi})^{\ast }=\partial _{i}\check{A}(x^{i},y^{3}).
\end{equation*}%
We also note that the equations for $n$-functions in (\ref{zerot1}) are
solved by any $n_{i}(x^{k})=\partial _{i}[\ ^{2}n(x^{k})].$ 

The above formulas allow us to write the quadratic element for
quasi-stationary solutions with $\widehat{\mathbf{T}}_{\ \alpha \beta
}^{\gamma }=0$ in a form similar to (\ref{qeltors}), but with a more special
class of generating data (labelled by "inverse hats"): 
\begin{eqnarray}
d\check{s}^{2} &=&\check{g}_{\alpha \beta }(x^{k},y^{3})du^{\alpha
}du^{\beta }=e^{\psi }[(dx^{1})^{2}+(dx^{2})^{2}]+\frac{[\check{\Psi}^{\ast
}]^{2}}{4(\ ^{v}\Upsilon \lbrack \check{\Psi}])^{2}\{h_{4}^{[0]}-\int dy^{3}[%
\check{\Psi}]^{\ast }/4\ ^{v}\Upsilon \lbrack \check{\Psi}]\}}\times
\label{qellc} \\
&&\{dy^{3}+[\partial _{i}(\check{A})]dx^{i}\}^{2}+\{h_{4}^{[0]}-\int dy^{3}%
\frac{[\check{\Psi}^{2}]^{\ast }}{4(\ ^{v}\Upsilon \lbrack \check{\Psi}])}%
\}\{dt+\partial _{i}[\ ^{2}n]dx^{i}\}^{2}.  \notag
\end{eqnarray}

\subsection{N-adapted propagators and locality of divergences}

\label{apendpropag}

Expanding at second order in perturbations the action (\ref{bfvpathint3}),
the propagators can be calculated similarly to \cite{bbd24} but for
N-adapted 3+1 decomposed perturbations (\ref{chicoeff}). For a Wick rotation
in the Fourier space $(k^{i^{\prime }},\omega ),$ we consider indices $%
i^{\prime },j^{\prime },...=1,2,3$ with space metrics of Euclidean signature
when the upper-lower cases are not distinguished. A respective N-adapted
frame formalism with abstract indices for orthonormalized frame
decompositions is applied.

Let us consider the projector $P_{i^{\prime }j^{\prime }}=\frac{1}{2}(\delta
_{i^{\prime }j^{\prime }}-\hat{k}_{i^{\prime }}\hat{k}_{j^{\prime }})$ and
introduce the values:%
\begin{eqnarray*}
M_{i^{\prime }j^{\prime }k^{\prime }l^{\prime }} &=&P_{i^{\prime }k^{\prime
}}P_{j^{\prime }l^{\prime }}+P_{i^{\prime }l^{\prime }}P_{j^{\prime
}k^{\prime }}-P_{i^{\prime }j^{\prime }}P_{k^{\prime }l^{\prime
}},Q_{i^{\prime }j^{\prime }k^{\prime }l^{\prime }}=\hat{k}_{i^{\prime }}%
\hat{k}_{k^{\prime }}P_{j^{\prime }l^{\prime }}+\hat{k}_{j^{\prime }}\hat{k}%
_{k^{\prime }}P_{i^{\prime }l^{\prime }}+\hat{k}_{i^{\prime }}\hat{k}%
_{l^{\prime }}P_{j^{\prime }k^{\prime }}+\hat{k}_{j^{\prime }}\hat{k}%
_{l^{\prime }}P_{i^{\prime }k^{\prime }}; \\
\ _{1}\mathcal{T} &\mathcal{=}&\frac{1}{\omega ^{2}+\beta _{3}k^{6}},\ _{2}%
\mathcal{T=}\frac{1}{\omega ^{2}+\widehat{\sigma }\nu k^{6}},\ _{3}\mathcal{%
T=}\frac{1}{\omega ^{2}-2\rho k^{6}},\ _{4}\mathcal{T=}\frac{1}{\omega
^{2}-4\rho \overline{\kappa }(1-\widehat{\lambda })k^{6}},
\end{eqnarray*}%
where $\nu =3\beta _{3}+8\beta _{3}-2(\alpha _{3})^{2}/\alpha _{4},$ $%
\widehat{\sigma }$ $=\widehat{\lambda }/(1-3\widehat{\lambda })$ and $%
\overline{\sigma }=(1-\widehat{\lambda })/(1-3\widehat{\lambda })$ are
chosen as in formulas (\ref{potz3}) and (\ref{chicoeff}). The
parametrization of such projectors is also chosen in a form to be compatible
with the results on ultraviolate divergences and renormalizability of
Lorentz-violating theories \cite{anselmi07,anselmi08,bbd24} and assuming
that infrared divergences have been regularized. Here we note that on
nonholonomic Lorentz manifolds, the effective anisotropic scaling is defined
by certain types of nonlinear off-diagonal interactions.

Using above formulas, the nonvanishing propagators are written in the form:%
\begin{eqnarray*}
\left\langle \overline{\mathcal{C}}_{i^{\prime }}\mathcal{C}_{j^{\prime
}}\right\rangle &=&-2P_{i^{\prime }j^{\prime }}\ _{2}\mathcal{T}-2\hat{k}%
_{i^{\prime }}\hat{k}_{j^{\prime }}\ _{4}\mathcal{T},\ \left\langle
p_{i^{\prime }j^{\prime }}p_{k^{\prime }l^{\prime }}\right\rangle =-\beta
_{3}M_{i^{\prime }j^{\prime }k^{\prime }l^{\prime }}k^{6}-2\nu P_{i^{\prime
}j^{\prime }}P_{j^{\prime }k^{\prime }}\ _{2}\mathcal{T}, \\
\left\langle h_{i^{\prime }j^{\prime }}h_{k^{\prime }l^{\prime
}}\right\rangle &=&4M_{i^{\prime }j^{\prime }k^{\prime }l^{\prime }}\ _{1}%
\mathcal{T}+8(\omega ^{2}+2\rho k^{6})Q_{i^{\prime }j^{\prime }k^{\prime
}l^{\prime }}(\ _{3}\mathcal{T)}^{2}+8(\widehat{\sigma }P_{i^{\prime
}j^{\prime }}P_{j^{\prime }k^{\prime }}+\overline{\sigma }\hat{k}_{k^{\prime
}}\hat{k}_{l^{\prime }}P_{i^{\prime }j^{\prime }}+\overline{\sigma }\hat{k}%
_{i^{\prime }}\hat{k}_{j^{\prime }}P_{k^{\prime }l^{\prime }})\ _{2}\mathcal{%
T} \\
&&+2\widehat{\lambda }\overline{\sigma }\ _{2}\mathcal{T}-4(1-\widehat{%
\lambda })^{-1}\hat{k}_{i^{\prime }}\hat{k}_{j^{\prime }}\hat{k}_{k^{\prime
}}\hat{k}_{l^{\prime }}[2\widehat{\lambda }\overline{\sigma }\ _{2}\mathcal{%
T+(}2\omega ^{2}\ _{2}\mathcal{T}-1\mathcal{)}\ _{4}\mathcal{T}],
\end{eqnarray*}%
\begin{eqnarray*}
\left\langle h_{i^{\prime }j^{\prime }}p_{k^{\prime }l^{\prime
}}\right\rangle &=&2\omega \lbrack M_{i^{\prime }j^{\prime }k^{\prime
}l^{\prime }}\ _{1}\mathcal{T}+2Q_{i^{\prime }j^{\prime }k^{\prime
}l^{\prime }}\ _{3}\mathcal{T}+2P_{i^{\prime }j^{\prime }}P_{j^{\prime
}k^{\prime }}\ _{2}\mathcal{T}+2\overline{\sigma }\hat{k}_{i^{\prime }}\hat{k%
}_{j^{\prime }}P_{k^{\prime }l^{\prime }}(\ _{2}\mathcal{T}-\ _{4}\mathcal{T}%
)+\hat{k}_{i^{\prime }}\hat{k}_{j^{\prime }}\hat{k}_{k^{\prime }}\hat{k}%
_{l^{\prime }}\ _{4}\mathcal{T}], \\
\left\langle h_{i^{\prime }j^{\prime }}n_{k^{\prime }}\right\rangle
&=&-16i\rho \omega k^{4}(P_{k^{\prime }i^{\prime }}\hat{k}_{j^{\prime }}(\
_{3}\mathcal{T)}^{2}+P_{k^{\prime }j^{\prime }}\hat{k}_{i^{\prime }}(\ _{3}%
\mathcal{T)}^{2}+\overline{\kappa }\hat{k}_{i^{\prime }}\hat{k}_{j^{\prime }}%
\hat{k}_{k^{\prime }}(\ _{4}\mathcal{T)}^{2}), \\
\left\langle p_{k^{\prime }l^{\prime }}n_{i^{\prime }}\right\rangle
&=&-4i\rho k^{4}(P_{i^{\prime }k^{\prime }}\hat{k}_{l^{\prime
}}+P_{i^{\prime }l^{\prime }}\hat{k}_{k^{\prime }})\ _{3}\mathcal{T}+4i\rho 
\overline{\kappa }(2\widehat{\lambda }P_{k^{\prime }l^{\prime }}-(1-\widehat{%
\lambda })\hat{k}_{k^{\prime }}\hat{k}_{l^{\prime }})\hat{k}_{i^{\prime
}}k^{4}\ _{4}\mathcal{T}, \\
\left\langle n_{i^{\prime }}n_{j^{\prime }}\right\rangle &=&-4\rho (\omega
^{2}+2\rho k^{6})k^{4}P_{i^{\prime }j^{\prime }}\ (\ _{3}\mathcal{T)}%
^{2}+4\rho \overline{\kappa }(\omega ^{2}+4\rho \overline{\kappa }k^{6}(1-%
\widehat{\lambda })\hat{k}_{k^{\prime }}\hat{k}_{l^{\prime }})k^{4}\hat{k}%
_{i^{\prime }}\hat{k}_{j^{\prime }}(\ _{4}\mathcal{T)}^{2},
\end{eqnarray*}%
\begin{eqnarray*}
\left\langle nh_{i^{\prime }j^{\prime }}\right\rangle &=&-\frac{4\alpha _{3}%
}{\alpha _{4}}(\widehat{\sigma }P_{i^{\prime }j^{\prime }}+\overline{\sigma }%
\hat{k}_{i^{\prime }}\hat{k}_{j^{\prime }})\ _{2}\mathcal{T},\left\langle
np_{i^{\prime }j^{\prime }}\right\rangle =-\frac{2\alpha _{3}}{\alpha _{4}}%
\omega P_{i^{\prime }j^{\prime }}\ _{2}\mathcal{T},\left\langle \pi
_{k^{\prime }}n_{j^{\prime }}\right\rangle =2\omega (P_{k^{\prime }j^{\prime
}}\ _{2}\mathcal{T+}\hat{k}_{k^{\prime }}\hat{k}_{j^{\prime }}\ _{4}\mathcal{%
T}), \\
\left\langle nn\right\rangle &=&\frac{2(\alpha _{3})^{2}}{(\alpha _{4})^{2}}%
\overline{\sigma },\left\langle \pi _{k^{\prime }}h_{i^{\prime }j^{\prime
}}\right\rangle =-4i(P_{i^{\prime }j^{\prime }}\hat{k}_{k^{\prime }}\ _{3}%
\mathcal{T}+P_{i^{\prime }k^{\prime }}\hat{k}_{j^{\prime }}\ _{3}\mathcal{T}+%
\hat{k}_{k^{\prime }}\hat{k}_{i^{\prime }}\hat{k}_{j^{\prime }}\ _{4}%
\mathcal{T}),\left\langle \mathcal{AA}\right\rangle =\left\langle \mathcal{A}%
n\right\rangle =\left\langle \overline{\acute{\eta}}\acute{\eta}%
\right\rangle =\frac{-k^{-6}}{\alpha _{4}}.
\end{eqnarray*}

We note that in the action (\ref{bfvpathint3}) the time derivatives arise in
terms that are of second order in prturbations and (as a consequence)
vertices do not depend on $\omega .$ This means that for integration on $%
\omega $ we need to consider only above presented propagators. Such details
are presented in \cite{bbd22,bbd23,bbd24}. At the perturbative level, our
N-adapted approach to quantizing off-diagonal configurations in GR
reproduces those results but using a general non-flat background with
decompositions defined by (\ref{chicoeff}).

\end{document}